\documentclass[%
    aps,
    prd,
    letterpaper,
    superscriptaddress,
    preprintnumbers,
    floatfix,
    twocolumn,
    nofootinbib,
    showpacs,
    hyphens
]{revtex4-1}  

\usepackage[normalem]{ulem}
\usepackage{subfigure}
\usepackage{graphicx}
\usepackage{dcolumn}
\usepackage{bm}
\usepackage{hyperref}
\usepackage{xcolor}
\usepackage{amsmath,amsfonts,amssymb,amsthm}
\newcommand{\ket}[1]{\left| #1 \right>} 
\newcommand{\bra}[1]{\left< #1 \right|} 
\newcommand{\braket}[2]{\left< #1 \vphantom{#2} \right|
 \left. #2 \vphantom{#1} \right>} 
\newcommand{\mbraket}[3]{\left< #1 \vphantom{#2#3} \right|
 #2 \left| #3 \vphantom{#1#2} \right>} 

\usepackage{slashed}

\usepackage{duckuments}

\newcommand{\bs}{\boldsymbol}
\newcommand{\eul}{\gamma_{\text{E}}}

\newcommand{\Ham}{H}
\newcommand{\Vop}{V}
\newcommand{\Top}{T}

\newcommand{\bbbb}{bb\overline{b}\overline{b}}
\newcommand{\bbtt}{bb\overline{t}\overline{t}}
\newcommand{\cccc}{cc\overline{c}\overline{c}}

\DeclareMathOperator{\Ps2}{\text{Ps}_2}
\DeclareMathOperator{\alphaem}{\alpha_{\text{em}}}
\DeclareMathOperator{\alphaemSq}{\alpha_{\text{em}}^2}

\begin{document}

\preprint{FERMILAB-PUB-23-660-T}

\title{Tetraquarks made of sufficiently unequal-mass heavy quarks are bound in QCD}
\author{Beno\^{i}t Assi}
\affiliation{Fermi National Accelerator Laboratory, Batavia, IL, 60510}
\author{Michael~L.~Wagman}
\affiliation{Fermi National Accelerator Laboratory, Batavia, IL, 60510}

\date{\today}

\begin{abstract}
Tetraquarks, bound states composed of two quarks and two antiquarks, have been the subject of intense study but are challenging to understand from first principles. We apply variational and Green's function Monte Carlo methods to compute tetraquark ground-state energies in potential nonrelativistic QCD using a wide range of color and spatial wavefunctions. We find no evidence for bound tetraquarks composed of equal-mass quarks and antiquarks.
Conversely, we find clear evidence for the existence of bound tetraquarks for sufficiently unequal quark/antiquark mass ratios at all overall mass scales where our effective theory results are applicable. We predict the critical mass ratios for bound state formation and study tetraquark bound states' spatial and color structure at leading order and next-to-leading order in potential nonrelativistic QCD.
\end{abstract}

\maketitle

\section{Introduction}
\label{sec:intro}
Tetraquarks were first proposed decades ago to explain the structure of the $a_0(980)$ and $f_0(980)$ resonances~\cite{Jaffe:1976ig,Jaffe:1976ih}. Recent experiments hint at tetraquark candidates among the exotic XYZ states, which are hypothesized to be composed of two heavy $b$ or $c$ quarks and two light quarks~\cite{Brambilla:2019esw,Chen:2022asf,ParticleDataGroup:2022pth}.
Several frameworks for describing tetraquarks have been proposed, but modeling their dynamics is complicated because it involves both short- and long-distance quantum chromodynamics (QCD)~\cite{Drenska:2010kg,Brambilla:2010cs,Brambilla:2014jmp,Esposito:2014rxa,Lebed:2016hpi,Chen:2016qju,Liu:2019zoy}.
Lattice QCD studies of XYZ states are challenging due to their position in the spectrum and proximity to multi-hadron thresholds and are being actively investigated~\cite{Bicudo:2012qt,Brown:2012tm,Bicudo:2015vta,Francis:2016hui,Francis:2018jyb,Junnarkar:2018twb,Leskovec:2019ioa,Hudspith:2020tdf,Mohanta:2020eed,Bicudo:2021qxj,Padmanath:2022cvl,Meinel:2022lzo,Bicudo:2022cqi,Lyu:2023xro,Hudspith:2023loy,Aoki:2023nzp,Padmanath:2023rdu,Alexandrou:2024iwi}.

With bound states comprised of heavy quarks, QCD dynamics are more straightforward due to the large hierarchy between the quark mass $m_Q$ and the Landau-pole scale $\Lambda_{\rm QCD}$ and can be studied using effective field theory (EFT)~\cite{Bodwin:1994jh,Caswell:1985ui,Pineda:1997bj,Pineda:1998kj,Brambilla:1999xf}.
Quark velocities are small in such systems, $v \ll 1$, leading to a clear hierarchy of scales: \(m_Q\gg p\sim m_Qv\gg E\sim m_Q v^2\)~\cite{Caswell:1985ui}.
Integrating out the hard scale \(m_Q\) leads to nonrelativistic QCD (NRQCD)~\cite{Bodwin:1994jh,Caswell:1985ui,Pineda:1997bj,Pineda:1998kj}, while further integrating out the soft scale \(p_Q\sim m_Q v\) leads to potential NRQCD (pNRQCD)~\cite{Brambilla:1999xf}.
This soft scale sets the typical bound state size, which is analogous to the Bohr radius of the hydrogen atom. In the weak coupling regime of pNRQCD~\cite{Pineda:2011dg}, dynamics at the soft scale are incorporated by solving the time-independent Schr{\"o}dinger equation with a potential that incorporates all NRQCD effects that are enhanced for small $p / m_Q$ and must be treated nonperturbatively.

Fully-heavy tetraquark states provide a theoretically simple starting point for understanding exotic states directly from QCD.
A variety of phenomenological potential models have been proposed and studied for fully-heavy tetraquarks~\cite{Ader:1981db,Richard:1992cb,Bai:2016int,Berezhnoy:2011xn,Carlson:1987hh,Chen:2016jxd,Heller:1986bt,Karliner:2016zzc,Wang:2018poa,Wu:2016vtq}, but these potential models cannot be reliably connected to QCD.
Systematically improvable calculations rooted in QCD for fully heavy tetraquarks have been studied more recently in lattice NRQCD~\cite{Hughes:2017xie},
which did not find evidence for the existence of bound fully-heavy tetraquarks, and single-gluon exchange models~\cite{Czarnecki:2017vco,Anwar:2017toa}, which suggest different conclusions for the existence of equal-mass fully-heavy tetraquarks.
Unequal-mass fully-heavy tetraquarks have also been studied phenomenologically and are suggested to be bound for sufficiently unequal masses based on physical arguments in simple diquark models~\cite{Ader:1981db,Vijande:2007ix}.
A one-gluon-exchange model predicts the existence of a critical mass ratio $0.15$ that is sufficiently unequal to permit bound fully-heavy tetraquarks~\cite{Czarnecki:2017vco}.

Studies of heavy quarkonium have shown that pNRQCD can accurately describe properties of fully heavy quark and antiquark bound states, including masses and decay widths~\cite{Brambilla:1999xj,Kniehl:2002br,Brambilla:1999xf,Pineda:2011dg,Pineda:1997hz}.
The potentials needed to describe more complex systems such as baryons have been studied more recently~\cite{Brambilla:2005yk,Brambilla:2009cd,Brambilla:2013vx,Assi:2023cfo}.
Variational methods have subsequently been used to bound fully-heavy baryon masses~\cite{Jia:2006gw,Llanes-Estrada:2011gwu,Assi:2023cfo}, and Green's function Monte Carlo (GFMC) methods used to solve quantum many-body problems in nuclear and condensed matter physics~\cite{Carlson:2014vla,Yan_2017,Gandolfi:2020pbj} were further used to compute baryon masses in pNRQCD in Ref.~\cite{Assi:2023cfo}.
This work applies the same quantum Monte Carlo (QMC) methods to study fully heavy tetraquarks in pNRQCD.
In particular, we study fully heavy tetraquarks with equal and unequal quark/antiquark masses at both leading-order (LO) and next-to-leading-order (NLO) in pNRQCD. We scan across the entire Hilbert space of color wavefunctions and study a variety of spatial wavefunctions using GFMC methods to isolate the ground state via imaginary-time evolution and search for the presence of tetraquark bound states.

We further validate our computational methods by examining di-positronium (\(\mathrm{Ps}_2\)) molecules in potential nonrelativistic quantum electrodynamics (pNRQED) and their unequal mass counterparts, which approach the H$_2$ molecule in the limit of asymptotically large mass ratios.
Both \(\mathrm{Ps}_2\) and H$_2$ molecules have been studied extensively and provide benchmarks for us to evaluate the effectiveness of the variational and GFMC methods used here. 
We perform a comparative study with foundational works on bound states in QED. Key references include the pioneering efforts by Hylleraas and others on the helium atom and hydrogen molecules~\cite{Hylleraas:1928, Hylleraas:1930, Hylleraas:1931, James:1933}, as well as early applications of GFMC to molecular systems in QED~\cite{Anderson:1975,Mentch:1981,Lee:1983}. We further adopt the wavefunction employed in these works as trial states for the GFMC evolution and find that our results match the state-of-the-art bounds obtained from purely variational techniques.

This paper is organized as follows: our framework and methods are introduced in Section~\ref{sec:methods}, outlining the theoretical background and computational approaches. Following this, in Section~\ref{sec:trial}, we explore the assortment of trial wavefunctions tested, highlighting their construction and importance in examining tetraquark dynamics. We then validate our methods and present results in pNRQED in Section~\ref{sec:QEDvalid}. The findings regarding tetraquark binding energies are presented in Section~\ref{sec:QCDresults}, and we then conclude in Section~\ref{sec:conclusion}.


\section{Framework and Methods}
\label{sec:methods}
\subsection{pNRQCD}
\label{sec:pnrqcd}

The pNRQCD Hamiltonian is given by
\begin{equation}
  \Ham = \Top + \Vop^{\psi\chi} + \Vop^{\psi\psi} + \Vop^{\chi\chi} + \ldots,
\end{equation}
where $\Top$ is the nonrelativistic kinetic energy operator defined in terms of heavy quark fields $\psi(\bs{r})$ and antiquark fields $\chi(\bs{r})$ by 
\begin{equation}
\Top = -\int d^3 \bs{r}\ \left[\psi_i^\dagger(\bs{r}) \frac{\nabla^2}{2m_Q} \psi_i(\bs{r}) + \chi_i^\dagger(\bs{r}) \frac{\nabla^2}{2m_Q} \chi_i(\bs{r})\right],
\end{equation}
where $i$ is a color index, and two-component spinor indices are suppressed.
The potential terms in $H$ are computed in a joint expansion in powers of $1/m_Q$, and the strong coupling constant $\alpha_s$ evaluated at scales proportional to $m_Q$.
Here, we include only the leading terms in $1/m_Q$ and refer to terms in the $\alpha_s$ expansion as leading order (LO), next-to-leading-order (NLO), etc.
The quark-antiquark potential operator $\Vop^{\psi\chi}$ is given by 
\begin{equation}
    \begin{split}
        V^{\psi\chi} &= \int d^3\bs{r}_1d^3\bs{r}_2\,  \psi^{\dagger}_i(\bs{r}_1)\chi_j(\bs{r}_2) \chi^{\dagger}_k(\bs{r}_2)\psi_l(\bs{r}_1) \\
       &\hspace{10pt} \times \left[ \frac{1}{3}\delta_{ij}\delta_{kl} V^{\psi\chi}_{\mathbf{1}}(\bs{r}_{12})  + 2 T^a_{ji} T^a_{kl} V^{\psi\chi}_{\text{Adj}}(\bs{r}_{12}) \right], \label{eq:Lpsichi}
    \end{split}
\end{equation}
where $V^{\psi\chi}_{\mathbf{1}}(\bs{r}_{12})$ and  $V^{\psi\chi}_{\text{Adj}}(\bs{r}_{12})$ are color-singlet and color-adjoint potentials respectively that 
are proportional to $1/r$ at LO and 
are presented at NNLO in Refs.~\cite{Kniehl:2002br,Anzai:2013tja,Assi:2023cfo}; see Appendix~\ref{app:pot} for a summary of the LO and NLO potentials used here. The $T^a$ are $\mathfrak{su}(3)$ generators normalized as $\text{Tr}[T^a T^b] = \frac{1}{2} \delta^{ab}$.
The quark-quark potential is given by 
      \begin{align}\label{eq:Lpsipsi}
        &V^{\psi\psi} = \int d^3\bs{r}_1d^3\bs{r}_2\,  \psi^{\dagger}_i(\bs{r}_1)\psi^{\dagger}_j(\bs{r}_2) \psi_k(\bs{r}_2)\psi_l(\bs{r}_1) \\\nonumber
      &\times \left[ \frac{1}{4} \epsilon_{ijo}\epsilon_{klo}  V^{\psi\psi}_{\text{A}}(\bs{r}_{12})  +  \frac{1}{4}  \left(  \delta_{il}\delta_{jk}+\delta_{jl}\delta_{ik}\right)   V^{\psi\psi}_{\text{S}}(\bs{r}_{12}) \right],
    \end{align}   
where $V^{\psi\psi}_{\text{A}}(\bs{r}_{12})$ and $V^{\psi\psi}_{\text{S}}(\bs{r}_{12})$ involve color-antisymmetric and color-symmetric products of quark fields, respectively,
and are presented at NNLO in Ref.~\cite{Assi:2023cfo};  see Appendix~\ref{app:pot} for a summary of the LO and NLO potentials used here. 
The antiquark-antiquark potential is identical by charge conjugation, and $V^{\chi\chi}$ is obtained from Eq.~\eqref{eq:Lpsipsi} via the replacement $\psi \rightarrow \chi$.
Three- and four-quark potentials enter $\Ham$ at NNLO~\cite{Brambilla:2009cd,Assi:2023cfo}.
Effects from ultra-soft modes lead to the appearance of additional non-potential terms in $\Ham$, but these do not enter until N${}^3$LO~\cite{Brambilla:1999xj,Pineda:1997bj,Pineda:1998kn,Kniehl:1999ud}.


Heavy quarkonium states only involve the color-singlet $\psi\chi$ potential, which is attractive and leads to a hydrogen-like spectrum of $Q\overline{Q}$ bound states.
Triply-heavy baryon states only involve the color-antisymmetric $\psi\psi$ potential, which is attractive and leads to the appearance of $QQQ$ bound states.
Tetraquark states involve these two attractive potentials and also the color-adjoint $\psi\chi$ and color-symmetric $\psi\psi$ / $\chi\chi$ potentials, both of which are repulsive.
For example, the action of the quark-antiquark potential on a heavy tetraquark state $\ket{\psi_{i}(\bm{r}_1),\chi^\dagger_{j}(\bm{r}_{2}) \psi_{k}(\bm{r}_3),\chi^\dagger_{l}(\bm{r}_{4}) } \delta_{ij}\delta_{kl}$ is given by 
\begin{equation}
\begin{split}
  &\Vop^{\psi\chi} \ket{\psi_{i}(\bm{r}_1),\chi^\dagger_{j}(\bm{r}_{2}) \psi_{k}(\bm{r}_3),\chi^\dagger_{l}(\bm{r}_{4}) } \delta_{ij}\delta_{kl}  \\
    &= \ket{\psi_{i}(\bm{r}_1),\chi^\dagger_{j}(\bm{r}_{2}) \psi_{k}(\bm{r}_3),\chi^\dagger_{l}(\bm{r}_{4}) } \\
    &\hspace{10pt} \times \left\lbrace \delta_{ij} \delta_{kl} \left[ V^{\psi\chi}_{\mathbf{1}}(\bs{r}_{12}) + V^{\psi\chi}_{\mathbf{1}}(\bs{r}_{34}) \right] \right. \\
    &\hspace{30pt} + \frac{1}{3}\delta_{il}\delta_{jk} \left[ V^{\psi\chi}_{\mathbf{1}}(\bs{r}_{14}) + V^{\psi\chi}_{\mathbf{1}}(\bs{r}_{23}) \right] \\
    &\hspace{30pt} \left. + 2 T^a_{il} T^a_{jk} \left[ V^{\psi\chi}_{\text{Adj}}(\bs{r}_{14}) + V^{\psi\chi}_{\text{Adj}}(\bs{r}_{23}) \right] \right\rbrace 
\end{split} \label{eq:tet}
\end{equation}
The contribution proportional to $\delta_{ij} \delta_{kl}$ represents a product of the potentials appearing for two quarkonium states. In contrast, the other contributions describe interactions between the two quark-antiquark pairs analogous to atomic van der Waals forces.
This work studies whether the combination of attractive and repulsive pNRQCD van der Waals interactions arising in Eq.~\eqref{eq:tet} plus those arising from $V^{\psi\psi}$ and $V^{\chi\chi}$ lead to the appearance of fully-heavy tetraquark bound states.

The eigenvalues of $\Ham$, denoted $\Delta E$ below, are the masses of pNRQCD energy eigenstates minus the rest masses $m_Q$ of their constituent heavy quarks/antiquarks.
The definition of $m_Q$ and choice of renormalization scheme and scale $\mu$ will modify $\Delta E(m_Q,\mu)$ such that the masses of pNRQCD energy eigenstates are scheme- and scale-independent up to perturbative truncation effects.
We use the quark mass scheme and results for $m_b$ and $m_c$ obtained in Ref.~\cite{Assi:2023cfo} in which the ``pole masses'' $m_Q$ appearing in $\Ham$ are obtained by solving
\begin{equation}
    M_{Q\overline{Q}} = 2 m_Q + \Delta E_{Q\overline{Q}},
\end{equation}
using experimental results for $M_{Q\overline{Q}}$,
where  $\Delta E_{Q\overline{Q}}(m_Q, \mu)$ is computed using $\mu = \mu_p(m_Q)$ obtained numerically as the solution to
\begin{equation}
     \mu_p(m_Q)\equiv 4 \alpha_s(\mu_p(m_Q)) m_Q. \label{eq:mup}
\end{equation}

\subsection{Quantum Monte Carlo}
\label{sec:qmc}

For an arbitrary trial state $\ket{\Psi_T(\bs{\omega})}$ with parameters \( \bs{\omega} = (\omega_1,\ldots) \), the variational principle dictates that $\Delta E \leq \mbraket{\Psi_T(\bs{\omega})}{\Ham}{\Psi_T(\bs{\omega})}$.
Numerical minimization of $\mbraket{\Psi_T(\bs{\omega})}{\Ham}{\Psi_T(\bs{\omega})}$ can therefore be used to determine the best ground-state approximation within a parameterized family of wavefunctions~\cite{toulouse2007optimization}.
We evaluate these matrix elements using 
wavefunctions $\Psi_T(\bs{R};\bs{\omega}) \equiv \braket{\bs{R}}{\Psi_T(\bs{\omega})}$ that depend on spatial coordinates $\bs{R} \equiv (\bs{r}_1,\ldots,\bs{r}_{N_Q})$,
\begin{equation}
  \mbraket{\Psi_T}{\Ham}{\Psi_T}  = \frac{\int d^3\bs{R}\, \Psi_T(\bs{R})^*H(\bs{R}) \Psi_T(\bs{R})}{\int d^3\bs{R}\, |\Psi_T(\bs{R})|^2 }, \label{eq:me}
\end{equation}
where \( \mbraket{\bs{R}}{\Ham}{\bs{R}'} = H(\bs{R}) \delta(\bs{R}-\bs{R}') \), states are assumed to be normalized as $\braket{\Psi_T}{\Psi_T} = 1$, and dependence on $\bs{\omega}$ is suppressed for brevity.
We  use Monte Carlo methods to stochastically approximate Eq.~\eqref{eq:me} by sampling $\bs{R}$ from a probability distribution proportional to $|\Psi_T(\bs{R})|^2$ and then obtaining $\mbraket{\Psi_T}{H}{\Psi_T}$ as the sample mean of $\Psi_T(\bs{R})^*H(\bs{R}) \Psi_T(\bs{R})$ for this ensemble.

The accuracy of ground-state determinations using this variational Monte Carlo (VMC) approach is limited by the expressivity of a given family of trial wavefunctions, and we therefore adopt the standard QMC strategy of using optimal trial wavefunctions obtained using VMC as the foundation for subsequent GFMC calculation~\cite{Carlson:2014vla,Gandolfi:2020pbj}. GFMC employs imaginary-time \(\tau\) evolution (analogous to lattice QCD calculations) to dampen the excited-state components of \(\ket{\Psi_T}\) and formally allows the ground-state for a set of quantum numbers to be obtained from any trial wavefunction with the same quantum numbers as $\lim_{\tau \to \infty} e^{-\Ham \tau}\ket{\Psi_T}$.
The imaginary-time evolution operator $e^{-\Ham\tau}$ cannot be straightforwardly constructed for arbitrary $\tau$, but it can be approximated by splitting $\tau$ into $N_\tau$ intervals of size \(\delta\tau=\tau/N_\tau\) for \(N_\tau \gg 1\) and using the Lie-Trotter product formula:
\begin{align}
    \mbraket{\bs{R}}{e^{-\Ham\tau}}{\Psi_T} \approx {}&\int\prod_{i=0}^{N_{\tau}-1}d\bs{R}_i\bra{\bs{R}_{N_{\tau}}}e^{-H\delta\tau}\ket{\bs{R}_{N-1}}\times \cdots \nonumber\\{}&
    \times \bra{\bs{R}_1}e^{-H\delta\tau}\ket{\bs{R}_{0}}\braket{\bs{R}_0}{\Psi_T},
\end{align}
with equality obtained in the $N_\tau \rightarrow \infty$ limit.
The Green's functions $G_{\delta\tau}(\bs{R},\bs{R}' ) \equiv \mbraket{\bs{R}}{e^{-\Ham\delta\tau}}{\bs{R}'}$ are approximated with the Trotter-Suzuki expansion 
\begin{equation}
\begin{split}
    &G_{\delta\tau}(\bs{R},\bs{R}' ) \equiv \mbraket{\bs{R}}{e^{-\Ham\delta\tau}}{\bs{R}'} \\
    &\hspace{20pt} \approx e^{-V(\bs{R})\delta \tau/2} \mbraket{\bs{R}}{e^{-T \delta\tau}}{\bs{R}'} e^{-V(\bs{R}')\delta \tau/2},
    \end{split}\label{eq:VK}
\end{equation}
where the kinetic piece $\mbraket{\bs{R}}{e^{-T \delta\tau}}{\bs{R}'}$ is proportional to a Gaussian $e^{-(\bs{R}-\bs{R}')^2/\lambda^2}$ with \(\lambda^2 = 2 \delta \tau / m_Q\)~\cite{Carlson:2014vla,Gandolfi:2020pbj}.
Therefore, GFMC evolution for each Trotter step can be achieved by sampling $\bs{R} - \bs{R}'$ from a Gaussian distribution and computing the action of the potential on coordinate-space states.
We further employ strategies to improve the precision of GFMC by randomly choosing between updates with $\pm(\bs{R} - \bs{R}')$ as detailed in Ref.~\cite{Gandolfi:2020pbj}.
The kinetic piece is diagonal in color, while for a state built from $N_Q$ heavy quark/antiquark fields, the potential is represented as a $3N_Q \times 3N_Q$ color matrix, and we approximate the matrix exponentials in Eq.~\eqref{eq:VK} using a second-order Taylor expansion.

Calculations of Hamiltonian matrix elements after imaginary-time evolution provide effective energies
\begin{equation}
  \Delta E(\tau) = \mbraket{\Psi_T}{H e^{-H \tau}}{\Psi_T}.
\end{equation}
These effective energies approach the ground-state energies $\Delta E$ in the $\tau \rightarrow \infty$ limit and include additional exponentially-suppressed contributions from excited states at finite $\tau$ as summarized in Ref.~\cite{Assi:2023cfo}.
The contributions are necessarily positive, and therefore $\Delta E(\tau)$ provides a variational bound that must approach the ground-state energy from above.

\section{Trial wavefunctions}
\label{sec:trial}

\subsection{Quarkonium trial wavefunctions}

This work uses Coulomb ground-state wavefunctions as trial wavefunctions for quarkonium states.
Annihilation operators for these trial states are defined by
\begin{equation}
    \mathcal{O}^{Q\overline{Q}}(\bs{R}; a) = \psi_i(\bs{r}_1) \chi^\dagger_j(\bs{r}_2) \delta_{ij} \Psi_T(\bs{R},a), \label{eq:meson_op}
\end{equation}
with
\begin{equation}
\Psi_T(\bs{R};a) \propto ^{-|\bs{r}_{12}|/a}, \label{eq:meson_wvfn}
\end{equation}
where the ``Bohr radius'' $a$ is a variational parameter that should be chosen to minimize the expectation value of the Hamiltonian in order to approximate the ground state.
At LO the pNRQCD color-singlet potential is simply a Coulomb potential $V^{\psi\chi}_{\mathbf{1}}(|\bs{r}|,\mu) = C_F \alpha_s(\mu)/|\bs{r}|$ where $C_F = 4/3$, and therefore the exact ground-state wavefunction is obtained using $a = 2/(C_F \alpha_s(\mu) m_Q)$.
For QED applications exact ground-state wavefunctions are simply obtained by replacing $C_F \alpha_s(\mu)$ with $\alphaem$ and omitting the color indices in Eq.~\eqref{eq:meson_op}.

Beyond LO in pNRQCD it is convenient to describe the potential using a (distance-dependent) effective coupling
\begin{equation}
\alpha_V(|\bs{r}|,\mu) \equiv - \frac{|\bs{r}| }{C_F }V^{\psi\chi}_{\mathbf{1}}(|\bs{r}|,\mu),
\label{eq:alphav}
\end{equation}
that is equal to $\alpha_s(\mu)$ at LO.
A physically-motivated choice for the Bohr radius is therefore to replace $\alpha_s(\mu)$ with $\alpha_V(|\bs{r}|,\mu)$ in the LO Bohr radius formula with $|\bs{r}|$ taken to be the average inter-particle distance.
However, because the average inter-particle distance is not known \emph{a priori}, it is convenient to parameterize instead the $|\bs{r}|$ dependence of $\alpha_V(|\bs{r}|,\mu)$ by the size of the logarithmic terms $L_\mu \equiv \ln( |\bs{r}| \mu e^{\gamma_E})$ that give rise to this dependence.
Such a parameterization is achieved by defining~\cite{Assi:2023cfo}
\begin{equation}
  a(L_{\mu}) \equiv \frac{2}{\alpha_V(|\bs{r}|,\mu = e^{L_{\mu}-\gamma_E}/|\bs{r}|) C_F m_Q}.
  \label{eq:aL}
\end{equation}
The VMC and GFMC results of Ref.~\cite{Assi:2023cfo} demonstrate that $L_\mu = 0$ is approximately optimal for a wide range of masses (using the scale choice $\mu = 4\alpha_s(\mu) m_Q$) and that
$\Psi_T^{Q_i\overline{Q}_j}(\bs{R};a(L_\mu = 0))$ provides a suitable trial wavefunction for heavy quarkonium at NLO and NNLO.
For unequal-mass systems with masses $m_1$ and $m_2$, a suitable trial wavefunction is obtained analogously by replacing $m_Q$ in Eq.~\eqref{eq:aL} with
\begin{equation}
    m_{12} \equiv \frac{2 m_1 m_2}{m_1 + m_2}, \label{eq:m12}
\end{equation}
With this modification, $L_\mu = 0$ is again approximately optimal in the unequal-mass case~\cite{Assi:2023cfo}.
This choice will be used for the quarkonium calculations in this work and will further inform the construction of the tetraquark trial wavefunction.

\subsection{Hylleraas molecular trial wavefunctions}

Variational studies by Hylleraas beginning in the 1920s used simple trial wavefunctions to demonstrate the existence of molecular bound states such as the hydrogen ion H$_2^-$ and hydrogen molecule H$_2$~\cite{Hylleraas:1928,Hylleraas:1930,Hylleraas:1931}.
These molecular trial wavefunctions are constructed from products of Coulomb ground-state wavefunctions with smaller Bohr radii describing correlations between the constituents of each atom and larger Bohr radii describing weaker correlations between the constituents of different atoms.
These early studies also included polynomial functions of coordinate separations to increase expressivity and obtain tighter variational bounds.
Here, we consider only the simplest versions of Hylleraas molecular trial wavefunctions built from products of exponentials and rely on GFMC imaginary-time evolution to converge to the true ground-state energy as pioneered for QED applications in Refs.~\cite{Anderson:1975,Mentch:1981,Lee:1983}.

The operator describing a system of two positive charges, denoted $\mu^+$, and two negative charges, denoted $e^-$, with this trial wavefunction takes the form
\begin{equation}
\begin{split}
    \mathcal{O}^{\mu^+ \mu^+ e^- e^- }_{H}(\bs{R}; a, b, c) &= \mu^+(\bs{r}_1) \mu^+(\bs{r}_2) e^-(\bs{r}_3) e^-(\bs{r}_4)\\
    &\hspace{10pt} \times \Psi_H(\bs{R}; a, b, c),
    \end{split}
\end{equation}
where the Hylleraas molecular trial wavefunction is defined by
\begin{equation}
\begin{split}
    \Psi_H(\bs{R}; a, b, c) &\propto  e^{-|\bs{r}_{13}|/a} e^{-|\bs{r}_{24}|/a} e^{-|\bs{r}_{14}|/b} e^{-|\bs{r}_{23}|/b}\\
    &\hspace{10pt} \times   e^{-|\bs{r}_{12}|/c} e^{-|\bs{r}_{34}|/c},
    \label{eq:tetra_wvfn_H}
\end{split}
\end{equation}
with $a,b,c$ variational parameters.
In particular, a molecular configuration in which $\mu^+(\bs{r}_1) e^-(\bs{r}_3)$ and $\mu^+(\bs{r}_2)e^-(\bs{r}_4)$ are chosen to be the strongly-correlated atomic constituents is obtained by choosing $a$ on the order of atomic Bohr radius and taking $b \gg a$ to describe inter-atomic correlations within the molecule.
The $c$ parameter describes inter-atomic correlations between same-sign charges that experience repulsive interactions.
Physical intuition suggests these should be weaker than correlations between opposite-sign charges, so $c \gg b$. Although these correlations were not included in early applications of Hylleraas molecular trial wavefunctions~\cite{James:1933,Hylleraas:1947zza}, they have been shown to improve the convergence of subsequent GFMC calculations~\cite{Lee:1983}.
Hylleraas molecular trial wavefunctions for tetraquarks can be defined analogously, but the color state of the system must also be incorporated; this is deferred to Sec.~\ref{subsec:potentials} below.

\subsection{Born-Oppenheimer trial wavefunctions}


For four-particle systems with extremely unequal masses such as H$_2$, the dynamics of heavy and light particles can be approximately factorized, and experimental binding-energy results can be reproduced by calculations using the Born-Oppenheimer approximation~\cite{Born:1927rpw} in which the protons are treated as static charges with a fixed separation~\cite{Hylleraas:1928,Hylleraas:1930,Hylleraas:1931,James:1933}.
Although the Born-Oppenheimer approximation will not be used below, the success of these approximate calculations motivates the introduction of trial wavefunctions that incorporate these physical features.
Operators describing $\mu^+ \mu^+ e^- e^-$ systems with such Born-Oppenheimer trial wavefunctions are denoted
\begin{equation}
\begin{split}
    \mathcal{O}^{\mu^+ \mu^+ e^- e^- }_{BO}(\bs{R}; a, b, R_0) &= \mu^+(\bs{r}_1) \mu^+(\bs{r}_2) e^-(\bs{r}_3) e^-(\bs{r}_4)\\
    &\hspace{10pt} \times \Psi_{BO}(\bs{R}; a, b, R_0),
    \end{split}
\end{equation}
with wavefunctions
\begin{equation}
\begin{split}
    \Psi_{BO}(\bs{R}; a, b, R_0) &\propto \delta^3(\bs{r}_{12} - R_0 \hat{e}_z ) e^{-|\bs{r}_{13}|/a} e^{-|\bs{r}_{24}|/a} \\
    &\hspace{10pt} \times  e^{-|\bs{r}_{14}|/b} e^{-|\bs{r}_{23}|/b},
    \label{eq:tetra_wvfn_BO}
\end{split}
\end{equation}
that include exponential correlations between opposite-charged particles analogous to Eq.~\eqref{eq:tetra_wvfn_H} but replace the correlations betwen same-sign particles with a $\delta$-function that fixes the separation between the two positive charges ($m_{\mu} \geq m_e$ will be assumed below; the opposite case is related by charge conjugation).
In Eq.~\eqref{eq:tetra_wvfn_BO}, $R_0$ is a variational parameter specifying this separation.

Note that the constraint $\bs{r}_{12} = R_0 \hat{e}_z$ is not preserved by the action of the Hamiltonian when $m_{\mu}$ is finite, and therefore GFMC calculations using this Born-Oppenheimer trial wavefunction will not be limited to configurations satisfying this constraint.
The use of $\Psi_{BO}(\bs{R}; a, b, R_0)$ as a trial wavefunction is therefore distinct from the imposition of the Born-Oppenheimer approximation to the Hamiltonian and does not bias the approach of GFMC results to the true ground-state energy for infinite imaginary-time evolution.

\subsection{Tetraquark color wavefunctions}\label{subsec:potentials}

Trial wavefunctions for four-quark states must specify both the spatial and color configuration of the quarks.
Operators involving the Hylleraas molecular trial wavefunctions defined in Eq.~\eqref{eq:tetra_wvfn_H} can be defined as
\begin{equation}
\begin{split}
    \mathcal{O}^{QQ\overline{Q}\overline{Q}}_{H,\rho}(\bs{R}; a, b, c) &= \psi_i(\bs{r}_1) \psi_j(\bs{r}_2) \chi^\dagger_k(\bs{r}_3) \chi^\dagger_l(\bs{r}_4)\\
    &\hspace{10pt} \times T^{\rho}_{ijkl} \Psi_H(\bs{R}; a, b, c),
    \end{split}
    \label{eq:tetra_wvfn_H_col}
\end{equation}
where $T^{\rho}_{ijkl}$ are tensors specifying the color configuration with $\rho$ labeling different tensor choices.
The color indices of $Q_i Q_j$ can be either symmetrized or antisymmetrized to project to the $\mathbf{6}$ and $\overline{\mathbf{3}}$ irreps respectively, and similarly $\overline{Q}_k \overline{Q}_l$ can be projected to the $\overline{\mathbf{6}}$ and $\mathbf{3}$ irreps.
The only ways to construct a color-singlet four-quark state are to combine these diquark irreps as $\mathbf{6} \otimes \overline{\mathbf{6}}$ or $\overline{\mathbf{3}} \otimes \mathbf{3}$,
and there are, therefore two orthogonal color tensors associated with color-singlet four-quark states,~\cite{Jaffe:1976ig,Jaffe:1976ih}
\begin{equation}
\begin{split}
    T^{\mathbf{3}\otimes\overline{\mathbf{3}}}_{ijkl} &= \frac{1}{2\sqrt{3}}(\delta^{ik} \delta^{jl} - \delta^{il} \delta^{jk}), \\
    T^{\mathbf{6}\otimes\overline{\mathbf{6}}}_{ijkl} &= \frac{1}{2\sqrt{6}}(\delta^{ik} \delta^{jl} + \delta^{il} \delta^{jk}). \label{eq:color_basis}
\end{split}
\end{equation}
It will also prove instructive below to consider the color tensor constructed by first projecting $Q_i \overline{Q}_k$ and $Q_j \overline{Q}_l$ into color-singlet irreps as
\begin{equation}
    T^{\mathbf{1}\otimes\mathbf{1}}_{ijkl} = \frac{1}{3}\delta^{ik} \delta^{jl}.
\end{equation}
This tensor is not independent from the complete basis in Eq.~\eqref{eq:color_basis}, in particular $T^{\mathbf{1}\otimes\mathbf{1}}_{ijkl} = (T^{\mathbf{3}\otimes\overline{\mathbf{3}}}_{ijkl} + \sqrt{2} T^{\mathbf{6}\otimes\overline{\mathbf{6}}}_{ijkl})/\sqrt{3}$.

A generic color state can be expressed as a linear combination of $\mathbf{6} \otimes \overline{\mathbf{6}}$ and $\overline{\mathbf{3}} \otimes \mathbf{3}$ basis states.
The color tensor for a generic unit-normalized state can therefore be written as
\begin{equation}
    T^\rho_{ijkl} = \cos(\theta_{\mathbf{3}\otimes\overline{\mathbf{3}}}^\rho)  T^{\mathbf{3}\otimes\overline{\mathbf{3}}}_{ijkl} + \sin(\theta_{\mathbf{3}\otimes\overline{\mathbf{3}}}^\rho) T^{\mathbf{6}\otimes\overline{\mathbf{6}}}_{ijkl}.
\end{equation}
The single angular parameter $\theta_{\mathbf{3}\otimes\overline{\mathbf{3}}}^\rho$ can therefore be used to describe the color state of an arbitrary $QQ\overline{Q}\overline{Q}$ system.
This parameter can be explicitly computed as
\begin{equation}
    \theta_{\mathbf{3}\otimes\overline{\mathbf{3}}}^\rho = \text{arccos}\left( T^\rho_{ijkl} T^{\mathbf{3}\otimes\overline{\mathbf{3}}}_{ijkl} \right). \label{eq:color_angle}
\end{equation}

The pNRQCD potential in a given color state can be obtained by contracting the corresponding color tensor with the potential tensor, defined for the quark-antiquark case as the tensor appearing in Eq.~\eqref{eq:Lpsichi},
\begin{equation}
\begin{split}
  V^{\psi\chi}_{ijkl}(\bs{R},\mu) &= \frac{1}{3}\delta_{ij}\delta_{kl} V^{\psi\chi}_{\mathbf{1}}(\bs{R},\mu) \\
    &\hspace{10pt} + 2 T^a_{ji} T^a_{kl} V^{\psi\chi}_{\text{Adj}}(\bs{R},\mu).
\end{split}
\end{equation}
For the quark-quark case Eq.~\eqref{eq:Lpsipsi} gives
\begin{equation}
\begin{split}
    V^{\psi\psi}_{ijkl}(\bs{R},\mu) 
    &= \frac{1}{4}\epsilon_{ijm}\epsilon_{klm} V^{\psi\psi}_{A}(\bs{R},\mu)  \\
    &\hspace{10pt} +  \frac{1}{4}(\delta_{ik}\delta_{jl} + \delta_{il}\delta_{jk}) V^{\psi\psi}_{S}(\bs{R},\mu).
    \end{split}
\end{equation}
The antiquark-antiquark case is identical, that is $V^{\chi\chi}_{ijkl}(\bs{R}) = V^{\psi\psi}_{ijkl}(\bs{R})$.
The pNRQCD matrix element
\begin{equation}
  V^\rho(\bs{R},\mu) \equiv \mbraket{0}{\mathcal{O}^{QQ\overline{Q}\overline{Q}}_{H,\rho}(\bs{R})\, V \, \left( \mathcal{O}^{QQ\overline{Q}\overline{Q}}_{H,\rho}(\bs{R}) \right)^\dagger}{0},
\end{equation}
is then given by a sum of Wick contractions, resulting in
\begin{equation}
\begin{split}
    V^\rho =& T^\rho_{jmin} V^{\psi\chi}_{ijkl}T^\rho_{lmkn} + T^\rho_{mjin} V^{\psi\chi}_{ijkl} T^\rho_{mlkn} \\
    & +  T^\rho_{jmni} V^{\psi\chi}_{ijkl}  T^\rho_{lmnk} +  T^\rho_{mjni} V^{\psi\chi}_{ijkl}  T^\rho_{mlnk} \\
    &+ T^\rho_{ijmn} V^{\psi\psi}_{ijkl} T^\rho_{klmn} + T^\rho_{jimn} V^{\psi\psi}_{ijkl} T^\rho_{klmn} \\
    &+ T^\rho_{mnij} V^{\psi\psi}_{ijkl} T^\rho_{mnkl} + T^\rho_{mnji} V^{\psi\psi}_{ijkl} T^\rho_{mnlk},
    \end{split} \label{eq:contractions}
\end{equation}
where the $\bs{R}$ and $\mu$ arguments of the potential are suppressed for clarity.

The result for $V^\rho(\bs{R},\mu)$ in a general color state, which can be written as linear combinations of $\mathbf{3}\otimes\overline{\mathbf{3}}$ and $\mathbf{6}\otimes\overline{\mathbf{6}}$, has already been derived for the LO potential in Ref.~\cite{Huang:2020dci} and is reproduced by Eq.~\eqref{eq:contractions}.
The results for the particular color states described above are used to motivate the spatial structure of the trial wavefunction to be used in conjunction with each color state and are therefore explicitly presented below.
For the $\mathbf{3}\otimes\overline{\mathbf{3}}$ case, the potential is given at LO and NLO by
\begin{equation}
\begin{split}
    V^{\mathbf{3}\otimes\overline{\mathbf{3}}}(\bs{R},\mu) =& -\frac{2}{3}\left( \frac{\alpha_V(|\bs{r}_{12}|,\mu)}{|\bs{r_{12}|}} + \frac{\alpha_V(|\bs{r}_{34}|,\mu)}{|\bs{r_{34}|}} \right) \\
    &-\frac{1}{3}\left( \frac{\alpha_V(|\bs{r}_{13}|,\mu)}{|\bs{r_{13}|}} + \frac{\alpha_V(|\bs{r}_{14}|,\mu)}{|\bs{r_{14}|}} \right. \\
    &\hspace{30pt} \left.+ \frac{\alpha_V(|\bs{r}_{23}|,\mu)}{|\bs{r_{23}|}} + \frac{\alpha_V(|\bs{r}_{24}|,\mu)}{|\bs{r_{24}|}} \right).
    \end{split} \label{eq:V33bar}
\end{equation}
The attraction within the two diquarks is half as strong as the color-singlet $Q\overline{Q}$ potential, and therefore, the Bohr radii for these pairs should be twice the $Q\overline{Q}$ Bohr radius.
The potential between the quarks and antiquarks is smaller by another factor of two, and therefore one expects that a suitable trial wavefunction for this configuration is given by $\mathcal{O}^{QQ\overline{Q}\overline{Q}}_{H\mathbf{3}\otimes\overline{\mathbf{3}}}(\bs{R}; \lambda_1 a(L_{\mu}), \lambda_1 a(L_{\mu}) , \lambda_2 a(L_{\mu}))$
with $\lambda_1 = 4$ and $\lambda_2 = 2$.
The $\mathbf{6}\otimes\overline{\mathbf{6}}$ case results in
\begin{equation}
\begin{split}
    V^{\mathbf{6}\otimes\overline{\mathbf{6}}}(\bs{R},\mu) =& \frac{1}{3}\left( \frac{\alpha_V(|\bs{r}_{12}|,\mu)}{|\bs{r_{12}|}} + \frac{\alpha_V(|\bs{r}_{34}|,\mu)}{|\bs{r_{34}|}} \right) \\
    &-\frac{5}{6}\left( \frac{\alpha_V(|\bs{r}_{13}|,\mu)}{|\bs{r_{13}|}} + \frac{\alpha_V(|\bs{r}_{14}|,\mu)}{|\bs{r_{14}|}} \right. \\
    &\hspace{30pt} \left.+ \frac{\alpha_V(|\bs{r}_{23}|,\mu)}{|\bs{r_{23}|}} + \frac{\alpha_V(|\bs{r}_{24}|,\mu)}{|\bs{r_{24}|}} \right),
    \end{split}
\end{equation}
which analogously motivates operators of the form
 $\mathcal{O}^{QQ\overline{Q}\overline{Q}}_{H,\mathbf{6}\otimes\overline{\mathbf{6}}}(\bs{R}; \lambda_1 a(L_{\mu}), \lambda_1 a(L_{\mu}) , \lambda_2 a(L_{\mu}))$ with $\lambda_2 = 8/5$ and $\lambda_1 \gg \lambda_2$.

For the $\mathbf{1} \otimes \mathbf{1}$ case describing a pair of color-singlet mesons, the analogous result is
\begin{equation}
\begin{split}
    V^{\mathbf{1}\otimes \mathbf{1}}(\bs{R},\mu) =& -\frac{4}{3}\left( \frac{\alpha_V(|\bs{r}_{13}|,\mu)}{|\bs{r_{13}|}} +  \frac{\alpha_V(|\bs{r}_{24}|,\mu)}{|\bs{r_{24}|}} \right).
    \end{split}
\end{equation}
The symmetric/antisymmetric and singlet/adjoint potentials between the constituents of the two mesons cancel exactly, as noted in Ref.~\cite{Huang:2020dci}.
This is in contrast to the case of four-body systems in QED, where such potentials are associated with van der Waals forces and give rise to molecular bound states.
The operator $\mathcal{O}^{QQ\overline{Q}\overline{Q}}_{H,\mathbf{1}\otimes\mathbf{1}}(\bs{R}; \lambda_1 a(L_{\mu}), \lambda_2 a(L_{\mu}), \lambda_2 a(L_{\mu}))$ with $\lambda_1 = 1$ and $\lambda_2 \rightarrow \infty$ will therefore be associated with an exact eigenstate of the pNRQCD Hamiltonian at LO and an approximate local minimum at NLO.
If fully-heavy tetraquarks are bound at LO or NLO in pNRQCD, they must have a different color structure than a meson-meson product.

At NNLO, additional representation-dependent terms appear in the potential.
In particular, these terms modify the conclusion that there are net zero inter-meson potentials for products of color-singlet mesons.
\begin{equation}
\begin{split}
    V^{\mathbf{1}\otimes \mathbf{1}}(\bs{R},\mu) =& -\frac{4}{3}\left( \frac{\alpha_V(|\bs{r}_{13}|,\mu)}{|\bs{r_{13}|}} +  \frac{\alpha_V(|\bs{r}_{24}|,\mu)}{|\bs{r_{24}|}} \right) \\
    &-  \delta V(\mu) \left( \frac{\alpha_V(|\bs{r}_{12}|,\mu)}{|\bs{r_{12}|}} + \frac{\alpha_V(|\bs{r}_{14}|,\mu)}{|\bs{r_{14}|}} \right. \\
    &\hspace{30pt} \left.+ \frac{\alpha_V(|\bs{r}_{23}|,\mu)}{|\bs{r_{23}|}} + \frac{\alpha_V(|\bs{r}_{34}|,\mu)}{|\bs{r_{34}|}} \right),
    \end{split}
\end{equation}
where
\begin{equation}
  \delta V(\mu) = \frac{4 \pi^2(12 - \pi^2)\alpha_s(\mu)^2}{3(4\pi)^2}.
\end{equation}
Since $\delta V(\mu) > 0$, these additional terms give rise to attractive long-range inter-meson forces that might lead to bound tetraquark states.
However, additional $QQ\overline{Q}$ and $QQ\overline{Q}\overline{Q}$ potentials also arise at NNLO whose form has not yet been determined~\cite{Assi:2023cfo}.
To predict whether bound tetraquarks with a $\mathbf{1}\otimes \mathbf{1}$ color configuration are present at NNLO, it is necessary to determine and include these beyond two-body effects.
Corrections to the potential suppressed by $1/m_Q$ and $1/m_Q^2$ can also be numerically comparable and treated as NNLO since $r \sim 1 / (\alpha_s m_Q)$, and they should be included in future studies.

\section{QED validation}
\label{sec:QEDvalid}
\subsection{Positronium molecules}


Di-positronium bound states ${\rm Ps}_2$ comprised of $e^+e^+e^-e^-$ provide a simple analog of tetraquarks in QED that has been extensively studied theoretically with variational methods~\cite{Wheeler:1946xth,Hylleraas:1947zza,Sharma:1968,Ho:1986zz,Frolov:1996kq,Vijande:2007ix,Czarnecki:2009yeo,Bai:2016int,Aslam:2021uqu}, as well as GFMC~\cite{Lee:1983}, and whose signatures have been detected experimentally~\cite{Cassidy:2005tb,Cassidy:2012}. For a contemporary review of computational methods and experimental studies for positronium molecules, see Ref.~\cite{emami2021review}.

The static potential describing nonrelativistic QED interactions is known to all orders in the coupling $\alphaem$ and, in the absence of relativistic degrees of freedom, is given by~\cite{dyson1952divergence,pineda1998potential},
\begin{equation}
    V_{\rm QED}(\bs{r}_{ij}) =-\alphaem\frac{q_iq_j}{r_{ij}},
\end{equation}
where $q_i,\ q_j$ represent the charges of particles with positions $\bs{r}_i$ and $\bs{r}_j$. As in pNRQCD, the spin-dependent corrections arise at higher orders in $1/m_e$ in pNRQED power counting~\cite{Pineda:1998kn}, which we do not consider in this work. 
The positronium ground-state wavefunction is given exactly by Eq.~\eqref{eq:meson_wvfn} with $a = a_0^{e^+e^-} \equiv 2/(\alphaem m_e)$, where $m_e$ is the electron mass. The positronium binding energy is $\Delta E_{e^+e^-} = -\alphaemSq m_e/4$.

We use the Hylleraas molecular trial wavefunction $\Psi_H(\bs{R}; a,b,c)$ defined in Eq.~\eqref{eq:tetra_wvfn_H} as our $\Ps2$ trial wavefunction for VMC and GFMC. 
To verify our VMC methods, we first consider the variational wavefunction described in Ref.~\cite{Hylleraas:1947zza},
\begin{equation}
\begin{split}
  &\Psi_H^{e^+e^+e^-e^-}(\bs{R}; \beta) \propto  e^{(-|\bs{r}_{13}|-|\bs{r}_{24}|-|\bs{r}_{14}|-|\bs{r}_{23}|)  m_e/(2 k(\beta)\alphaem)}
      \\
    &\hspace{20pt} \hspace{10pt}\times   \cosh\left[ \frac{\beta  m_e}{2k(\beta)\alphaem}  (|\bs{r}_{13}|+|\bs{r}_{24}|-|\bs{r}_{14}|-|\bs{r}_{23}|) \right],
       \label{eqn:hylqed}
\end{split}
\end{equation}
where $k(\beta)$ is defined through an optimized rescaling of the coordinates as detailed in Appendix~\ref{app:HO}.
This corresponds to a wavefunction of the form $\Psi_H(\bs{R}; a,b,c)$ for $a = (1+\beta)/(1-\beta)$, $b/a = 1/(k(\beta)(1-\beta))$, and $c \rightarrow \infty$.
As shown in Fig.~\ref{fig:Hylleraas}, VMC results agree with corresponding analytic results~\cite{Hylleraas:1947zza} presented in Appendix~\ref{app:HO}.
The optimal variational wavefunction has $\beta^2=1/2$, which corresponds to $a = 1.0003 a_0^{e^+e^-}$ and $b/a = 5.8284$, and the corresponding Hamiltonian matrix element is more negative than twice the $e^+e^-$ binding energy.
This corresponds to a $\Ps2$ bound state with a variational bound on the binding energy $B_{\Ps2} = -M_{\Ps2} + 2M_{e^+e^-} = -\Delta E_{\Ps2} + 2\Delta E_{e^+e^-}$ given by $B_{\Ps2} \geq 0.004 \alphaemSq m_e$~\cite{Hylleraas:1947zza}.

\begin{figure}[t]
  \includegraphics[width=\linewidth]{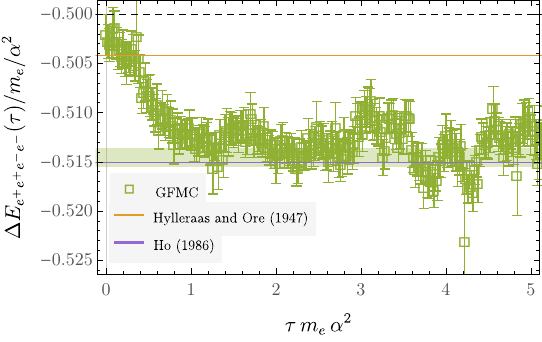}
   \caption{Effective energy for GFMC calculations of $\Ps2$ using the Hylleraas molecular trial wavefunction described in the main text. Variational bounds from Hylleraas and Ore~\cite{Hylleraas:1947zza} and Ho~\cite{Ho:1986zz} are compared.  
\label{fig:Ho}}
\end{figure}

This variationally optimized wavefunction can be used as a starting point for GFMC calculations.
It is straightforward to compute results for all values of $\alphaem$ and $m_e$ by working in atomic units in which energies correspond to $\Delta E / (\alphaemSq m_e)$, imaginary times correspond to $\tau \alphaemSq m_e$, and distances correspond to $\bs{r} m_e / \alphaem$.
Following Ref.~\cite{Lee:1983}, the inclusion of $e^-e^-$ and $e^+e^+$ correlations is found to lead to somewhat faster convergence towards the ground state during imaginary-time evolution, and $c/a = b/a = 5.8284$ is used along with $a = 1.0003 a_0^{e^+e^-}$ for the numerical results below.
The GFMC results in Fig.~\ref{fig:Ho} show that the effective energies using this trial wavefunction decrease with $\tau$ and plateau after $\tau \sim 1/(m_e \alphaemSq)$.
The Akaike information criterion~\cite{AkaikeAIC} (AIC)-weighted~\cite{Jay:2020jkz} average constant fit result, computed as described in Ref.~\cite{Assi:2023cfo}, for this results corresponds to $B_{\Ps2} = 0.015(1) \alphaemSq m_e = 0.40(3)$ eV.
This binding energy is in agreement with the state-of-the-art variational calculations involving sums of many Gaussian wavefunctions first performed by Ho in 1986~\cite{Ho:1986zz} and reproduced by others since~\cite{Frolov:1996kq,Vijande:2007ix,Bai:2016int}.\footnote{We have been unable to reproduce the deeper 1 eV binding energy reported by Ref.~\cite{Sharma:1968} using the wavefunction described in that work.}

 \begin{figure}[t]
  \includegraphics[width=\linewidth]{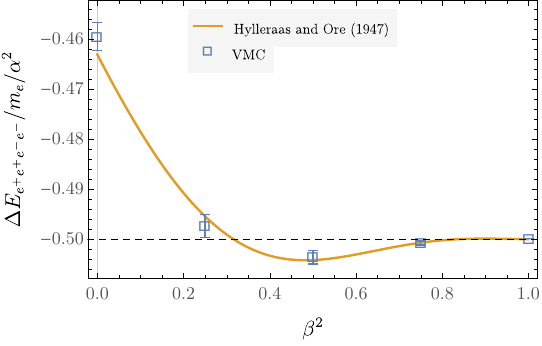}
   \caption{A comparison of analytic and VMC results for the Hylleraas and Ore wavefunction~\cite{Hylleraas:1947zza} for $\Ps2$ described in the main text with variational parameter $\beta$ as shown in Eq.~\eqref{eqn:hylqed}. The dashed line shows the threshold $2\Delta E_{e^+ e^-}$; energies below this threshold indicate the presence of a $\Ps2$ bound state. 
\label{fig:Hylleraas}}
\end{figure}

\subsection{Muonium and hydrogen molecules}\label{sec:qed_unequal}

Unequal-mass molecular systems have been studied since the early days of QED, a paradigmatic example being the hydrogen molecule H$_2$.
Variational calculations using the Born-Oppenheimer approximation in which the protons are treated as infinitely heavy static charges were shown to reproduce the experimentally measured H$_2$ binding energy~\cite{Hylleraas:1931,James:1933}.
Results for the binding energies of unequal-mass systems with finite mass ratios, denoted $\mu^+\mu^+e^-e^-$, should approach the well-known H$_2$ molecule result $B_{\text{H}_2} \approx 4.78$ eV~\cite{James:1933} for $m_e / m_\mu \rightarrow 0$.
Unequal-mass $\mu^+\mu^+e^-e^-$ systems have also been studied using variational methods in Ref.~\cite{Richard:1993zx} and shown to bind for arbitrary mass ratios.

 \begin{figure}[t]
   \includegraphics[width=\linewidth]{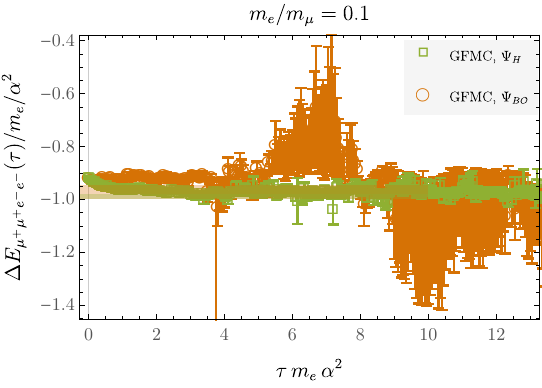} \\
  \includegraphics[width=\linewidth]{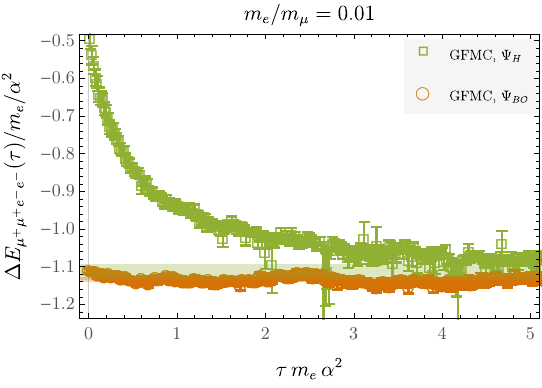}
   \caption{Effective energies for GFMC calculations using Hylleeraas molecular trial wavefunctions as well as Born-Oppenheimer trial wavefunctions for $\mu^+\mu^+ e^- e^-$ systems with two different values of $m_e/m_\mu$ indicated.  
\label{fig:BO}}
\end{figure}

\begin{figure}[t]
   \includegraphics[width=\linewidth]{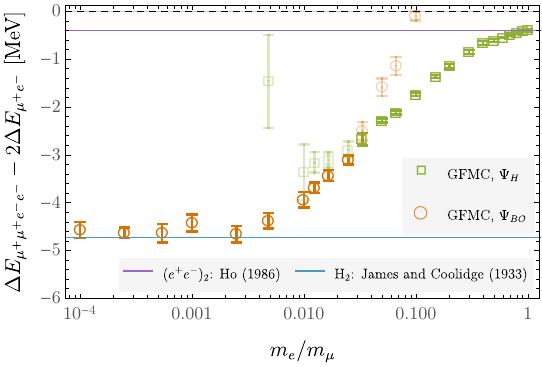} \\
   \includegraphics[width=\linewidth]{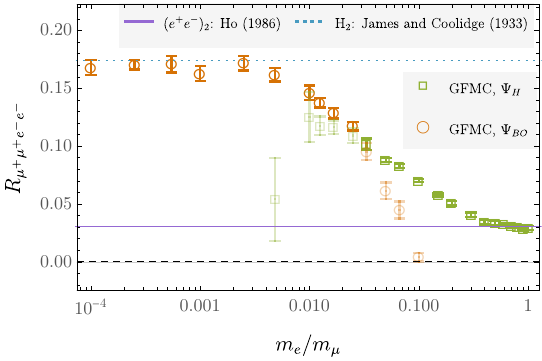} 
   \caption{Top: the $\mu^+\mu^+ e^-e^-$ binding energy as a function of $m_e/m_\mu$ obtained from GFMC calculations using Hylleraas (green) and Born-Oppenheimer (brown) trial wavefunction. The strongest variational bound obtained after GFMC imaginary-time evolution is shown with full opacity; weaker variational bounds are semi-transparent. Variational results are shown for comparison: $\Ps2$ results from Ho~\cite{Ho:1986zz} correspond to $m_e/m_\mu=1$, while H$_2$ molecule results from James and Coolidge~\cite{James:1933} correspond to $m_e/m_\mu = 0$.
   Bottom: the corresponding results for the binding energy ratio defined in Eq.~\eqref{eq:R_qed_def}.
\label{fig:qed_unequal}}
\end{figure}

For mass ratios $m_e / m_\mu \gtrsim 0.05$, GFMC calculations using Hylleraas molecular trial wavefunctions $\Psi_H$ achieve precise variational bounds that plateau after similar imaginary times $\tau \sim 1/(m_e \alpha^2)$ as the equal-mass $\Ps2$ case.
For this range of mass ratios, GFMC calculations using Born-Oppenheimer trial wavefunctions $\Psi_{BO}$ lead to weaker variational bounds with very slow imaginary-time evolution that become lost in noise before reaching the $\Psi_H$ bounds as shown in Figs.~\ref{fig:BO}-\ref{fig:qed_unequal}.
Conversely, for $m_e / m_\mu \lesssim 0.05$, stronger variational bounds are achieved using $\Psi_{BO}$ rather than $\Psi_H$.
Hylleraas trial wavefunctions require more imaginary-time evolution to reach a plateau for these extremely unequal masses, and the signal is lost to noise before a plateau is reached for $m_e / m_\mu \lesssim 0.01$.
Around $m_e / m_\mu \approx 0.05$ there is a region where the variational bounds obtained using both $\Psi_{BO}$ and $\Psi_H$ agree within uncertainties.

Performing GFMC calculations with $\Psi_{BO}$ and $\Psi_H$ for each $m_e / m_\mu$ with the same parameters as used for $\Ps2$ above and taking the strongest of the two variational bounds leads to results for $\Delta E_{\mu^+ \mu^+ e^- e^-} - 2\Delta E_{\mu^+ e^-}$ that smoothly change from $-4.64(19)$ eV at $m_e/m_\mu = 0.00054$ (the physical electron-to-proton mass ratio) and below to $-0.40(3)$ eV at $m_e/m_\mu = 1$ as shown in Fig.~\ref{fig:qed_unequal}.
Taking the physical value of the muon mass, corresponding to $m_e/m_\mu = 0.0048$, leads to a muonium molecule binding energy of $-4.38(16)$ eV.

These results can also be cast into a dimensionless form to facilitate comparisons with QCD systems below.
The ratio
\begin{equation}
    R_{\mu^+\mu^+ e^- e^-} = \frac{\Delta E_{\mu^+\mu^+ e^- e^-} - 2\Delta E_{\mu^+ e^-}}{2\Delta E_{\mu^+ e^-}}, \label{eq:R_qed_def}
\end{equation}
quantifies the relative strength of the molecular binding energy to the total binding energy of its atomic constituents.
This ratio vanishes for an unbound system, while $R_{\mu^+\mu^+ e^- e^-} > 0$ indicates the presence of a bound state.
The GFMC results for $\Delta E_{\mu^+\mu^+ e^- e^-}$ can be used to immediately obtain $R_{\mu^+\mu^+ e^- e^-}$ since $\Delta E_{\mu^+ e^-} = \alphaemSq m_{e \mu} / 4$, where $m_{e \mu} = 2 m_e m_\mu / (m_e + m_\mu)$ is the unequal-mass parameter defined in Eq.~\eqref{eq:m12}.
Results for $R_{\mu^+\mu^+ e^- e^-}$ as a function of $m_e / m_\mu$ are shown in Fig.~\ref{fig:qed_unequal} and smoothly decrease from 0.17(1) at $m_e / m_\mu \lesssim 0.001$ to 0.029(2) at $m_e / m_\mu = 1$.

\section{Tetraquark binding energy results}
\label{sec:QCDresults}

\subsection{Equal-mass heavy quarks}


To investigate fully-heavy equal-mass tetraquarks in pNRQCD, we search for optimal trial wavefunctions using VMC that are subsequently used as initial states for GFMC evolution. We consider a wide range of color wavefunctions, $T_{\rho}^{ijkl}$ in Eq.~\eqref{eq:tetra_wvfn_H_col}, as well as spatial wavefunctions that are separately optimized for each color wavefunction.

At LO, the quark mass only enters through the kinetic energy term, and it is possible to obtain results for arbitrary masses using an analog of the atomic units discussed for QED above.
In particular, using momentum in units of $C_F \alpha_s(\mu_p) m_Q$, energies in units of $C_F^2 \alpha_s(\mu_p)^2 m_Q$, and lengths and times in inverses of these units, respectively, provides results that can be trivially rescaled to make predictions for any value of $m_Q$.
At NLO, quark masses also implicitly enter the potential through logarithms involving $\mu_p$, and it is impossible to obtain results for all masses through simple rescalings.
However, the resulting mass dependence is found to be relatively mild and qualitatively similar results are found for a wide range of quark masses.
For concreteness, NLO results are presented here using $m_Q$ equal to the NLO value of $m_b$ obtained in Ref.~\cite{Assi:2023cfo} and the notations $QQ\overline{Q}\overline{Q}$ and $\bbbb$ will be used interchangeably for equal-mass tetraquark systems below.

 \begin{figure}[t]
   \includegraphics[width=\linewidth]{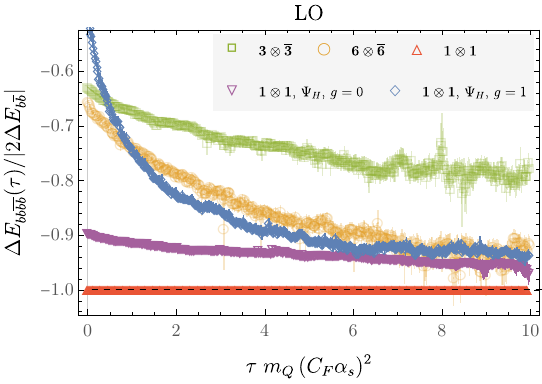} \\
   \includegraphics[width=\linewidth]{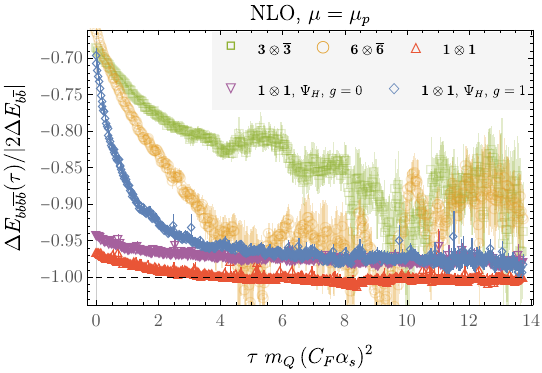} 
   \caption{
   Top: effective energies for equal-mass tetraquarks at LO in pNRQCD. Different colors/markers correspond to trial wavefunctions with different colored wavefunctions described in the main text. The minimum-energy spatial wavefunctions for $\mathbf{3\times \overline{3}}$, $\mathbf{6\times \overline{6}}$, and $\mathbf{1\times 1}$ in green, orange, and red, respectively. For $\mathbf{1\times 1}$ the minimum-energy spatial wavefunction is found to be a product of two quarkonium wavefunctions; additional results for $\mathbf{1\times 1}$ wavefunctions with spatial structure analogous to $\Ps2$ bound states in pNRQCD with symmetry factors $g$ defined in Eq.~\eqref{eqn:hyl_tetra} are shown in purple and blue. 
   Bottom: analogous results at NLO using $m_Q = m_b$ from Ref.~\cite{Assi:2023cfo} and the corresponding renormalization scale $\mu_p$ defined in Eq.~\eqref{eq:mup}.
\label{fig:equalmass_effmass} }
\end{figure}

 \begin{figure}[t]
   \includegraphics[width=\linewidth]{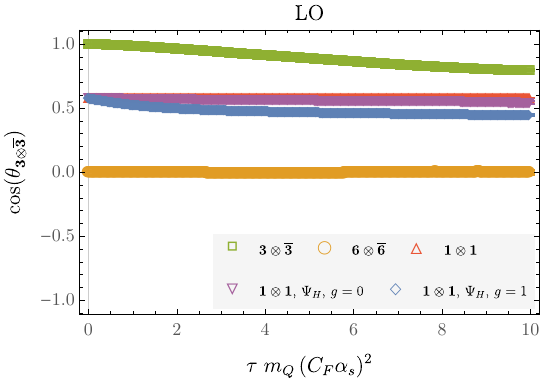} \\
   \includegraphics[width=\linewidth]{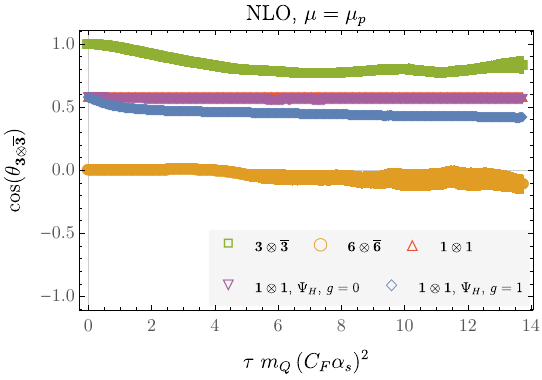} 
   \caption{ Top: the overlap between $\mathbf{3 \otimes \overline{3}}$ color states and GFMC-evolved wavefunctions defined in  Eq.~\eqref{eq:color_angle} for the same set of trial wavefunctions as in Fig.~\ref{fig:equalmass_effmass} at LO in pNRQCD. Bottom: analogous results at NLO. 
\label{fig:equalmass_effZ}}
\end{figure}

First, consider the $\mathbf{1\times 1}$ tetraquark color state with the Hylleraas molecular trial wavefunction Eq.~\eqref{eqn:hylqed}. Using parameters inspired by the QED analog, $\Ps2$, namely  $a=1.0003$ and $b/a=c/a=5.824$, gives rise to a GFMC effective energy well above the two-meson threshold for imaginary times of 1-10 times the Coulomb energy scale $m_Q (C_F \alpha_s)^2$  at both LO and NLO as shown in Fig.~\ref{fig:equalmass_effmass}. 
This already demonstrates that mesons in pNRQCD behave very differently from positronia in pNRQED. 
This is not surprising because, in QED, two atoms can be configured so that opposite charges are closer, resulting in a net attractive Van-der-Waals force. On the other hand,
two color-singlet heavy mesons have net zero inter-meson potentials as discussed in Sec.~\ref{subsec:potentials} and thus net-zero Van-der-Waals force.

Different quark/antiquark permutation structures for the trial wavefunction can be described by using trial wavefunctions of the form
\begin{equation}
\begin{split}
    \Psi_H(\bs{R};a,b,c,g) &= \frac{1}{\sqrt{2}}\left[ \Psi_H(\bs{r}_1,\bs{r}_2,\bs{r}_3,\bs{r}_3;a,b,c) \right. \\
    &\hspace{20pt} \left. + g\,  \Psi_H(\bs{r}_2,\bs{r}_1,\bs{r}_3,\bs{r}_3;a,b,c) \right].
    \end{split}
\label{eqn:hyl_tetra}
\end{equation}
Note that $\bs{r}_3 \leftrightarrow \bs{r}_4$ exchange is equivalent to $\bs{r}_1 \leftrightarrow \bs{r}_2$ exchange, and there is therefore no need to symmetrize identical quarks and antiquarks separately.
Symmetric spatial wavefunctions with $g=1$ can be combined with antisymmetric spin wavefunctions where the two quarks and two antiquarks are both combined into spin singlets to build totally antisymmetric wavefunctions.\footnote{Since the pNRQCD Hamiltonian is spin independent at $\mathcal{O}(1/m_Q^0)$, the spin state does not need to be explicitly included in GFMC evolution or calculation of Hamiltonian matrix elements.}
Spatial wavefunctions with $g=0$ do not have definite exchange symmetry; however, the combined color-space wavefunction is symmetric under $\psi_i(\bs{r}_1) \psi_j(\bs{r}_2) \leftrightarrow \psi_j(\bs{r}_2) \psi_i(\bs{r}_1)$ and therefore totally antisymmetric wavefunctions can again be constructed by combining with spin-singlet wavefunctions.
Results using $g \in \{0,1\}$ are both shown in Fig.~\ref{fig:equalmass_effmass}, and the $g=0$ case is seen to provide stronger variational bounds.
Spatial wavefunctions with $g=-1$ can be combined with symmetric spin-triplet wavefunctions, but these spatial wavefunctions have nodes and lead to much weaker variational energy bounds.

Upon scanning over the parameters of $\mathcal{O}^{QQ\overline{Q}\overline{Q}}_{H,\mathbf{1}\otimes\mathbf{1}}(\bs{R}; \lambda_1 a(L_{\mu}), \lambda_2 a(L_{\mu}), \lambda_2 a(L_{\mu}))$, the spatial wavefunction that minimizes the energy is found to have $\lambda_1 = 1$ and $\lambda_2 \rightarrow \infty$ at both LO and NLO.
This corresponds to an uncorrelated product of two single-meson wavefunctions and leads to GFMC effective energies consistent with the two-meson threshold in Fig.~\ref{fig:equalmass_effmass}. Thus, as in the equal-mass case, the meson product state remains an exact eigenstate, consistent with the fact that two color-singlet heavy mesons have net zero inter-meson potentials at LO and NLO. The two-meson product wavefunction is also an exact eigenstate at LO (although it is not necessarily the ground state).

Analogous scans for the parameters of $\mathcal{O}^{QQ\overline{Q}\overline{Q}}_{H\mathbf{3}\otimes\overline{\mathbf{3}}}(\bs{R}; \lambda_1 a(L_{\mu}), \lambda_1 a(L_{\mu}), \lambda_2 a(L_{\mu}))$ demonstrate that minimum-energy wavefunctions are obtained using the values $\lambda_1 = 4$ and $\lambda_2 = 2$ that correspond to the Bohr radii for each potential treated independently, as discussed in Sec.~\ref{subsec:potentials}. 
Similarly, minimum-energy wavefunctions for $\mathcal{O}^{QQ\overline{Q}\overline{Q}}_{H,\mathbf{6}\otimes\overline{\mathbf{6}}}(\bs{R}; \lambda_1 a(L_{\mu}), \lambda_1 a(L_{\mu}) , \lambda_2 a(L_{\mu}))$ are verified numerically to be achieved using $\lambda_2 = 8/5$ and $\lambda_1 \rightarrow \infty$.
These trial wavefunctions provide weaker variational bounds than those achieved using the $\mathbf{1}\otimes \mathbf{1}$ states as shown in Figs.~\ref{fig:equalmass_effmass}. We further tested the $\mathbf{3}\otimes\overline{\mathbf{3}}$ Gaussian spatial wavefunction presented in Ref.~\cite{Anwar:2017toa} and found that it provides similar variational bounds to $\mathbf{3}\otimes\overline{\mathbf{3}}$  Hylleraas molecular trial wavefunctions.

The most robust variational bounds found in this work arise from meson-meson product wavefunctions and are consistent with the two-meson threshold; therefore, this analysis provides no evidence supporting the binding of equal-mass tetraquarks.
However, excluding the possibility that an approximately orthogonal trial wavefunction could describe a bound state is difficult.
In particular, the imaginary-time evolution used here leads to only small changes in the color state of the trial wavefunction, as shown in Fig.~\ref{fig:equalmass_effZ}.
This highlights the importance of using trial wavefunctions that span the entire color space to explore Hilbert space effectively.
We therefore performed a scan over the color dimension of Hilbert space by preparing trial wavefunctions with values of
$\theta_{\mathbf{3}\otimes\overline{\mathbf{3}}}$ ranging from $-\pi$ to $\pi$. However, the strongest variational bounds are again found to be obtained using the $\mathbf{1} \otimes \mathbf{1}$ wavefunctions already described above.


\subsection{Unequal-mass heavy quarks}\label{sec:uneq_results}

 \begin{figure}[t]
   \includegraphics[width=\linewidth]{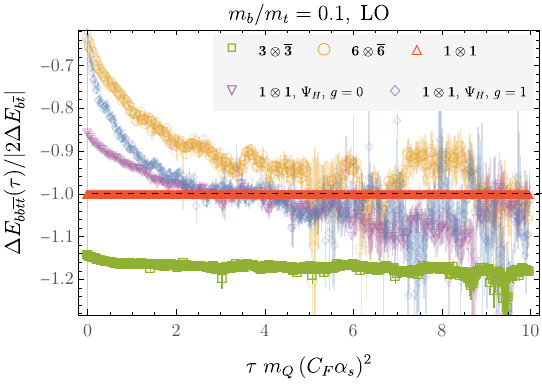} \\
   \includegraphics[width=\linewidth]{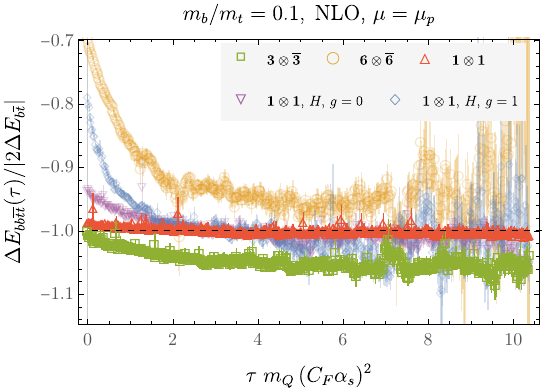} 
   \caption{
   Top: effective energies for unequal-mass tetraquarks with a quark/antiquark mass ratio of $0.1$ at LO in pNRQCD; details are as in Fig.~\ref{fig:equalmass_effmass}. Bottom: Analogous results at NLO.
\label{fig:unequalmass_effmass} }
\end{figure}

 \begin{figure}[t]
   \includegraphics[width=\linewidth]{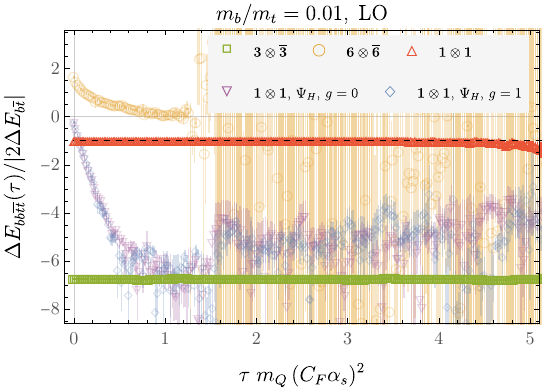} \\
   \includegraphics[width=\linewidth]{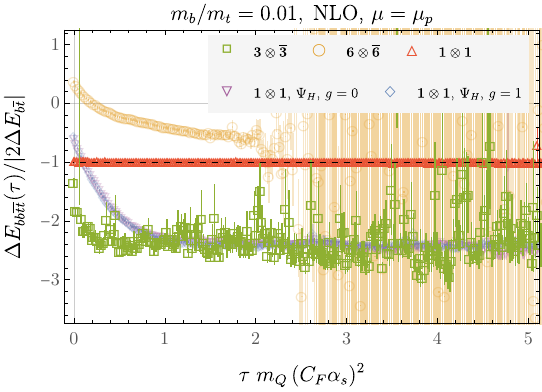} 
   \caption{Top: effective energies for unequal-mass tetraquarks with a quark/antiquark mass ratio of $0.01$ at LO in pNRQCD; details are as in Fig.~\ref{fig:equalmass_effmass}. Bottom: Analogous results at NLO.
\label{fig:moreunequalmass_effmass}}
\end{figure}

 \begin{figure}[t]
   \includegraphics[width=\linewidth]{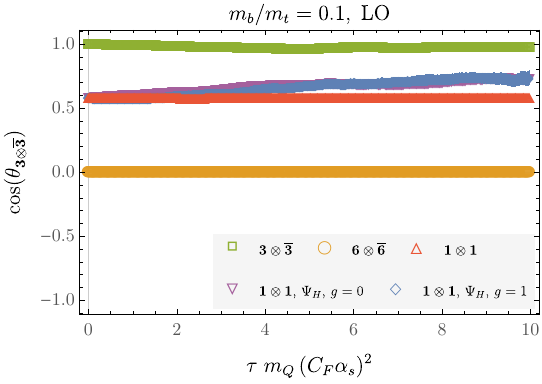} \\
   \includegraphics[width=\linewidth]{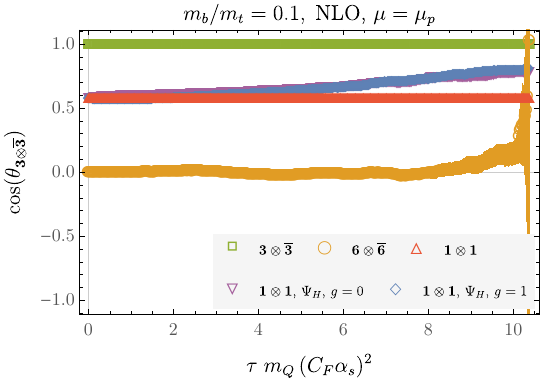} 
   \caption{Top: the overlap between $\mathbf{3 \otimes \overline{3}}$ color states and GFMC-evolved wavefunctions for unequal-mass tetraquarks with a quark/antiquark mass ratio of $0.1$ at LO in pNRQCD; details are as in Fig.~\ref{fig:equalmass_effZ}. Bottom: Analogous results at NLO. 
\label{fig:unequalmass_effZ}}
\end{figure}

 \begin{figure}[t]
   \includegraphics[width=\linewidth]{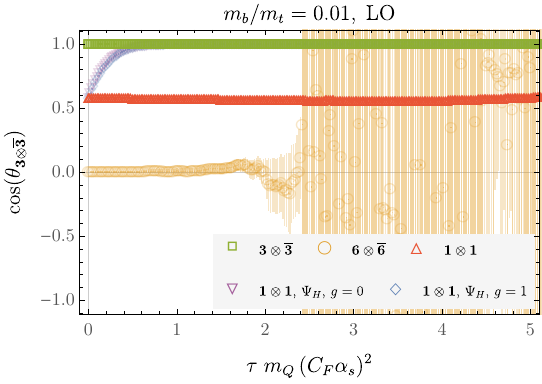} \\
   \includegraphics[width=\linewidth]{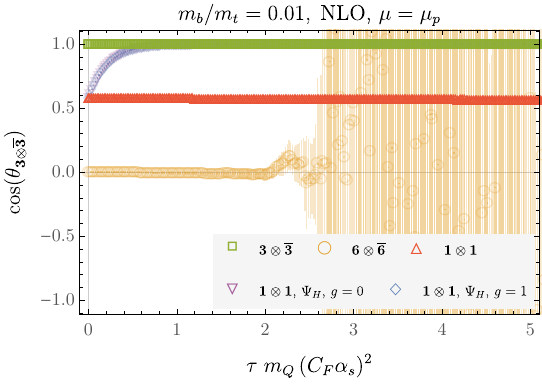} 
   \caption{Top: the overlap between $\mathbf{3 \otimes \overline{3}}$ color states and GFMC-evolved wavefunctions for unequal-mass tetraquarks with a quark/antiquark mass ratio of $0.01$ at LO in pNRQCD; details are as in Fig.~\ref{fig:equalmass_effZ}. Bottom: Analogous results at NLO. 
\label{fig:moreunequalmass_effZ}}
\end{figure}

Unequal-mass systems in which the antiquarks are much heavier than the quarks,\footnote{Results for systems in which the quark and antiquark masses are exchanged are identical by charge conjugation symmetry.} denoted $\bbtt$ with $m_t > m_b \gg \Lambda_{\rm QCD}$ free parameters not necessarily equal to physical $b$ and $t$ quark masses, can be studied in pNRQCD using the same methods and trial wavefunction structures as the equal-mass case discussed above.
At LO, unequal-mass quarkonium binding energies can be calculated analytically using a reduced mass and are given by $\Delta_{b\overline{t}}^{(\text{LO})} = - C_F^2 \alpha_s^2 m_{bt} / 4$ in terms of the mass parameter $m_{bt} = 2m_b m_t/(m_b+m_t)$ defined in Eq.~\eqref{eq:m12}. Generalized atomic units defined with $m_Q$ replaced by $m_{b\overline{t}}$ can be used to obtain LO results for arbitrary quark masses.
At NLO, results are presented for concreteness with $m_b$ set equal to the physical $b$ quark mass.
At both LO and NLO, $m_b / m_t$ is treated as a free parameter.

We performed VMC and GFMC calculations for a wide range of $m_b / m_t$ using the same trial wavefunction structures described above.
As in the equal-mass case, the strongest variational bounds obtained using each color structure are consistent with those obtained using the scales suggested in Sec.~\ref{subsec:potentials}.
Unlike the equal-mass case, stronger variational bounds are obtained using $\mathbf{3} \otimes \overline{\mathbf{3}}$ operators instead of $\mathbf{1} \otimes \mathbf{1}$ operators for sufficiently small $m_b / m_t$.
As shown in Fig.~\ref{fig:unequalmass_effmass} for $m_b / m_t = 0.1$, these bounds are significantly below the meson-meson threshold at both LO and NLO.
This unambiguously demonstrates that tetraquark bound states are present in pNRQCD for these masses.
Results for meson-meson product operators of the form $\mathcal{O}^{\bbtt}_{H,\mathbf{1}\otimes\mathbf{1}}(\bs{R}; \lambda_1 a(L_{\mu}), \lambda_2 a(L_{\mu}), \lambda_2 a(L_{\mu}))$ with $\lambda_1 = 1$ and $\lambda_2 \gg 1$ are consistent with the two-meson threshold and approximately $\tau$ independent, indicating that this operator describes an approximate pNRQCD eigenstate but not the ground state for these masses.
Results for $\mathcal{O}^{\bbtt}_{H,\mathbf{1}\otimes\mathbf{1}}(\bs{R}; \lambda_1 a(L_{\mu}), \lambda_2 a(L_{\mu}), \lambda_2 a(L_{\mu}))$ with smaller values of $\lambda_2$ cross below the two-meson threshold for sufficiently large imaginary times, although they are significantly noisier and slower to plateau than results obtained using $\mathcal{O}^{\bbtt}_{H\mathbf{3}\otimes\overline{\mathbf{3}}}(\bs{R}; \lambda_1 a(L_{\mu}), \lambda_1 a(L_{\mu}) , \lambda_2 a(L_{\mu}))$ with $\lambda_1 = 4$ and $\lambda_2 = 2$.

Qualitatively similar results are obtained for more extreme mass ratios $m_b / m_t = 0.01$ as shown in Fig.~\ref{fig:moreunequalmass_effmass}.
It is noteworthy that Hylleraas molecular trial wavefunctions lead to precise GFMC results with mild $\tau$ dependence for these extreme mass ratios.
This contrasts sharply with the unequal-mass QED results discussed in Sec.~\ref{sec:qed_unequal} and suggests that tetraquark bound states are qualitatively different from molecular bound states in QED.

The color structure of tetraquark bound states can be studied by computing the color states obtained after GFMC evolution.
As shown in Figs.~\ref{fig:unequalmass_effZ}-\ref{fig:moreunequalmass_effZ}, results using meson product states and $\mathbf{6} \otimes \overline{\mathbf{6}}$ trial wavefunctions show negligible imaginary-time dependence.
However, results using QED $\Ps2$ parameters converge towards $\mathbf{3} \otimes \overline{\mathbf{3}}$ color states, slowly with $m_b / m_t = 0.1$ and much more quickly with $m_b / m_t = 0.01$.
This suggests that the pNRQCD ground state has a $\mathbf{3} \otimes \overline{\mathbf{3}}$ color structure. However, it also highlights the importance of considering trial wavefunctions that overlap more with $\mathbf{3} \otimes \overline{\mathbf{3}}$ states than with other pNRQCD eigenstates such as the meson-meson product state.

\begin{figure}[t]  \includegraphics[width=\linewidth]{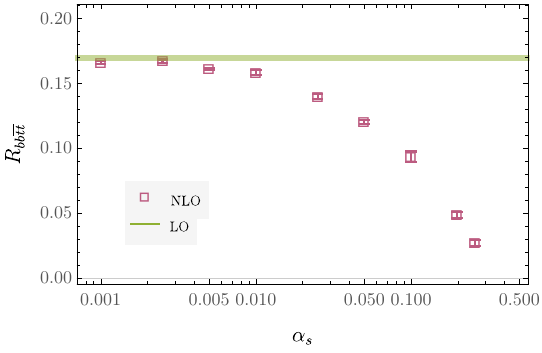}
   \caption{ The tetraquark binding-energy ratio defined in Eq.~\eqref{eq:Rtetra} for fixed $m_b/m_t = 0.1$ as a function of $\alpha_s(\mu_p)$, where $\mu_p$ is implicitly a function of $m_{bt}$. Points show fitted GFMC results at NLO using $\mathbf{3\otimes \overline{3}}$ trial wavefunctions, while the green band shows the corresponding fitted GFMC results at LO, which is independent of $m_{bt}$ for fixed $m_b/m_t$. 
\label{fig:unequalmass_vs_alpha} }
\end{figure}

The relative size of the tetraquark binding energy to the binding energies of its constituent mesons can be quantified by the ratio
\begin{equation}
    R_{\bbtt} = \frac{\Delta E_{\bbtt} - 2\Delta E_{b\overline{t}}}{2\Delta E_{b\overline{t}}}, \label{eq:Rtetra}
\end{equation}
which is a function of $m_b/m_t$ at LO and depends on both $m_b$ and $m_t$ at NLO.
For $m_b / m_t = 0.1$, this ratio equals $0.170(2)$, which is similar to the corresponding ratio for H$_2$ in QED.
The dependence of NLO results on $m_b$ for fixed $m_b/m_t = 0.1$ is shown in Fig.~\ref{fig:unequalmass_vs_alpha}.
For the physical $b$ quark mass, the ratio is significantly smaller, $0.050(2)$.
Masses leading to $\alpha_s \ll 0.1$ are required for NLO corrections to be small, indicating that perturbative convergence is slow.
This is consistent with expectations that renormalon effects on the static potential are large and numerical results for meson and baryon masses~\cite{Assi:2023cfo}.

\begin{figure}[t]  \includegraphics[width=\linewidth]{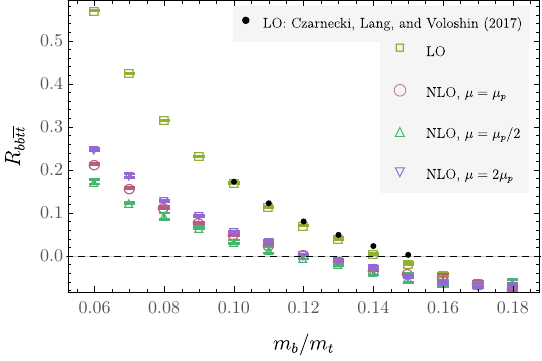} \\
   \includegraphics[width=\linewidth]{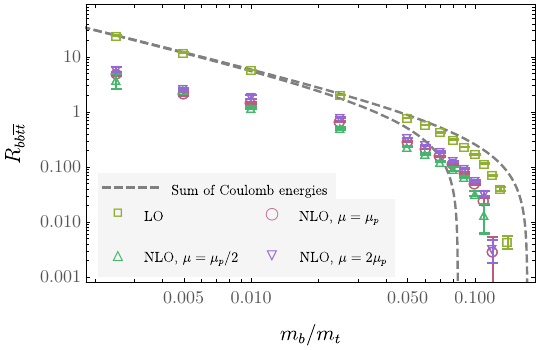} 
   \caption{ Top: tetraquark binding-energy ratio as a function of the quark/antiquark mass ratio $m_b/m_t$ at LO. At NLO, $m_b$ is fixed to reproduce the measured quarkonium spectrum~\cite{Assi:2023cfo}, and results with different values of $\mu$ are shown to indicate the size of renormalization scale variation.
   All points show fitted GFMC results using $\mathbf{3\otimes \overline{3}}$ trial wavefunctions.
   LO results are compared with the results of Czarnecki, Lang, and Voloshin~\cite{Czarnecki:2017vco}, which are also valid at LO in pNRQCD.
   Bottom: the same tetraquark binding-energy ratio for a wide range of $m_b/m_t$ on a log-log plot. The sum of Coulomb energies in Eq.~\eqref{eq:uneqapprox} is shown as a dashed line for comparison.
\label{fig:unequalmass_binding}}
\end{figure}

The dependence of $R_{\bbtt}$ on $m_b/m_t$ is shown in Fig.~\ref{fig:unequalmass_binding}.
LO results can be compared with variational results for unequal-mass tetraquarks with a one-gluon-exchange potential from Ref.~\cite{Czarnecki:2017vco}.
Results agree within $1\sigma$ statistical uncertainties for $m_b/m_t \in [0.1,0.13]$.
The threshold for tetraquark binding at LO is $m_b / m_t \lesssim 0.15$; right at the threshold, the energies obtained from our GFMC calculations are slightly higher than those obtained by applying variational methods to sums of 200 Gaussian wavefunctions in Ref.~\cite{Czarnecki:2017vco}.
It is likely that our GFMC results have not totally converged to the ground state for masses right at the threshold, which may indicate that the structure of near-threshold bound states is more complicated than that of our Hylleraas molecular trial wavefunctions.
Our LO results are completely consistent with those of Ref.~\cite{Czarnecki:2017vco} when interpreted as variational bounds.

\begin{figure}[t]   \includegraphics[width=\linewidth]{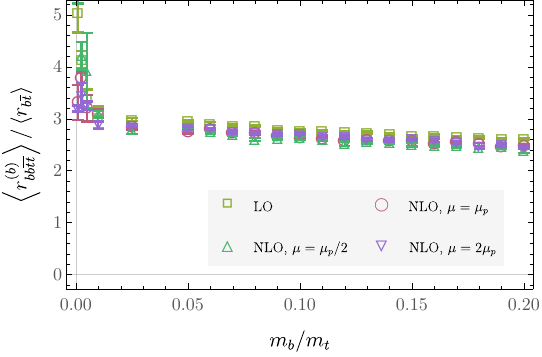} \\
   \includegraphics[width=\linewidth]{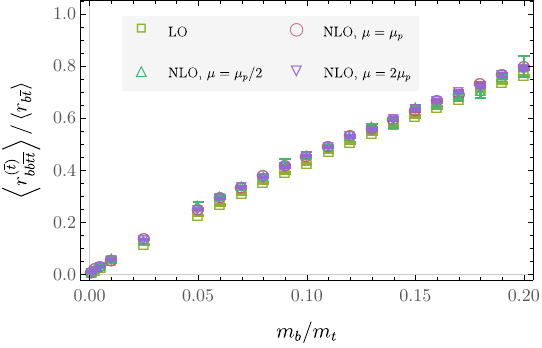} 
   \caption{ Top: the average radius of the lighter tetraquark constituent, as defined in Eq.~\eqref{eq:rAveDef}, divided by the average $b\overline{t}$ meson radius. Points are obtained from fitted GFMC results with $\mathbf{3\otimes \overline{3}}$ trial wavefunctions. Bottom: the corresponding average radius of the heavier tetraquark constituent dived by the average $b\overline{t}$ meson radius. }
\label{fig:unequalmass_Rs}
\end{figure}

Although tetraquark binding energies are quantitatively sensitive to the order of the pNRQCD potential, the existence of tetraquark bound states for sufficiently unequal mass ratios is a robust prediction at LO and NLO.
At NLO, the threshold for tetraquark binding is $m_b / m_t \lesssim 0.12$.
This partially explains why NLO corrections to $R_{\bbtt}$ are fractionally large at $m_b / m_t = 0.1$; near-threshold the absolute magnitude of $R_{\bbtt}$ is small.
A qualitatively similar dependence of $R_{\bbtt}$ on $m_b/m_t$ is seen at LO and NLO as shown in Fig.~\ref{fig:unequalmass_binding}.

It is also instructive to compare pNRQCD results to a simple diquark model.
For a $\mathbf{3} \otimes \overline{\mathbf{3}}$ color state, the $bb$ and $\overline{t}\overline{t}$ pairs experience attractive Coulomb interactions that can give rise to diquark bound states.
The sum of these Coulomb binding energies and those associated with the corresponding quark-antiquark potentials in this state gives a crude approximation to the $\bbtt$ energy,
\begin{equation}
\begin{split}
    \Delta E_{\bbtt}^{(\text{LO})} &\sim -\frac{m_t(C_F\alpha_s/2)^2}{4}-\frac{m_b(C_F\alpha_s/2)^2}{4} \\
    &\hspace{20pt} -4 \frac{ m_{bt} (C_F\alpha_s/4)^2}{4},
\end{split}
\label{eq:uneqapprox}
\end{equation}
where $C_F / 2$ is the color factor associated with the antisymmetric quark-quark potential and $C_F / 4$ is the color factor associated with quark-antiquark potentials in this color state as in Eq.~\eqref{eq:V33bar}.
This can be compared to twice the quarkonium binding energy,
\begin{equation}
\begin{split}
    2\Delta E_{b\overline{t}}^{(\text{LO})} = - \frac{m_{bt} (C_F\alpha_s)^2}{2}.
\end{split}
\end{equation}
For $m_t \sim m_b$ the larger color factor in the color-singlet channel leads to a smaller $|\Delta E_{\bbtt}|$ than $|2\Delta E_{b\overline{t}}|$ in the diquark approximation.
However, for $m_t \gg m_b$ this diquark model predicts $|\Delta E_{\bbtt}| \gg |2\Delta E_{b\overline{t}}|$ and therefore the presence of bound tetraquarks.
Remarkably, the crude approximation in Eq.~\eqref{eq:uneqapprox} quantitatively reproduces pNRQCD results at LO for $m_b / m_t \lesssim 0.01$ as shown in Fig.~\ref{fig:unequalmass_binding}.
Note, however, that a naive replacement of $\alpha_s$ by $\alpha_V$ in Eq.~\eqref{eq:uneqapprox} would predict that tetraquarks are more deeply bound at NLO; this is not what is observed in pNRQCD, which indicates that this crude diquark model does not capture logarithmic effects present at NLO.

\begin{table*}[!t]
\begin{ruledtabular}
\begin{tabular}{c|llll} 
            Flavor & $M_{QQ\overline{QQ}}^{(\text{LO})}$ [GeV] & $M_{QQ\overline{QQ}}^{(\text{NLO})}$ [GeV] & $B_{QQ\overline{QQ}}^{(\text{LO})}$ [MeV] & $B_{QQ\overline{QQ}}^{(\text{NLO})}$ [MeV] \\\\
            $bb\overline{tt}$ 
            & 354.251(1)(279) & 354.019(3)(204) & 518(1)(179) & 476(3)(19) \\
            $bc\overline{tt}$ 
            & 350.853(1)(423) & 350.840(4)(102) & 743(1)(315) & 516(4)(135) \\
            $cc\overline{tt}$ 
            & 347.494(1)(551) & 347.675(12)(500) & 995(1)(477) & 622(12)(240)  \\
  \end{tabular}
  \caption{ 
  Masses and binding energies for tetraquarks comprised of $\{c, b, t\}$ quarks, defined with respect to the lightest meson-meson threshold as in Eqs.~\eqref{eq:Bbbtt}-\eqref{eq:Bcbbt}, that are predicted to be stable in QCD (despite likely decaying before hadronizing in the full Standard Model) according to pNRQCD calculations at LO and NLO as indicated. In all cases the tetraquark energy is determined by applying GFMC to a $\mathbf{3}\otimes\overline{\mathbf{3}}$ diquark-molecule trial wavefunction as described in Sec.~\ref{sec:uneq_results}.  Masses correspond to 
  $m_t = 172.76$ GeV~\cite{ParticleDataGroup:2020ssz}
  along with
  $m_b^{(\text{LO})} = 4.770$ GeV, $m_b^{(\text{NLO})} = 4.868$ GeV,
  $m_c^{(\text{LO})} = 1.562$ GeV, and $m_c^{(\text{NLO})} = 1.654$ GeV
  from Ref.~\cite{Assi:2023cfo}.
  The first uncertainty is the GFMC statistical uncertainty and the second uncertainty is half the difference between results with $\mu = 2\mu_p$ and results with $\mu = 1/2 \mu_p$.
  Applying analogous methods to other systems does not provide any evidence for the binding of equal-mass, $bt\overline{tt}$, $ct\overline{tt}$, $bb\overline{bt}$, $bc\overline{bt}$, $cc\overline{bt}$, $bb\overline{ct}$, $bc\overline{ct}$, $cc\overline{ct}$, $bb\overline{bc}$, $bc\overline{bc}$, $cc\overline{bc}$, or $cc\overline{bb}$ tetraquarks.
  }
  \label{tab:tetra}
   \end{ruledtabular}
\end{table*}

Studying the average inter-quark distances for the lighter and heavier constituents can provide further insight into the structure of tetraquark bound states in pNRQCD.
The average distance between the center of mass and a particular constituent can defined for a system of $N$ heavy quarks/antiquarks as
\begin{equation}
    r_{Q_1\ldots Q_N} = \frac{1}{N} \sum_{i=1}^N \left| \bm{r}_i - \bm{R}_{\text{CM}} \right|,
\end{equation}
where
\begin{equation}
    \bm{R}_{\text{CM}} = \frac{ \sum_{i=1}^N m_i \bm{r}_i}{\sum_{j=1}^N m_j}. \label{eq:RCM}
\end{equation}
The average quarkonium size defined in this sense can be computed analytically at LO as $\left< r_{b\overline{t}}^{(\text{LO})} \right> = 3/(2 \alpha_s m_{bt})$ and numerically at NLO by fitting the average $b\overline{t}$ distance after GFMC evolution.
It is useful to further define the average tetraquark sizes of the lighter and heavier constituents separately.
For a generic system of $K$ heavy quarks/antiquarks of type $Q$ and $N-K$ heavy quarks/antiquarks of type $Q'$ such as distance is defined by
\begin{equation}
    r_{Q_1\ldots Q_K Q'_1 \ldots Q'_{N-K}}^{(Q)} = \frac{1}{K} \sum_{i=1}^K \left| \bm{r}_i - \bm{R}_{\text{CM}} \right|, \label{eq:rAveDef}
\end{equation}
where $\bm{R}_{\text{CM}}$ is defined in terms of a sum over all $N$ constituents as in Eq.~\eqref{eq:RCM}.
The average tetraquark size for the lighter constituents, $\left< r^{(b)}_{\bbtt} \right>$, is found to be 2-3 times larger than $\left< r_{b\overline{t}} \right>$ at both LO and NLO and depends weakly on $m_b / m_t$, as shown in the top plot of Fig.~\ref{fig:unequalmass_Rs}.
Conversely, the average tetraquark size for the heavier constituents, $\left< r^{(\overline{t})}_{\bbtt} \right>$, decreases approximately linearly with $m_b / m_t$ for small values of this ratio, as shown in the bottom plot of Fig.~\ref{fig:unequalmass_Rs}.
This indicates that bound tetraquarks in pNRQCD can be described as a spatially compact heavy diquark core with the lighter constituents forming a more spatially diffuse cloud around this core.

In the full Standard Model, the top quark decays weakly at too fast a rate to experimentally detect signatures of top quark hadronization~\cite{Bigi:1986jk}.
This makes it unlikely that there are any phenomenological signatures of the $\bbtt$ tetraquarks discussed here.
Nevertheless, fully-heavy tetraquarks comprised of $\{c, b, t\}$ quarks provide stable QCD bound states whose masses are well-defined.
The binding energies defined by the difference between these tetraquark masses and the closest quarkonium-quarkonium threshold provide benchmarks that can be used to compare various approximations to QCD and other hadronic models.

The $\bbtt$ tetraquarks described above bind with physical values of $m_t = 172.76$ GeV~\cite{ParticleDataGroup:2020ssz} and $m_b$ given by pNRQCD fits to 1S quarkonium masses in Ref.~\cite{Assi:2023cfo} 
The binding energy defined by 
\begin{equation}\label{eq:Bbbtt}
    B_{\bbtt} = 2 \Delta E_{b\overline{t}} - \Delta E_{\bbtt},
\end{equation}
is given at NLO by $B_{\bbtt}^{(\text{NLO})} = 476(3)(19)$ MeV, where the first uncertainty is statistical and the much larger second uncertainty is the difference between results with $\mu = 2\mu_p$ and $\mu = \mu_p /2$.
LO results are shown in Table~\ref{tab:tetra}, along with the corresponding results for $cc\overline{tt}$ and $cb\overline{tt}$ tetraquarks.
Binding energies increase as the lighter quark mass is decreased, although scale uncertainties make it difficult to draw firm conclusions for states involving charm quark, with $B_{cb\overline{tt}}^{(\text{NLO})}~=~516(4)(135)$ MeV and $B_{cc\overline{tt}}^{(\text{NLO})}~=~622(12)(240)$ MeV.

Other flavor combinations of $\{c, b, t\}$ quarks can be studied using the same GFMC methods; however, our results do not provide any evidence for the binding of $bt\overline{tt}$, $ct\overline{tt}$, $bb\overline{bt}$, $bc\overline{bt}$, $cc\overline{bt}$, $bb\overline{ct}$, $bc\overline{ct}$, $cc\overline{ct}$, $bb\overline{bc}$, $bc\overline{bc}$, $cc\overline{bc}$, or $cc\overline{bb}$ tetraquarks.
Several of these combinations involving the appearance of quark-antiquark pairs with the same flavor.
In these cases the lowest quarkonium-quarkonium threshold must be taken into account, for example 
\begin{equation}\label{eq:Bcbbt}
    B_{cb\overline{bt}} = \Delta E_{b\overline{b}} + \Delta E_{c\overline{t}} - \Delta E_{cb\overline{bt}},
\end{equation}
since the combination $(b\overline{b})(c\overline{t})$ is more deeply bound than $(b\overline{t})(c\overline{b})$.
The same considerations apply to ``Z-type'' tetraquarks such as $bt\overline{bt}$.
As discussed in Appendix~\ref{app:Z}, our GFMC results do not provide any evidence for bound Z-type fully-heavy tetraquarks in pNRQCD. Although the constituent quarks we study are both heavy but differ in mass significantly, the lack of evidence for Z-type for this system is in-line with previous estimates and calculations of doubly-heavy tetraquarks~\cite{Ader:1981db,Braaten:2020nwp,Bicudo:2022cqi,Berwein:2024ztx}.
Since our trial wavefunctions span the entire color sector of Hilbert space, this suggests that Z-type tetraquarks and these combinations of fully-heavy $\{c,b,t\}$ tetraquarks are unbound at LO and NLO in pNRQCD unless the spatial structure of such a bound state is very different from the molecular structures studied for various color configurations here.
It is noteworthy that all flavor combinations without a flavor-singlet quark-antiquark pair that do not lead to the appearance of bound tetraquarks in our calculations --- $cc\overline{bt}$, $bb\overline{ct}$, and $cc\overline{bb}$ --- have ratios of reduced masses for the $QQ$ and $\overline{QQ}$ pairs that are larger than the 0.12-0.15 critical ratio where binding is expected based on the $\bbtt$ calculations above.

\section{Discussion}
\label{sec:conclusion}

In this work, we have used quantum Monte Carlo methods to study the ground-state energies of four heavy-quark $QQ\overline{Q}\overline{Q}$ systems in pNRQCD, a systematically improvable EFT of QCD.
We do not find evidence that equal-mass tetraquarks exist as stable QCD bound states for asymptotically heavy quark masses at LO and NLO. 
Variational and GFMC calculations are performed for a scan over the one-dimensional space of color wavefunctions and a variety of spatial wavefunctions. The results suggest that equal-mass tetraquarks are unbound, although these methods cannot exclude the possibility that there is a bound state whose wavefunction is approximately orthogonal to those studied here.
Conversely, variational and GFMC calculations provide robust evidence that unequal-mass tetraquarks with a sufficiently large ratio between the quark and antiquark masses form bound states.
Results for the average radii and bound-state color structure suggest that the two heavy constituents form a spatially compact, color-antisymmetric diquark core whose average radius is suppressed by the mass ratio in comparison with the average radii of the lighter constituents.
The critical mass ratio below which bound state formation is observed to occur is 0.15 at LO and 0.12 at NLO.

Experimental searches for tetraquark states at the LHC and other colliders have observed a large number of tetraquark states. The majority of these are heavy-heavy-light-light tetraquarks~\cite{Belle:2011aa,BESIII:2022joj,Brambilla:2019esw,LHCb:2016axx}. More recently, the detection of three $T_{\cccc}$ states whose energies are significantly above two charm-meson mass thresholds have recently been seen at the LHC~\cite{LHCb:2020bwg,CMS:2023owd}.
Signals for the observations of tetraquark bound states and resonances have been extensively studied theoretically~\cite{Alexandrou:2024iwi,Anwar:2017toa,Brambilla:2019esw,Eichten:2017ual,Esposito:2021ptx,Francis:2016hui,Junnarkar:2018twb,Vega-Morales:2017pmm}, but sub-threshold states corresponding to tetraquark bound states have not been definitively observed.

Heavy-heavy-light-light tetraquarks are known from heavy-quark symmetry to be bound for asymptotically large heavy-quark masses~\cite{Manohar:1992nd,Eichten:2017ffp}.
The picture is more complicated for finite quark masses, and for example lattice QCD studies find clear evidence for bound $ud\overline{b}\overline{b}$ tetraquarks~\cite{Francis:2016hui,Francis:2018jyb,Junnarkar:2018twb,Leskovec:2019ioa,Hudspith:2020tdf,Mohanta:2020eed,Bicudo:2021qxj,Meinel:2022lzo,Hudspith:2023loy,Aoki:2023nzp,Alexandrou:2024iwi} but not for bound $uu\overline{b}\overline{b}$ tetraquarks~\cite{Junnarkar:2018twb}.

The results of this work suggest a simpler picture for fully-heavy tetraquarks.
We find no evidence for bound equal-mass fully-heavy tetraquarks, which is consistent with a previous lattice NRQCD study~\cite{Hughes:2017xie} and a variational calculation valid at LO in pNRQCD~\cite{Czarnecki:2017vco} that searched for $T_{\bbbb}$ bound states and found no evidence for their existence.
Conversely, unequal-mass tetraquarks are found to bind for quark/antiquark mass ratios $\lesssim 0.15$ at LO, in agreement with Ref.~\cite{Czarnecki:2017vco}, and mass ratios $\lesssim 0.12$ at NLO.
These results apply to all spin and flavor combinations of $QQ \overline{Q}'\overline{Q}'$ tetraquarks, since heavy-quark spin symmetry is exact at NLO in pNRQCD, and suggest all such fully-heavy tetraquarks are bound for sufficiently unequal quark/antiquark masses.
Our LO and NLO results do not suggest the existence of $T_{\bbbb}$, $T_{cc\overline{b}\overline{b}}$, or $T_{\cccc}$ bound states, which is consistent with their experimental non-observation.
Bound states only exist for tetraquarks including top quarks, namely $T_{bb\overline{tt}}$, $T_{bc\overline{tt}}$, and $T_{cc\overline{tt}}$, which decay before they hadronize.
It would be interesting to extend the calculations here to study tetraquark resonances in these channels that could be associated with the $T_{\cccc}$ states recently seen at the LHC~\cite{LHCb:2020bwg,CMS:2023owd}.

For masses where fully-heavy tetraquarks bind, our results indicate that the average radii of the heavier tetraquark constituents is approximately proportional to the inverse of the heavier quark mass, while the average radii of the lighter tetraquark constituents is approximately proportional to the inverse of the reduced mass.
Together with results for the bound-state color structure indicating that the heavy constituents are in an attractive color-antisymmetric configuration, this suggests that unequal-mass fully-heavy tetraquarks can be understood as spatially compact heavy diquark cores bound by iterated one-gluon-exchange to which the relatively lighter constituents are more weakly bound.
This qualitative picture is consistent with long-standing models of tetraquark structure~\cite{Jaffe:1976ig,Jaffe:1976ih,Ader:1981db}.

Applying the QMC methods used here to pNRQCD, including NNLO potentials, would allow for more precise studies of fully-heavy tetraquarks and other exotic hadrons.
Determining four-body potentials present at NNLO~\cite{Assi:2023cfo} will be essential to enable complete NNLO calculations.
In addition, $1/m_Q^2$-suppressed effects are expected to be comparable to NNLO effects. They will introduce features such as spin-dependent potentials that will add technical complications and could alter our conclusions for non-asymptotic quark masses, including physical $b$-quark masses.
Such $1/m_Q^2$-suppressed effects have been included in lattice NRQCD calculations, however, that also find no evidence for $T_{\bbbb}$ bound states~\cite{Hughes:2017xie}.
The significant differences between pNRQCD results at LO and NLO observed here and in Ref.~\cite{Assi:2023cfo} also suggest that applying resummation and renormalon subtraction~\cite{Beneke:1998ui,Beneke:1998rk, Brambilla:2017hcq,Kronfeld:2024qao,Pineda:2000gza,Pineda:2011dg} to the pNRQCD potential could lead to significant precision improvements.

Applying QMC methods to pNRQCD further opens a new avenue for studying newly discovered or undiscovered exotic hadrons using computationally efficient and systematically improvable EFT approximations to QCD.
Future studies will provide insight into the structure of exotic hadrons comprised of heavy quarks and illuminate which aspects of the complex dynamics of QCD are essential for forming multi-hadron bound states and resonances.

\begin{acknowledgments}
We thank Muhammad Naeem Anwar, Matthew Baumgart, Eric Braaten, Nora Brambilla, William Detmold, Estia Eichten, Majid Ekhterachian, Anthony Grebe, Yingsheng Huang, Stefan Stelzl, Daniel Stolarski, Antonio Vairo, and Ruth Van de Water for helpful discussions and insightful comments. 
This manuscript has been authored by Fermi Research Alliance, LLC under Contract No. DE-AC02-07CH11359 with the U.S. Department of Energy, Office of Science, Office of High Energy Physics. This work was performed in part at the Aspen Center for Physics, which is supported by National Science Foundation grant PHY-2210452.
\end{acknowledgments}

\appendix

\section{Summary of pNRQCD potentials}\label{app:pot}

The pNRQCD Lagrangian defined in Eqs.~\eqref{eq:Lpsichi}-\eqref{eq:Lpsipsi} depends on color-singlet and color-octet quark-antiquark potentials, $V^{\psi\chi}_{\mathbf{1}}$ and $V^{\psi\chi}_{\mathbf{8}}$, as well as color-antisymmetric, and color-symmetric quark-quark potentials, $V^{\psi\psi}_A$ and $V^{\psi\psi}_S$.
These potential are expanded in powers of $1/m_Q$, with the first correction to the $\mathcal{O}(m_Q^0)$ potential arising at $\mathcal{O}(m_Q^{-2})$.
For the NLO calculations performed here it is sufficient to consider the $\mathcal{O}(m_Q^0)$ potential, which is given by~\cite{Kniehl:2002br,Anzai:2013tja,Assi:2023cfo}
\begin{align}\label{eq:pots1}
V^{\psi\chi}_{\mathbf{1}}(|\bs{r}|,\mu)  &= - \left( \frac{4}{3} \right) \frac{\alpha_V(|\bs{r}|,\mu) }{|\bs{r}| }, \\
V^{\psi\chi}_{\mathbf{8}}(|\bs{r}|,\mu)  &= \left( \frac{1}{6} \right) \frac{\alpha_V(|\bs{r}|,\mu) }{|\bs{r}| }, \\
V^{\psi\psi}_{A}(|\bs{r}|,\mu)  &= -  \left( \frac{2}{3} \right) \frac{\alpha_V(|\bs{r}|,\mu) }{|\bs{r}| }, \\
V^{\psi\psi}_{S}(|\bs{r}|,\mu)  &= \left( \frac{1}{3} \right) \frac{\alpha_V(|\bs{r}|,\mu) }{|\bs{r}| },
\label{eq:potsS}
\end{align}
where $\alpha_V(r,\mu)$ has the perturbative expansion
\begin{align}
   \alpha_{V_{\mathbf{1}}}(r,\mu) = \alpha_s(\mu)\left(1+\sum_{n}\left(\frac{\alpha_s(\mu)}{4\pi}\right)^n \tilde{a}_n(r,\mu)\right).
   \label{eqn:alphav1}
\end{align}
The perturbative coefficients $\tilde{a}_n(r,\mu)$ up to $\text{N}^3\text{LO}$ ($n=3$) are given in Ref.~\cite{Kniehl:2002br}; in this work only the NLO coefficient
\begin{align}
    \tilde{a}_1(r,\mu)={}&a_1+2\beta_0\ln(r\mu e^{\eul}),
\end{align}
is required, where 
\begin{align}
    \beta_0 &= 11 - \frac{2}{3} N_f, \hspace{10pt} a_1 = \frac{31}{3} - \frac{10}{9} N_f,
\end{align}
depend on the number of active light-quark flavors, $N_f$.

\begin{figure}[t]
   \includegraphics[width=\linewidth]{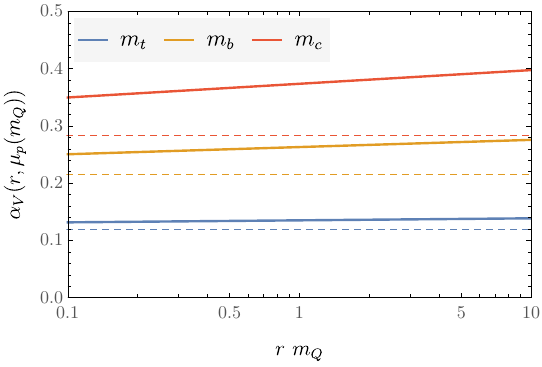} 
   \caption{ The effective coupling  $\alpha_V(r,\mu_p(m_Q))$ governing the strength of the pNRQCD potentials in Eqs.~\eqref{eq:pots1}-\eqref{eq:potsS} as a function of $r$ for the values of $m_Q$ indicated is shown for LO and NLO as dashed and solid lines, respectively. }
\label{fig:alphaV}
\end{figure}

This work uses a renormalization scale $\mu_p$ related to the quark mass by $\mu_p(m_Q) = 4 \alpha_s(\mu_p(m_Q)) m_Q$. 
Fig.~\ref{fig:alphaV} shows $\alpha_V(r,\mu_p(m_Q))$ as a function of $r$ for several values of $m_Q$ in order to provide intuition for the strength and shape of the pNRQCD potentials used here.

\section{Z-type tetraquark binding energy results}\label{app:Z}

Fully-heavy tetraquarks with configurations $tb\overline{tb}$, called Z-type tetraquarks~\cite{Ader:1981db,Braaten:2020nwp,Bicudo:2022cqi,Berwein:2024ztx}, can be studied using pNRQCD analogously to $bb\overline{tt}$ tetraquarks.
The binding energy of a Z-type tetraquark is defined analogously to Eq.~\eqref{eq:Bbbtt}:
\begin{equation}\label{eq:BZ}
    B_{tb\overline{tb}} = \Delta E_{t\overline{t}} + \Delta E_{b\overline{b}} - \Delta E_{tb\overline{tb}},
\end{equation}
where $\Delta E_{t\overline{t}}$ and $\Delta E_{b\overline{b}}$ are the ground-state energies of the corresponding quarkonia. Our trial wavefunctions span the color space and include various spatial configurations as described in the main text.

\begin{figure}[tb]
 \includegraphics[width=\linewidth]{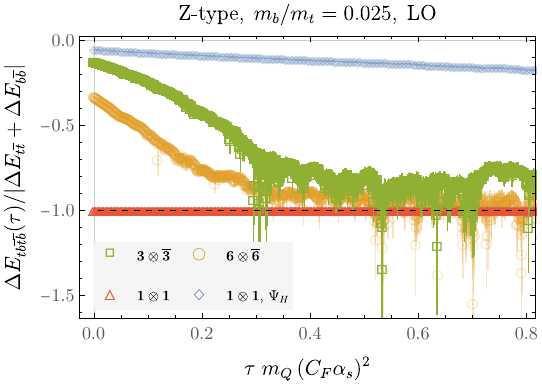} \\
   \includegraphics[width=\linewidth]{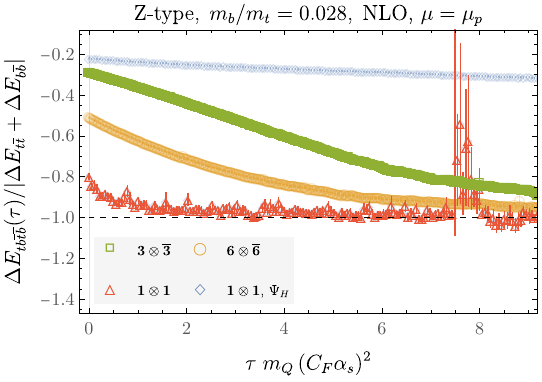} 
   \caption{
   Top: example effective energies for highly unequal-mass Z-type tetraquarks at LO in pNRQCD. Different colors/markers correspond to trial wavefunctions with different colored wavefunctions described in the main text.  
   Bottom: analogous results at NLO using $m_b$ from Ref.~\cite{Assi:2023cfo} with the corresponding renormalization scale $\mu_p$ defined in Eq.~\eqref{eq:mup} and $m_t=172.76\,\rm{GeV}$.
\label{fig:Ztype_effmass} }
\end{figure}

The computed binding energies for these Z-type tetraquarks are consistent with zero within statistical uncertainties at both LO and NLO in pNRQCD, indicating that these systems are unbound, as illustrated in Fig.~\ref{fig:Ztype_effmass}. The effective energies as a function of imaginary time $\tau$ approach the sum of the quarkonium energies for all trial states studies, though much faster for some trial states than others.
Even for mass ratios as small as $m_b / m_t = 0.025$, our results do not provide any evidence in favor of Z-type tetraquark binding at LO or NLO in pNRQCD.

\section{Hylleraas and Ore wavefunction}\label{app:HO}

The variational wavefunction used in Ref.~\cite{Hylleraas:1947zza} includes a free parameter $k$ that rescales the coordinates as $\bs{r} \rightarrow \bs{r}/k$.
The optimal value of this parameter that minimizes the energy can be obtained analytically and corresponds to $k(\beta)$ in Eq.~\eqref{eqn:hylqed}.
The corresponding energy is denoted
\begin{equation}
  \begin{split}
    \Delta E(\beta) &= k(\beta)^2 M(\beta) - k(\beta) L(\beta),
  \end{split}
\end{equation}
where $M(\beta)$ and $L(\beta)$ are matrix elements of the kinetic and potential operators, respectively, computed in Ref.~\cite{Hylleraas:1947zza} and presented below.
This is minimized by
\begin{equation}
  k(\beta) = \frac{L(\beta)}{2 M(\beta)}, \label{eq:kbeta}
\end{equation}
with the result that
\begin{equation}
  \begin{split}
    \Delta E(\beta) &= - \frac{L(\beta)^2}{4M(\beta)}.
  \end{split}
\end{equation}
The variational wavefunction used in Ref.~\cite{Hylleraas:1947zza} is therefore given by Eq.~\eqref{eqn:hylqed} with $k(\beta)$ given by Eq.~\eqref{eq:kbeta} and~\cite{Hylleraas:1947zza}
\begin{equation}
  \begin{split}
    L(\beta) &= \frac{19}{6} - \frac{5 \beta^4 - 18 \beta^2 + 21}{ 4 (\beta^2 - 1)^3}  \\
    &+ \frac{ 5 \beta^8 - 8 \beta^6  - 7 \beta^4  + 2 \beta^2  + 2 (-1 + \beta^2)^4 \ln(1 - \beta^2)}{ 8 \beta^6 (\beta^2 - 1)^2}, \\
    M(\beta) &= \frac{21}{4} - \frac{3\beta^2}{2} - \frac{\beta^2 - 6\beta^2 + 21 }{ 8 (\beta^2 -1 )^3}.
  \end{split}
\end{equation}

\bibliography{pnrqcd}

\begin{thebibliography}{109}%
\makeatletter
\providecommand \@ifxundefined [1]{%
 \@ifx{#1\undefined}
}%
\providecommand \@ifnum [1]{%
 \ifnum #1\expandafter \@firstoftwo
 \else \expandafter \@secondoftwo
 \fi
}%
\providecommand \@ifx [1]{%
 \ifx #1\expandafter \@firstoftwo
 \else \expandafter \@secondoftwo
 \fi
}%
\providecommand \natexlab [1]{#1}%
\providecommand \enquote  [1]{``#1''}%
\providecommand \bibnamefont  [1]{#1}%
\providecommand \bibfnamefont [1]{#1}%
\providecommand \citenamefont [1]{#1}%
\providecommand \href@noop [0]{\@secondoftwo}%
\providecommand \href [0]{\begingroup \@sanitize@url \@href}%
\providecommand \@href[1]{\@@startlink{#1}\@@href}%
\providecommand \@@href[1]{\endgroup#1\@@endlink}%
\providecommand \@sanitize@url [0]{\catcode `\\12\catcode `\$12\catcode `\&12\catcode `\#12\catcode `\^12\catcode `\_12\catcode `\%12\relax}%
\providecommand \@@startlink[1]{}%
\providecommand \@@endlink[0]{}%
\providecommand \url  [0]{\begingroup\@sanitize@url \@url }%
\providecommand \@url [1]{\endgroup\@href {#1}{\urlprefix }}%
\providecommand \urlprefix  [0]{URL }%
\providecommand \Eprint [0]{\href }%
\providecommand \doibase [0]{http://dx.doi.org/}%
\providecommand \selectlanguage [0]{\@gobble}%
\providecommand \bibinfo  [0]{\@secondoftwo}%
\providecommand \bibfield  [0]{\@secondoftwo}%
\providecommand \translation [1]{[#1]}%
\providecommand \BibitemOpen [0]{}%
\providecommand \bibitemStop [0]{}%
\providecommand \bibitemNoStop [0]{.\EOS\space}%
\providecommand \EOS [0]{\spacefactor3000\relax}%
\providecommand \BibitemShut  [1]{\csname bibitem#1\endcsname}%
\let\auto@bib@innerbib\@empty
\bibitem [{\citenamefont {Jaffe}(1977{\natexlab{a}})}]{Jaffe:1976ig}%
  \BibitemOpen
  \bibfield  {author} {\bibinfo {author} {\bibfnamefont {R.~L.}\ \bibnamefont {Jaffe}},\ }\href {\doibase 10.1103/PhysRevD.15.267} {\bibfield  {journal} {\bibinfo  {journal} {Phys. Rev. D}\ }\textbf {\bibinfo {volume} {15}},\ \bibinfo {pages} {267} (\bibinfo {year} {1977}{\natexlab{a}})}\BibitemShut {NoStop}%
\bibitem [{\citenamefont {Jaffe}(1977{\natexlab{b}})}]{Jaffe:1976ih}%
  \BibitemOpen
  \bibfield  {author} {\bibinfo {author} {\bibfnamefont {R.~L.}\ \bibnamefont {Jaffe}},\ }\href {\doibase 10.1103/PhysRevD.15.281} {\bibfield  {journal} {\bibinfo  {journal} {Phys. Rev. D}\ }\textbf {\bibinfo {volume} {15}},\ \bibinfo {pages} {281} (\bibinfo {year} {1977}{\natexlab{b}})}\BibitemShut {NoStop}%
\bibitem [{\citenamefont {Brambilla}\ \emph {et~al.}(2020)\citenamefont {Brambilla}, \citenamefont {Eidelman}, \citenamefont {Hanhart}, \citenamefont {Nefediev}, \citenamefont {Shen}, \citenamefont {Thomas}, \citenamefont {Vairo},\ and\ \citenamefont {Yuan}}]{Brambilla:2019esw}%
  \BibitemOpen
  \bibfield  {author} {\bibinfo {author} {\bibfnamefont {N.}~\bibnamefont {Brambilla}}, \bibinfo {author} {\bibfnamefont {S.}~\bibnamefont {Eidelman}}, \bibinfo {author} {\bibfnamefont {C.}~\bibnamefont {Hanhart}}, \bibinfo {author} {\bibfnamefont {A.}~\bibnamefont {Nefediev}}, \bibinfo {author} {\bibfnamefont {C.-P.}\ \bibnamefont {Shen}}, \bibinfo {author} {\bibfnamefont {C.~E.}\ \bibnamefont {Thomas}}, \bibinfo {author} {\bibfnamefont {A.}~\bibnamefont {Vairo}}, \ and\ \bibinfo {author} {\bibfnamefont {C.-Z.}\ \bibnamefont {Yuan}},\ }\href {\doibase 10.1016/j.physrep.2020.05.001} {\bibfield  {journal} {\bibinfo  {journal} {Phys. Rept.}\ }\textbf {\bibinfo {volume} {873}},\ \bibinfo {pages} {1} (\bibinfo {year} {2020})},\ \Eprint {http://arxiv.org/abs/1907.07583} {arXiv:1907.07583 [hep-ex]} \BibitemShut {NoStop}%
\bibitem [{\citenamefont {Chen}\ \emph {et~al.}(2023)\citenamefont {Chen}, \citenamefont {Chen}, \citenamefont {Liu}, \citenamefont {Liu},\ and\ \citenamefont {Zhu}}]{Chen:2022asf}%
  \BibitemOpen
  \bibfield  {author} {\bibinfo {author} {\bibfnamefont {H.-X.}\ \bibnamefont {Chen}}, \bibinfo {author} {\bibfnamefont {W.}~\bibnamefont {Chen}}, \bibinfo {author} {\bibfnamefont {X.}~\bibnamefont {Liu}}, \bibinfo {author} {\bibfnamefont {Y.-R.}\ \bibnamefont {Liu}}, \ and\ \bibinfo {author} {\bibfnamefont {S.-L.}\ \bibnamefont {Zhu}},\ }\href {\doibase 10.1088/1361-6633/aca3b6} {\bibfield  {journal} {\bibinfo  {journal} {Rept. Prog. Phys.}\ }\textbf {\bibinfo {volume} {86}},\ \bibinfo {pages} {026201} (\bibinfo {year} {2023})},\ \Eprint {http://arxiv.org/abs/2204.02649} {arXiv:2204.02649 [hep-ph]} \BibitemShut {NoStop}%
\bibitem [{\citenamefont {Workman}\ \emph {et~al.}(2022)\citenamefont {Workman} \emph {et~al.}}]{ParticleDataGroup:2022pth}%
  \BibitemOpen
  \bibfield  {author} {\bibinfo {author} {\bibfnamefont {R.~L.}\ \bibnamefont {Workman}} \emph {et~al.} (\bibinfo {collaboration} {Particle Data Group}),\ }\href {\doibase 10.1093/ptep/ptac097} {\bibfield  {journal} {\bibinfo  {journal} {PTEP}\ }\textbf {\bibinfo {volume} {2022}},\ \bibinfo {pages} {083C01} (\bibinfo {year} {2022})}\BibitemShut {NoStop}%
\bibitem [{\citenamefont {Drenska}\ \emph {et~al.}(2010)\citenamefont {Drenska}, \citenamefont {Faccini}, \citenamefont {Piccinini}, \citenamefont {Polosa}, \citenamefont {Renga},\ and\ \citenamefont {Sabelli}}]{Drenska:2010kg}%
  \BibitemOpen
  \bibfield  {author} {\bibinfo {author} {\bibfnamefont {N.}~\bibnamefont {Drenska}}, \bibinfo {author} {\bibfnamefont {R.}~\bibnamefont {Faccini}}, \bibinfo {author} {\bibfnamefont {F.}~\bibnamefont {Piccinini}}, \bibinfo {author} {\bibfnamefont {A.}~\bibnamefont {Polosa}}, \bibinfo {author} {\bibfnamefont {F.}~\bibnamefont {Renga}}, \ and\ \bibinfo {author} {\bibfnamefont {C.}~\bibnamefont {Sabelli}},\ }\href {\doibase 10.1393/ncr/i2010-10059-8} {\bibfield  {journal} {\bibinfo  {journal} {Riv. Nuovo Cim.}\ }\textbf {\bibinfo {volume} {33}},\ \bibinfo {pages} {633} (\bibinfo {year} {2010})},\ \Eprint {http://arxiv.org/abs/1006.2741} {arXiv:1006.2741 [hep-ph]} \BibitemShut {NoStop}%
\bibitem [{\citenamefont {Brambilla}\ \emph {et~al.}(2011)\citenamefont {Brambilla} \emph {et~al.}}]{Brambilla:2010cs}%
  \BibitemOpen
  \bibfield  {author} {\bibinfo {author} {\bibfnamefont {N.}~\bibnamefont {Brambilla}} \emph {et~al.},\ }\href {\doibase 10.1140/epjc/s10052-010-1534-9} {\bibfield  {journal} {\bibinfo  {journal} {Eur. Phys. J. C}\ }\textbf {\bibinfo {volume} {71}},\ \bibinfo {pages} {1534} (\bibinfo {year} {2011})},\ \Eprint {http://arxiv.org/abs/1010.5827} {arXiv:1010.5827 [hep-ph]} \BibitemShut {NoStop}%
\bibitem [{\citenamefont {Brambilla}\ \emph {et~al.}(2014)\citenamefont {Brambilla} \emph {et~al.}}]{Brambilla:2014jmp}%
  \BibitemOpen
  \bibfield  {author} {\bibinfo {author} {\bibfnamefont {N.}~\bibnamefont {Brambilla}} \emph {et~al.},\ }\href {\doibase 10.1140/epjc/s10052-014-2981-5} {\bibfield  {journal} {\bibinfo  {journal} {Eur. Phys. J. C}\ }\textbf {\bibinfo {volume} {74}},\ \bibinfo {pages} {2981} (\bibinfo {year} {2014})},\ \Eprint {http://arxiv.org/abs/1404.3723} {arXiv:1404.3723 [hep-ph]} \BibitemShut {NoStop}%
\bibitem [{\citenamefont {Esposito}\ \emph {et~al.}(2015)\citenamefont {Esposito}, \citenamefont {Guerrieri}, \citenamefont {Piccinini}, \citenamefont {Pilloni},\ and\ \citenamefont {Polosa}}]{Esposito:2014rxa}%
  \BibitemOpen
  \bibfield  {author} {\bibinfo {author} {\bibfnamefont {A.}~\bibnamefont {Esposito}}, \bibinfo {author} {\bibfnamefont {A.~L.}\ \bibnamefont {Guerrieri}}, \bibinfo {author} {\bibfnamefont {F.}~\bibnamefont {Piccinini}}, \bibinfo {author} {\bibfnamefont {A.}~\bibnamefont {Pilloni}}, \ and\ \bibinfo {author} {\bibfnamefont {A.~D.}\ \bibnamefont {Polosa}},\ }\href {\doibase 10.1142/S0217751X15300021} {\bibfield  {journal} {\bibinfo  {journal} {Int. J. Mod. Phys. A}\ }\textbf {\bibinfo {volume} {30}},\ \bibinfo {pages} {1530002} (\bibinfo {year} {2015})},\ \Eprint {http://arxiv.org/abs/1411.5997} {arXiv:1411.5997 [hep-ph]} \BibitemShut {NoStop}%
\bibitem [{\citenamefont {Lebed}\ \emph {et~al.}(2017)\citenamefont {Lebed}, \citenamefont {Mitchell},\ and\ \citenamefont {Swanson}}]{Lebed:2016hpi}%
  \BibitemOpen
  \bibfield  {author} {\bibinfo {author} {\bibfnamefont {R.~F.}\ \bibnamefont {Lebed}}, \bibinfo {author} {\bibfnamefont {R.~E.}\ \bibnamefont {Mitchell}}, \ and\ \bibinfo {author} {\bibfnamefont {E.~S.}\ \bibnamefont {Swanson}},\ }\href {\doibase 10.1016/j.ppnp.2016.11.003} {\bibfield  {journal} {\bibinfo  {journal} {Prog. Part. Nucl. Phys.}\ }\textbf {\bibinfo {volume} {93}},\ \bibinfo {pages} {143} (\bibinfo {year} {2017})},\ \Eprint {http://arxiv.org/abs/1610.04528} {arXiv:1610.04528 [hep-ph]} \BibitemShut {NoStop}%
\bibitem [{\citenamefont {Chen}\ \emph {et~al.}(2016)\citenamefont {Chen}, \citenamefont {Chen}, \citenamefont {Liu},\ and\ \citenamefont {Zhu}}]{Chen:2016qju}%
  \BibitemOpen
  \bibfield  {author} {\bibinfo {author} {\bibfnamefont {H.-X.}\ \bibnamefont {Chen}}, \bibinfo {author} {\bibfnamefont {W.}~\bibnamefont {Chen}}, \bibinfo {author} {\bibfnamefont {X.}~\bibnamefont {Liu}}, \ and\ \bibinfo {author} {\bibfnamefont {S.-L.}\ \bibnamefont {Zhu}},\ }\href {\doibase 10.1016/j.physrep.2016.05.004} {\bibfield  {journal} {\bibinfo  {journal} {Phys. Rept.}\ }\textbf {\bibinfo {volume} {639}},\ \bibinfo {pages} {1} (\bibinfo {year} {2016})},\ \Eprint {http://arxiv.org/abs/1601.02092} {arXiv:1601.02092 [hep-ph]} \BibitemShut {NoStop}%
\bibitem [{\citenamefont {Liu}\ \emph {et~al.}(2019)\citenamefont {Liu}, \citenamefont {Chen}, \citenamefont {Chen}, \citenamefont {Liu},\ and\ \citenamefont {Zhu}}]{Liu:2019zoy}%
  \BibitemOpen
  \bibfield  {author} {\bibinfo {author} {\bibfnamefont {Y.-R.}\ \bibnamefont {Liu}}, \bibinfo {author} {\bibfnamefont {H.-X.}\ \bibnamefont {Chen}}, \bibinfo {author} {\bibfnamefont {W.}~\bibnamefont {Chen}}, \bibinfo {author} {\bibfnamefont {X.}~\bibnamefont {Liu}}, \ and\ \bibinfo {author} {\bibfnamefont {S.-L.}\ \bibnamefont {Zhu}},\ }\href {\doibase 10.1016/j.ppnp.2019.04.003} {\bibfield  {journal} {\bibinfo  {journal} {Prog. Part. Nucl. Phys.}\ }\textbf {\bibinfo {volume} {107}},\ \bibinfo {pages} {237} (\bibinfo {year} {2019})},\ \Eprint {http://arxiv.org/abs/1903.11976} {arXiv:1903.11976 [hep-ph]} \BibitemShut {NoStop}%
\bibitem [{\citenamefont {Bicudo}\ and\ \citenamefont {Wagner}(2013)}]{Bicudo:2012qt}%
  \BibitemOpen
  \bibfield  {author} {\bibinfo {author} {\bibfnamefont {P.}~\bibnamefont {Bicudo}}\ and\ \bibinfo {author} {\bibfnamefont {M.}~\bibnamefont {Wagner}} (\bibinfo {collaboration} {European Twisted Mass}),\ }\href {\doibase 10.1103/PhysRevD.87.114511} {\bibfield  {journal} {\bibinfo  {journal} {Phys. Rev. D}\ }\textbf {\bibinfo {volume} {87}},\ \bibinfo {pages} {114511} (\bibinfo {year} {2013})},\ \Eprint {http://arxiv.org/abs/1209.6274} {arXiv:1209.6274 [hep-ph]} \BibitemShut {NoStop}%
\bibitem [{\citenamefont {Brown}\ and\ \citenamefont {Orginos}(2012)}]{Brown:2012tm}%
  \BibitemOpen
  \bibfield  {author} {\bibinfo {author} {\bibfnamefont {Z.~S.}\ \bibnamefont {Brown}}\ and\ \bibinfo {author} {\bibfnamefont {K.}~\bibnamefont {Orginos}},\ }\href {\doibase 10.1103/PhysRevD.86.114506} {\bibfield  {journal} {\bibinfo  {journal} {Phys. Rev. D}\ }\textbf {\bibinfo {volume} {86}},\ \bibinfo {pages} {114506} (\bibinfo {year} {2012})},\ \Eprint {http://arxiv.org/abs/1210.1953} {arXiv:1210.1953 [hep-lat]} \BibitemShut {NoStop}%
\bibitem [{\citenamefont {Bicudo}\ \emph {et~al.}(2015)\citenamefont {Bicudo}, \citenamefont {Cichy}, \citenamefont {Peters}, \citenamefont {Wagenbach},\ and\ \citenamefont {Wagner}}]{Bicudo:2015vta}%
  \BibitemOpen
  \bibfield  {author} {\bibinfo {author} {\bibfnamefont {P.}~\bibnamefont {Bicudo}}, \bibinfo {author} {\bibfnamefont {K.}~\bibnamefont {Cichy}}, \bibinfo {author} {\bibfnamefont {A.}~\bibnamefont {Peters}}, \bibinfo {author} {\bibfnamefont {B.}~\bibnamefont {Wagenbach}}, \ and\ \bibinfo {author} {\bibfnamefont {M.}~\bibnamefont {Wagner}},\ }\href {\doibase 10.1103/PhysRevD.92.014507} {\bibfield  {journal} {\bibinfo  {journal} {Phys. Rev. D}\ }\textbf {\bibinfo {volume} {92}},\ \bibinfo {pages} {014507} (\bibinfo {year} {2015})},\ \Eprint {http://arxiv.org/abs/1505.00613} {arXiv:1505.00613 [hep-lat]} \BibitemShut {NoStop}%
\bibitem [{\citenamefont {Francis}\ \emph {et~al.}(2017)\citenamefont {Francis}, \citenamefont {Hudspith}, \citenamefont {Lewis},\ and\ \citenamefont {Maltman}}]{Francis:2016hui}%
  \BibitemOpen
  \bibfield  {author} {\bibinfo {author} {\bibfnamefont {A.}~\bibnamefont {Francis}}, \bibinfo {author} {\bibfnamefont {R.~J.}\ \bibnamefont {Hudspith}}, \bibinfo {author} {\bibfnamefont {R.}~\bibnamefont {Lewis}}, \ and\ \bibinfo {author} {\bibfnamefont {K.}~\bibnamefont {Maltman}},\ }\href {\doibase 10.1103/PhysRevLett.118.142001} {\bibfield  {journal} {\bibinfo  {journal} {Phys. Rev. Lett.}\ }\textbf {\bibinfo {volume} {118}},\ \bibinfo {pages} {142001} (\bibinfo {year} {2017})},\ \Eprint {http://arxiv.org/abs/1607.05214} {arXiv:1607.05214 [hep-lat]} \BibitemShut {NoStop}%
\bibitem [{\citenamefont {Francis}\ \emph {et~al.}(2019)\citenamefont {Francis}, \citenamefont {Hudspith}, \citenamefont {Lewis},\ and\ \citenamefont {Maltman}}]{Francis:2018jyb}%
  \BibitemOpen
  \bibfield  {author} {\bibinfo {author} {\bibfnamefont {A.}~\bibnamefont {Francis}}, \bibinfo {author} {\bibfnamefont {R.~J.}\ \bibnamefont {Hudspith}}, \bibinfo {author} {\bibfnamefont {R.}~\bibnamefont {Lewis}}, \ and\ \bibinfo {author} {\bibfnamefont {K.}~\bibnamefont {Maltman}},\ }\href {\doibase 10.1103/PhysRevD.99.054505} {\bibfield  {journal} {\bibinfo  {journal} {Phys. Rev. D}\ }\textbf {\bibinfo {volume} {99}},\ \bibinfo {pages} {054505} (\bibinfo {year} {2019})},\ \Eprint {http://arxiv.org/abs/1810.10550} {arXiv:1810.10550 [hep-lat]} \BibitemShut {NoStop}%
\bibitem [{\citenamefont {Junnarkar}\ \emph {et~al.}(2019)\citenamefont {Junnarkar}, \citenamefont {Mathur},\ and\ \citenamefont {Padmanath}}]{Junnarkar:2018twb}%
  \BibitemOpen
  \bibfield  {author} {\bibinfo {author} {\bibfnamefont {P.}~\bibnamefont {Junnarkar}}, \bibinfo {author} {\bibfnamefont {N.}~\bibnamefont {Mathur}}, \ and\ \bibinfo {author} {\bibfnamefont {M.}~\bibnamefont {Padmanath}},\ }\href {\doibase 10.1103/PhysRevD.99.034507} {\bibfield  {journal} {\bibinfo  {journal} {Phys. Rev. D}\ }\textbf {\bibinfo {volume} {99}},\ \bibinfo {pages} {034507} (\bibinfo {year} {2019})},\ \Eprint {http://arxiv.org/abs/1810.12285} {arXiv:1810.12285 [hep-lat]} \BibitemShut {NoStop}%
\bibitem [{\citenamefont {Leskovec}\ \emph {et~al.}(2019)\citenamefont {Leskovec}, \citenamefont {Meinel}, \citenamefont {Pflaumer},\ and\ \citenamefont {Wagner}}]{Leskovec:2019ioa}%
  \BibitemOpen
  \bibfield  {author} {\bibinfo {author} {\bibfnamefont {L.}~\bibnamefont {Leskovec}}, \bibinfo {author} {\bibfnamefont {S.}~\bibnamefont {Meinel}}, \bibinfo {author} {\bibfnamefont {M.}~\bibnamefont {Pflaumer}}, \ and\ \bibinfo {author} {\bibfnamefont {M.}~\bibnamefont {Wagner}},\ }\href {\doibase 10.1103/PhysRevD.100.014503} {\bibfield  {journal} {\bibinfo  {journal} {Phys. Rev. D}\ }\textbf {\bibinfo {volume} {100}},\ \bibinfo {pages} {014503} (\bibinfo {year} {2019})},\ \Eprint {http://arxiv.org/abs/1904.04197} {arXiv:1904.04197 [hep-lat]} \BibitemShut {NoStop}%
\bibitem [{\citenamefont {Hudspith}\ \emph {et~al.}(2020)\citenamefont {Hudspith}, \citenamefont {Colquhoun}, \citenamefont {Francis}, \citenamefont {Lewis},\ and\ \citenamefont {Maltman}}]{Hudspith:2020tdf}%
  \BibitemOpen
  \bibfield  {author} {\bibinfo {author} {\bibfnamefont {R.~J.}\ \bibnamefont {Hudspith}}, \bibinfo {author} {\bibfnamefont {B.}~\bibnamefont {Colquhoun}}, \bibinfo {author} {\bibfnamefont {A.}~\bibnamefont {Francis}}, \bibinfo {author} {\bibfnamefont {R.}~\bibnamefont {Lewis}}, \ and\ \bibinfo {author} {\bibfnamefont {K.}~\bibnamefont {Maltman}},\ }\href {\doibase 10.1103/PhysRevD.102.114506} {\bibfield  {journal} {\bibinfo  {journal} {Phys. Rev. D}\ }\textbf {\bibinfo {volume} {102}},\ \bibinfo {pages} {114506} (\bibinfo {year} {2020})},\ \Eprint {http://arxiv.org/abs/2006.14294} {arXiv:2006.14294 [hep-lat]} \BibitemShut {NoStop}%
\bibitem [{\citenamefont {Mohanta}\ and\ \citenamefont {Basak}(2020)}]{Mohanta:2020eed}%
  \BibitemOpen
  \bibfield  {author} {\bibinfo {author} {\bibfnamefont {P.}~\bibnamefont {Mohanta}}\ and\ \bibinfo {author} {\bibfnamefont {S.}~\bibnamefont {Basak}},\ }\href {\doibase 10.1103/PhysRevD.102.094516} {\bibfield  {journal} {\bibinfo  {journal} {Phys. Rev. D}\ }\textbf {\bibinfo {volume} {102}},\ \bibinfo {pages} {094516} (\bibinfo {year} {2020})},\ \Eprint {http://arxiv.org/abs/2008.11146} {arXiv:2008.11146 [hep-lat]} \BibitemShut {NoStop}%
\bibitem [{\citenamefont {Bicudo}\ \emph {et~al.}(2021)\citenamefont {Bicudo}, \citenamefont {Peters}, \citenamefont {Velten},\ and\ \citenamefont {Wagner}}]{Bicudo:2021qxj}%
  \BibitemOpen
  \bibfield  {author} {\bibinfo {author} {\bibfnamefont {P.}~\bibnamefont {Bicudo}}, \bibinfo {author} {\bibfnamefont {A.}~\bibnamefont {Peters}}, \bibinfo {author} {\bibfnamefont {S.}~\bibnamefont {Velten}}, \ and\ \bibinfo {author} {\bibfnamefont {M.}~\bibnamefont {Wagner}},\ }\href {\doibase 10.1103/PhysRevD.103.114506} {\bibfield  {journal} {\bibinfo  {journal} {Phys. Rev. D}\ }\textbf {\bibinfo {volume} {103}},\ \bibinfo {pages} {114506} (\bibinfo {year} {2021})},\ \Eprint {http://arxiv.org/abs/2101.00723} {arXiv:2101.00723 [hep-lat]} \BibitemShut {NoStop}%
\bibitem [{\citenamefont {Padmanath}\ and\ \citenamefont {Prelovsek}(2022)}]{Padmanath:2022cvl}%
  \BibitemOpen
  \bibfield  {author} {\bibinfo {author} {\bibfnamefont {M.}~\bibnamefont {Padmanath}}\ and\ \bibinfo {author} {\bibfnamefont {S.}~\bibnamefont {Prelovsek}},\ }\href {\doibase 10.1103/PhysRevLett.129.032002} {\bibfield  {journal} {\bibinfo  {journal} {Phys. Rev. Lett.}\ }\textbf {\bibinfo {volume} {129}},\ \bibinfo {pages} {032002} (\bibinfo {year} {2022})},\ \Eprint {http://arxiv.org/abs/2202.10110} {arXiv:2202.10110 [hep-lat]} \BibitemShut {NoStop}%
\bibitem [{\citenamefont {Meinel}\ \emph {et~al.}(2022)\citenamefont {Meinel}, \citenamefont {Pflaumer},\ and\ \citenamefont {Wagner}}]{Meinel:2022lzo}%
  \BibitemOpen
  \bibfield  {author} {\bibinfo {author} {\bibfnamefont {S.}~\bibnamefont {Meinel}}, \bibinfo {author} {\bibfnamefont {M.}~\bibnamefont {Pflaumer}}, \ and\ \bibinfo {author} {\bibfnamefont {M.}~\bibnamefont {Wagner}},\ }\href {\doibase 10.1103/PhysRevD.106.034507} {\bibfield  {journal} {\bibinfo  {journal} {Phys. Rev. D}\ }\textbf {\bibinfo {volume} {106}},\ \bibinfo {pages} {034507} (\bibinfo {year} {2022})},\ \Eprint {http://arxiv.org/abs/2205.13982} {arXiv:2205.13982 [hep-lat]} \BibitemShut {NoStop}%
\bibitem [{\citenamefont {Bicudo}(2023)}]{Bicudo:2022cqi}%
  \BibitemOpen
  \bibfield  {author} {\bibinfo {author} {\bibfnamefont {P.}~\bibnamefont {Bicudo}},\ }\href {\doibase 10.1016/j.physrep.2023.10.001} {\bibfield  {journal} {\bibinfo  {journal} {Phys. Rept.}\ }\textbf {\bibinfo {volume} {1039}},\ \bibinfo {pages} {1} (\bibinfo {year} {2023})},\ \Eprint {http://arxiv.org/abs/2212.07793} {arXiv:2212.07793 [hep-lat]} \BibitemShut {NoStop}%
\bibitem [{\citenamefont {Lyu}\ \emph {et~al.}(2023)\citenamefont {Lyu}, \citenamefont {Aoki}, \citenamefont {Doi}, \citenamefont {Hatsuda}, \citenamefont {Ikeda},\ and\ \citenamefont {Meng}}]{Lyu:2023xro}%
  \BibitemOpen
  \bibfield  {author} {\bibinfo {author} {\bibfnamefont {Y.}~\bibnamefont {Lyu}}, \bibinfo {author} {\bibfnamefont {S.}~\bibnamefont {Aoki}}, \bibinfo {author} {\bibfnamefont {T.}~\bibnamefont {Doi}}, \bibinfo {author} {\bibfnamefont {T.}~\bibnamefont {Hatsuda}}, \bibinfo {author} {\bibfnamefont {Y.}~\bibnamefont {Ikeda}}, \ and\ \bibinfo {author} {\bibfnamefont {J.}~\bibnamefont {Meng}},\ }\href {\doibase 10.1103/PhysRevLett.131.161901} {\bibfield  {journal} {\bibinfo  {journal} {Phys. Rev. Lett.}\ }\textbf {\bibinfo {volume} {131}},\ \bibinfo {pages} {161901} (\bibinfo {year} {2023})},\ \Eprint {http://arxiv.org/abs/2302.04505} {arXiv:2302.04505 [hep-lat]} \BibitemShut {NoStop}%
\bibitem [{\citenamefont {Hudspith}\ and\ \citenamefont {Mohler}(2023)}]{Hudspith:2023loy}%
  \BibitemOpen
  \bibfield  {author} {\bibinfo {author} {\bibfnamefont {R.~J.}\ \bibnamefont {Hudspith}}\ and\ \bibinfo {author} {\bibfnamefont {D.}~\bibnamefont {Mohler}},\ }\href {\doibase 10.1103/PhysRevD.107.114510} {\bibfield  {journal} {\bibinfo  {journal} {Phys. Rev. D}\ }\textbf {\bibinfo {volume} {107}},\ \bibinfo {pages} {114510} (\bibinfo {year} {2023})},\ \Eprint {http://arxiv.org/abs/2303.17295} {arXiv:2303.17295 [hep-lat]} \BibitemShut {NoStop}%
\bibitem [{\citenamefont {Aoki}\ \emph {et~al.}(2023)\citenamefont {Aoki}, \citenamefont {Aoki},\ and\ \citenamefont {Inoue}}]{Aoki:2023nzp}%
  \BibitemOpen
  \bibfield  {author} {\bibinfo {author} {\bibfnamefont {T.}~\bibnamefont {Aoki}}, \bibinfo {author} {\bibfnamefont {S.}~\bibnamefont {Aoki}}, \ and\ \bibinfo {author} {\bibfnamefont {T.}~\bibnamefont {Inoue}},\ }\href {\doibase 10.1103/PhysRevD.108.054502} {\bibfield  {journal} {\bibinfo  {journal} {Phys. Rev. D}\ }\textbf {\bibinfo {volume} {108}},\ \bibinfo {pages} {054502} (\bibinfo {year} {2023})},\ \Eprint {http://arxiv.org/abs/2306.03565} {arXiv:2306.03565 [hep-lat]} \BibitemShut {NoStop}%
\bibitem [{\citenamefont {Padmanath}\ \emph {et~al.}(2024)\citenamefont {Padmanath}, \citenamefont {Radhakrishnan},\ and\ \citenamefont {Mathur}}]{Padmanath:2023rdu}%
  \BibitemOpen
  \bibfield  {author} {\bibinfo {author} {\bibfnamefont {M.}~\bibnamefont {Padmanath}}, \bibinfo {author} {\bibfnamefont {A.}~\bibnamefont {Radhakrishnan}}, \ and\ \bibinfo {author} {\bibfnamefont {N.}~\bibnamefont {Mathur}},\ }\href {\doibase 10.1103/PhysRevLett.132.201902} {\bibfield  {journal} {\bibinfo  {journal} {Phys. Rev. Lett.}\ }\textbf {\bibinfo {volume} {132}},\ \bibinfo {pages} {201902} (\bibinfo {year} {2024})},\ \Eprint {http://arxiv.org/abs/2307.14128} {arXiv:2307.14128 [hep-lat]} \BibitemShut {NoStop}%
\bibitem [{\citenamefont {Alexandrou}\ \emph {et~al.}(2024)\citenamefont {Alexandrou}, \citenamefont {Finkenrath}, \citenamefont {Leontiou}, \citenamefont {Meinel}, \citenamefont {Pflaumer},\ and\ \citenamefont {Wagner}}]{Alexandrou:2024iwi}%
  \BibitemOpen
  \bibfield  {author} {\bibinfo {author} {\bibfnamefont {C.}~\bibnamefont {Alexandrou}}, \bibinfo {author} {\bibfnamefont {J.}~\bibnamefont {Finkenrath}}, \bibinfo {author} {\bibfnamefont {T.}~\bibnamefont {Leontiou}}, \bibinfo {author} {\bibfnamefont {S.}~\bibnamefont {Meinel}}, \bibinfo {author} {\bibfnamefont {M.}~\bibnamefont {Pflaumer}}, \ and\ \bibinfo {author} {\bibfnamefont {M.}~\bibnamefont {Wagner}},\ }\href {\doibase 10.1103/PhysRevD.110.054510} {\bibfield  {journal} {\bibinfo  {journal} {Phys. Rev. D}\ }\textbf {\bibinfo {volume} {110}},\ \bibinfo {pages} {054510} (\bibinfo {year} {2024})},\ \Eprint {http://arxiv.org/abs/2404.03588} {arXiv:2404.03588 [hep-lat]} \BibitemShut {NoStop}%
\bibitem [{\citenamefont {Bodwin}\ \emph {et~al.}(1995)\citenamefont {Bodwin}, \citenamefont {Braaten},\ and\ \citenamefont {Lepage}}]{Bodwin:1994jh}%
  \BibitemOpen
  \bibfield  {author} {\bibinfo {author} {\bibfnamefont {G.~T.}\ \bibnamefont {Bodwin}}, \bibinfo {author} {\bibfnamefont {E.}~\bibnamefont {Braaten}}, \ and\ \bibinfo {author} {\bibfnamefont {G.~P.}\ \bibnamefont {Lepage}},\ }\href {\doibase 10.1103/PhysRevD.55.5853} {\bibfield  {journal} {\bibinfo  {journal} {Phys. Rev. D}\ }\textbf {\bibinfo {volume} {51}},\ \bibinfo {pages} {1125} (\bibinfo {year} {1995})},\ \bibinfo {note} {[Erratum: Phys.Rev.D 55, 5853 (1997)]},\ \Eprint {http://arxiv.org/abs/hep-ph/9407339} {arXiv:hep-ph/9407339} \BibitemShut {NoStop}%
\bibitem [{\citenamefont {Caswell}\ and\ \citenamefont {Lepage}(1986)}]{Caswell:1985ui}%
  \BibitemOpen
  \bibfield  {author} {\bibinfo {author} {\bibfnamefont {W.~E.}\ \bibnamefont {Caswell}}\ and\ \bibinfo {author} {\bibfnamefont {G.~P.}\ \bibnamefont {Lepage}},\ }\href {\doibase 10.1016/0370-2693(86)91297-9} {\bibfield  {journal} {\bibinfo  {journal} {Phys. Lett. B}\ }\textbf {\bibinfo {volume} {167}},\ \bibinfo {pages} {437} (\bibinfo {year} {1986})}\BibitemShut {NoStop}%
\bibitem [{\citenamefont {Pineda}\ and\ \citenamefont {Soto}(1998{\natexlab{a}})}]{Pineda:1997bj}%
  \BibitemOpen
  \bibfield  {author} {\bibinfo {author} {\bibfnamefont {A.}~\bibnamefont {Pineda}}\ and\ \bibinfo {author} {\bibfnamefont {J.}~\bibnamefont {Soto}},\ }\href {\doibase 10.1016/S0920-5632(97)01102-X} {\bibfield  {journal} {\bibinfo  {journal} {Nucl. Phys. B Proc. Suppl.}\ }\textbf {\bibinfo {volume} {64}},\ \bibinfo {pages} {428} (\bibinfo {year} {1998}{\natexlab{a}})},\ \Eprint {http://arxiv.org/abs/hep-ph/9707481} {arXiv:hep-ph/9707481} \BibitemShut {NoStop}%
\bibitem [{\citenamefont {Pineda}\ and\ \citenamefont {Soto}(1998{\natexlab{b}})}]{Pineda:1998kj}%
  \BibitemOpen
  \bibfield  {author} {\bibinfo {author} {\bibfnamefont {A.}~\bibnamefont {Pineda}}\ and\ \bibinfo {author} {\bibfnamefont {J.}~\bibnamefont {Soto}},\ }\href {\doibase 10.1103/PhysRevD.58.114011} {\bibfield  {journal} {\bibinfo  {journal} {Phys. Rev. D}\ }\textbf {\bibinfo {volume} {58}},\ \bibinfo {pages} {114011} (\bibinfo {year} {1998}{\natexlab{b}})},\ \Eprint {http://arxiv.org/abs/hep-ph/9802365} {arXiv:hep-ph/9802365} \BibitemShut {NoStop}%
\bibitem [{\citenamefont {Brambilla}\ \emph {et~al.}(2000)\citenamefont {Brambilla}, \citenamefont {Pineda}, \citenamefont {Soto},\ and\ \citenamefont {Vairo}}]{Brambilla:1999xf}%
  \BibitemOpen
  \bibfield  {author} {\bibinfo {author} {\bibfnamefont {N.}~\bibnamefont {Brambilla}}, \bibinfo {author} {\bibfnamefont {A.}~\bibnamefont {Pineda}}, \bibinfo {author} {\bibfnamefont {J.}~\bibnamefont {Soto}}, \ and\ \bibinfo {author} {\bibfnamefont {A.}~\bibnamefont {Vairo}},\ }\href {\doibase 10.1016/S0550-3213(99)00693-8} {\bibfield  {journal} {\bibinfo  {journal} {Nucl. Phys. B}\ }\textbf {\bibinfo {volume} {566}},\ \bibinfo {pages} {275} (\bibinfo {year} {2000})},\ \Eprint {http://arxiv.org/abs/hep-ph/9907240} {arXiv:hep-ph/9907240} \BibitemShut {NoStop}%
\bibitem [{\citenamefont {Pineda}(2012)}]{Pineda:2011dg}%
  \BibitemOpen
  \bibfield  {author} {\bibinfo {author} {\bibfnamefont {A.}~\bibnamefont {Pineda}},\ }\href {\doibase 10.1016/j.ppnp.2012.01.038} {\bibfield  {journal} {\bibinfo  {journal} {Prog. Part. Nucl. Phys.}\ }\textbf {\bibinfo {volume} {67}},\ \bibinfo {pages} {735} (\bibinfo {year} {2012})},\ \Eprint {http://arxiv.org/abs/1111.0165} {arXiv:1111.0165 [hep-ph]} \BibitemShut {NoStop}%
\bibitem [{\citenamefont {Ader}\ \emph {et~al.}(1982)\citenamefont {Ader}, \citenamefont {Richard},\ and\ \citenamefont {Taxil}}]{Ader:1981db}%
  \BibitemOpen
  \bibfield  {author} {\bibinfo {author} {\bibfnamefont {J.~P.}\ \bibnamefont {Ader}}, \bibinfo {author} {\bibfnamefont {J.~M.}\ \bibnamefont {Richard}}, \ and\ \bibinfo {author} {\bibfnamefont {P.}~\bibnamefont {Taxil}},\ }\href {\doibase 10.1103/PhysRevD.25.2370} {\bibfield  {journal} {\bibinfo  {journal} {Phys. Rev. D}\ }\textbf {\bibinfo {volume} {25}},\ \bibinfo {pages} {2370} (\bibinfo {year} {1982})}\BibitemShut {NoStop}%
\bibitem [{\citenamefont {Richard}(1992)}]{Richard:1992cb}%
  \BibitemOpen
  \bibfield  {author} {\bibinfo {author} {\bibfnamefont {J.~M.}\ \bibnamefont {Richard}},\ }in\ \href@noop {} {\emph {\bibinfo {booktitle} {{International Workshop on Quark Cluster Dynamics}}}}\ (\bibinfo {year} {1992})\ pp.\ \bibinfo {pages} {84--91}\BibitemShut {NoStop}%
\bibitem [{\citenamefont {Bai}\ \emph {et~al.}(2019)\citenamefont {Bai}, \citenamefont {Lu},\ and\ \citenamefont {Osborne}}]{Bai:2016int}%
  \BibitemOpen
  \bibfield  {author} {\bibinfo {author} {\bibfnamefont {Y.}~\bibnamefont {Bai}}, \bibinfo {author} {\bibfnamefont {S.}~\bibnamefont {Lu}}, \ and\ \bibinfo {author} {\bibfnamefont {J.}~\bibnamefont {Osborne}},\ }\href {\doibase 10.1016/j.physletb.2019.134930} {\bibfield  {journal} {\bibinfo  {journal} {Phys. Lett. B}\ }\textbf {\bibinfo {volume} {798}},\ \bibinfo {pages} {134930} (\bibinfo {year} {2019})},\ \Eprint {http://arxiv.org/abs/1612.00012} {arXiv:1612.00012 [hep-ph]} \BibitemShut {NoStop}%
\bibitem [{\citenamefont {Berezhnoy}\ \emph {et~al.}(2012)\citenamefont {Berezhnoy}, \citenamefont {Luchinsky},\ and\ \citenamefont {Novoselov}}]{Berezhnoy:2011xn}%
  \BibitemOpen
  \bibfield  {author} {\bibinfo {author} {\bibfnamefont {A.~V.}\ \bibnamefont {Berezhnoy}}, \bibinfo {author} {\bibfnamefont {A.~V.}\ \bibnamefont {Luchinsky}}, \ and\ \bibinfo {author} {\bibfnamefont {A.~A.}\ \bibnamefont {Novoselov}},\ }\href {\doibase 10.1103/PhysRevD.86.034004} {\bibfield  {journal} {\bibinfo  {journal} {Phys. Rev. D}\ }\textbf {\bibinfo {volume} {86}},\ \bibinfo {pages} {034004} (\bibinfo {year} {2012})},\ \Eprint {http://arxiv.org/abs/1111.1867} {arXiv:1111.1867 [hep-ph]} \BibitemShut {NoStop}%
\bibitem [{\citenamefont {Carlson}\ \emph {et~al.}(1988)\citenamefont {Carlson}, \citenamefont {Heller},\ and\ \citenamefont {Tjon}}]{Carlson:1987hh}%
  \BibitemOpen
  \bibfield  {author} {\bibinfo {author} {\bibfnamefont {J.}~\bibnamefont {Carlson}}, \bibinfo {author} {\bibfnamefont {L.}~\bibnamefont {Heller}}, \ and\ \bibinfo {author} {\bibfnamefont {J.~A.}\ \bibnamefont {Tjon}},\ }\href {\doibase 10.1103/PhysRevD.37.744} {\bibfield  {journal} {\bibinfo  {journal} {Phys. Rev. D}\ }\textbf {\bibinfo {volume} {37}},\ \bibinfo {pages} {744} (\bibinfo {year} {1988})}\BibitemShut {NoStop}%
\bibitem [{\citenamefont {Chen}\ \emph {et~al.}(2017)\citenamefont {Chen}, \citenamefont {Chen}, \citenamefont {Liu}, \citenamefont {Steele},\ and\ \citenamefont {Zhu}}]{Chen:2016jxd}%
  \BibitemOpen
  \bibfield  {author} {\bibinfo {author} {\bibfnamefont {W.}~\bibnamefont {Chen}}, \bibinfo {author} {\bibfnamefont {H.-X.}\ \bibnamefont {Chen}}, \bibinfo {author} {\bibfnamefont {X.}~\bibnamefont {Liu}}, \bibinfo {author} {\bibfnamefont {T.~G.}\ \bibnamefont {Steele}}, \ and\ \bibinfo {author} {\bibfnamefont {S.-L.}\ \bibnamefont {Zhu}},\ }\href {\doibase 10.1016/j.physletb.2017.08.034} {\bibfield  {journal} {\bibinfo  {journal} {Phys. Lett. B}\ }\textbf {\bibinfo {volume} {773}},\ \bibinfo {pages} {247} (\bibinfo {year} {2017})},\ \Eprint {http://arxiv.org/abs/1605.01647} {arXiv:1605.01647 [hep-ph]} \BibitemShut {NoStop}%
\bibitem [{\citenamefont {Heller}\ and\ \citenamefont {Tjon}(1987)}]{Heller:1986bt}%
  \BibitemOpen
  \bibfield  {author} {\bibinfo {author} {\bibfnamefont {L.}~\bibnamefont {Heller}}\ and\ \bibinfo {author} {\bibfnamefont {J.~A.}\ \bibnamefont {Tjon}},\ }\href {\doibase 10.1103/PhysRevD.35.969} {\bibfield  {journal} {\bibinfo  {journal} {Phys. Rev. D}\ }\textbf {\bibinfo {volume} {35}},\ \bibinfo {pages} {969} (\bibinfo {year} {1987})}\BibitemShut {NoStop}%
\bibitem [{\citenamefont {Karliner}\ \emph {et~al.}(2017)\citenamefont {Karliner}, \citenamefont {Nussinov},\ and\ \citenamefont {Rosner}}]{Karliner:2016zzc}%
  \BibitemOpen
  \bibfield  {author} {\bibinfo {author} {\bibfnamefont {M.}~\bibnamefont {Karliner}}, \bibinfo {author} {\bibfnamefont {S.}~\bibnamefont {Nussinov}}, \ and\ \bibinfo {author} {\bibfnamefont {J.~L.}\ \bibnamefont {Rosner}},\ }\href {\doibase 10.1103/PhysRevD.95.034011} {\bibfield  {journal} {\bibinfo  {journal} {Phys. Rev. D}\ }\textbf {\bibinfo {volume} {95}},\ \bibinfo {pages} {034011} (\bibinfo {year} {2017})},\ \Eprint {http://arxiv.org/abs/1611.00348} {arXiv:1611.00348 [hep-ph]} \BibitemShut {NoStop}%
\bibitem [{\citenamefont {Wang}\ and\ \citenamefont {Di}(2019)}]{Wang:2018poa}%
  \BibitemOpen
  \bibfield  {author} {\bibinfo {author} {\bibfnamefont {Z.-G.}\ \bibnamefont {Wang}}\ and\ \bibinfo {author} {\bibfnamefont {Z.-Y.}\ \bibnamefont {Di}},\ }\href {\doibase 10.5506/APhysPolB.50.1335} {\bibfield  {journal} {\bibinfo  {journal} {Acta Phys. Polon. B}\ }\textbf {\bibinfo {volume} {50}},\ \bibinfo {pages} {1335} (\bibinfo {year} {2019})},\ \Eprint {http://arxiv.org/abs/1807.08520} {arXiv:1807.08520 [hep-ph]} \BibitemShut {NoStop}%
\bibitem [{\citenamefont {Wu}\ \emph {et~al.}(2018)\citenamefont {Wu}, \citenamefont {Liu}, \citenamefont {Chen}, \citenamefont {Liu},\ and\ \citenamefont {Zhu}}]{Wu:2016vtq}%
  \BibitemOpen
  \bibfield  {author} {\bibinfo {author} {\bibfnamefont {J.}~\bibnamefont {Wu}}, \bibinfo {author} {\bibfnamefont {Y.-R.}\ \bibnamefont {Liu}}, \bibinfo {author} {\bibfnamefont {K.}~\bibnamefont {Chen}}, \bibinfo {author} {\bibfnamefont {X.}~\bibnamefont {Liu}}, \ and\ \bibinfo {author} {\bibfnamefont {S.-L.}\ \bibnamefont {Zhu}},\ }\href {\doibase 10.1103/PhysRevD.97.094015} {\bibfield  {journal} {\bibinfo  {journal} {Phys. Rev. D}\ }\textbf {\bibinfo {volume} {97}},\ \bibinfo {pages} {094015} (\bibinfo {year} {2018})},\ \Eprint {http://arxiv.org/abs/1605.01134} {arXiv:1605.01134 [hep-ph]} \BibitemShut {NoStop}%
\bibitem [{\citenamefont {Hughes}\ \emph {et~al.}(2018)\citenamefont {Hughes}, \citenamefont {Eichten},\ and\ \citenamefont {Davies}}]{Hughes:2017xie}%
  \BibitemOpen
  \bibfield  {author} {\bibinfo {author} {\bibfnamefont {C.}~\bibnamefont {Hughes}}, \bibinfo {author} {\bibfnamefont {E.}~\bibnamefont {Eichten}}, \ and\ \bibinfo {author} {\bibfnamefont {C.~T.~H.}\ \bibnamefont {Davies}},\ }\href {\doibase 10.1103/PhysRevD.97.054505} {\bibfield  {journal} {\bibinfo  {journal} {Phys. Rev. D}\ }\textbf {\bibinfo {volume} {97}},\ \bibinfo {pages} {054505} (\bibinfo {year} {2018})},\ \Eprint {http://arxiv.org/abs/1710.03236} {arXiv:1710.03236 [hep-lat]} \BibitemShut {NoStop}%
\bibitem [{\citenamefont {Czarnecki}\ \emph {et~al.}(2018)\citenamefont {Czarnecki}, \citenamefont {Leng},\ and\ \citenamefont {Voloshin}}]{Czarnecki:2017vco}%
  \BibitemOpen
  \bibfield  {author} {\bibinfo {author} {\bibfnamefont {A.}~\bibnamefont {Czarnecki}}, \bibinfo {author} {\bibfnamefont {B.}~\bibnamefont {Leng}}, \ and\ \bibinfo {author} {\bibfnamefont {M.~B.}\ \bibnamefont {Voloshin}},\ }\href {\doibase 10.1016/j.physletb.2018.01.034} {\bibfield  {journal} {\bibinfo  {journal} {Phys. Lett. B}\ }\textbf {\bibinfo {volume} {778}},\ \bibinfo {pages} {233} (\bibinfo {year} {2018})},\ \Eprint {http://arxiv.org/abs/1708.04594} {arXiv:1708.04594 [hep-ph]} \BibitemShut {NoStop}%
\bibitem [{\citenamefont {Anwar}\ \emph {et~al.}(2018)\citenamefont {Anwar}, \citenamefont {Ferretti}, \citenamefont {Guo}, \citenamefont {Santopinto},\ and\ \citenamefont {Zou}}]{Anwar:2017toa}%
  \BibitemOpen
  \bibfield  {author} {\bibinfo {author} {\bibfnamefont {M.~N.}\ \bibnamefont {Anwar}}, \bibinfo {author} {\bibfnamefont {J.}~\bibnamefont {Ferretti}}, \bibinfo {author} {\bibfnamefont {F.-K.}\ \bibnamefont {Guo}}, \bibinfo {author} {\bibfnamefont {E.}~\bibnamefont {Santopinto}}, \ and\ \bibinfo {author} {\bibfnamefont {B.-S.}\ \bibnamefont {Zou}},\ }\href {\doibase 10.1140/epjc/s10052-018-6073-9} {\bibfield  {journal} {\bibinfo  {journal} {Eur. Phys. J. C}\ }\textbf {\bibinfo {volume} {78}},\ \bibinfo {pages} {647} (\bibinfo {year} {2018})},\ \Eprint {http://arxiv.org/abs/1710.02540} {arXiv:1710.02540 [hep-ph]} \BibitemShut {NoStop}%
\bibitem [{\citenamefont {Vijande}\ \emph {et~al.}(2007)\citenamefont {Vijande}, \citenamefont {Valcarce},\ and\ \citenamefont {Richard}}]{Vijande:2007ix}%
  \BibitemOpen
  \bibfield  {author} {\bibinfo {author} {\bibfnamefont {J.}~\bibnamefont {Vijande}}, \bibinfo {author} {\bibfnamefont {A.}~\bibnamefont {Valcarce}}, \ and\ \bibinfo {author} {\bibfnamefont {J.~M.}\ \bibnamefont {Richard}},\ }\href {\doibase 10.1103/PhysRevD.76.114013} {\bibfield  {journal} {\bibinfo  {journal} {Phys. Rev. D}\ }\textbf {\bibinfo {volume} {76}},\ \bibinfo {pages} {114013} (\bibinfo {year} {2007})},\ \Eprint {http://arxiv.org/abs/0707.3996} {arXiv:0707.3996 [hep-ph]} \BibitemShut {NoStop}%
\bibitem [{\citenamefont {Brambilla}\ \emph {et~al.}(1999{\natexlab{a}})\citenamefont {Brambilla}, \citenamefont {Pineda}, \citenamefont {Soto},\ and\ \citenamefont {Vairo}}]{Brambilla:1999xj}%
  \BibitemOpen
  \bibfield  {author} {\bibinfo {author} {\bibfnamefont {N.}~\bibnamefont {Brambilla}}, \bibinfo {author} {\bibfnamefont {A.}~\bibnamefont {Pineda}}, \bibinfo {author} {\bibfnamefont {J.}~\bibnamefont {Soto}}, \ and\ \bibinfo {author} {\bibfnamefont {A.}~\bibnamefont {Vairo}},\ }\href {\doibase 10.1016/S0370-2693(99)01301-5} {\bibfield  {journal} {\bibinfo  {journal} {Phys. Lett. B}\ }\textbf {\bibinfo {volume} {470}},\ \bibinfo {pages} {215} (\bibinfo {year} {1999}{\natexlab{a}})},\ \Eprint {http://arxiv.org/abs/hep-ph/9910238} {arXiv:hep-ph/9910238} \BibitemShut {NoStop}%
\bibitem [{\citenamefont {Kniehl}\ \emph {et~al.}(2002)\citenamefont {Kniehl}, \citenamefont {Penin}, \citenamefont {Smirnov},\ and\ \citenamefont {Steinhauser}}]{Kniehl:2002br}%
  \BibitemOpen
  \bibfield  {author} {\bibinfo {author} {\bibfnamefont {B.~A.}\ \bibnamefont {Kniehl}}, \bibinfo {author} {\bibfnamefont {A.~A.}\ \bibnamefont {Penin}}, \bibinfo {author} {\bibfnamefont {V.~A.}\ \bibnamefont {Smirnov}}, \ and\ \bibinfo {author} {\bibfnamefont {M.}~\bibnamefont {Steinhauser}},\ }\href {\doibase 10.1016/S0550-3213(02)00403-0} {\bibfield  {journal} {\bibinfo  {journal} {Nucl. Phys. B}\ }\textbf {\bibinfo {volume} {635}},\ \bibinfo {pages} {357} (\bibinfo {year} {2002})},\ \Eprint {http://arxiv.org/abs/hep-ph/0203166} {arXiv:hep-ph/0203166} \BibitemShut {NoStop}%
\bibitem [{\citenamefont {Pineda}\ and\ \citenamefont {Yndurain}(1998)}]{Pineda:1997hz}%
  \BibitemOpen
  \bibfield  {author} {\bibinfo {author} {\bibfnamefont {A.}~\bibnamefont {Pineda}}\ and\ \bibinfo {author} {\bibfnamefont {F.~J.}\ \bibnamefont {Yndurain}},\ }\href {\doibase 10.1103/PhysRevD.58.094022} {\bibfield  {journal} {\bibinfo  {journal} {Phys. Rev. D}\ }\textbf {\bibinfo {volume} {58}},\ \bibinfo {pages} {094022} (\bibinfo {year} {1998})},\ \Eprint {http://arxiv.org/abs/hep-ph/9711287} {arXiv:hep-ph/9711287} \BibitemShut {NoStop}%
\bibitem [{\citenamefont {Brambilla}\ \emph {et~al.}(2005)\citenamefont {Brambilla}, \citenamefont {Vairo},\ and\ \citenamefont {Rosch}}]{Brambilla:2005yk}%
  \BibitemOpen
  \bibfield  {author} {\bibinfo {author} {\bibfnamefont {N.}~\bibnamefont {Brambilla}}, \bibinfo {author} {\bibfnamefont {A.}~\bibnamefont {Vairo}}, \ and\ \bibinfo {author} {\bibfnamefont {T.}~\bibnamefont {Rosch}},\ }\href {\doibase 10.1103/PhysRevD.72.034021} {\bibfield  {journal} {\bibinfo  {journal} {Phys. Rev. D}\ }\textbf {\bibinfo {volume} {72}},\ \bibinfo {pages} {034021} (\bibinfo {year} {2005})},\ \Eprint {http://arxiv.org/abs/hep-ph/0506065} {arXiv:hep-ph/0506065} \BibitemShut {NoStop}%
\bibitem [{\citenamefont {Brambilla}\ \emph {et~al.}(2010)\citenamefont {Brambilla}, \citenamefont {Ghiglieri},\ and\ \citenamefont {Vairo}}]{Brambilla:2009cd}%
  \BibitemOpen
  \bibfield  {author} {\bibinfo {author} {\bibfnamefont {N.}~\bibnamefont {Brambilla}}, \bibinfo {author} {\bibfnamefont {J.}~\bibnamefont {Ghiglieri}}, \ and\ \bibinfo {author} {\bibfnamefont {A.}~\bibnamefont {Vairo}},\ }\href {\doibase 10.1103/PhysRevD.81.054031} {\bibfield  {journal} {\bibinfo  {journal} {Phys. Rev. D}\ }\textbf {\bibinfo {volume} {81}},\ \bibinfo {pages} {054031} (\bibinfo {year} {2010})},\ \Eprint {http://arxiv.org/abs/0911.3541} {arXiv:0911.3541 [hep-ph]} \BibitemShut {NoStop}%
\bibitem [{\citenamefont {Brambilla}\ \emph {et~al.}(2013)\citenamefont {Brambilla}, \citenamefont {Karbstein},\ and\ \citenamefont {Vairo}}]{Brambilla:2013vx}%
  \BibitemOpen
  \bibfield  {author} {\bibinfo {author} {\bibfnamefont {N.}~\bibnamefont {Brambilla}}, \bibinfo {author} {\bibfnamefont {F.}~\bibnamefont {Karbstein}}, \ and\ \bibinfo {author} {\bibfnamefont {A.}~\bibnamefont {Vairo}},\ }\href {\doibase 10.1103/PhysRevD.87.074014} {\bibfield  {journal} {\bibinfo  {journal} {Phys. Rev. D}\ }\textbf {\bibinfo {volume} {87}},\ \bibinfo {pages} {074014} (\bibinfo {year} {2013})},\ \Eprint {http://arxiv.org/abs/1301.3013} {arXiv:1301.3013 [hep-ph]} \BibitemShut {NoStop}%
\bibitem [{\citenamefont {Assi}\ and\ \citenamefont {Wagman}(2023)}]{Assi:2023cfo}%
  \BibitemOpen
  \bibfield  {author} {\bibinfo {author} {\bibfnamefont {B.}~\bibnamefont {Assi}}\ and\ \bibinfo {author} {\bibfnamefont {M.~L.}\ \bibnamefont {Wagman}},\ }\href {\doibase 10.1103/PhysRevD.108.096004} {\bibfield  {journal} {\bibinfo  {journal} {Phys. Rev. D}\ }\textbf {\bibinfo {volume} {108}},\ \bibinfo {pages} {096004} (\bibinfo {year} {2023})},\ \Eprint {http://arxiv.org/abs/2305.01685} {arXiv:2305.01685 [hep-ph]} \BibitemShut {NoStop}%
\bibitem [{\citenamefont {Jia}(2006)}]{Jia:2006gw}%
  \BibitemOpen
  \bibfield  {author} {\bibinfo {author} {\bibfnamefont {Y.}~\bibnamefont {Jia}},\ }\href {\doibase 10.1088/1126-6708/2006/10/073} {\bibfield  {journal} {\bibinfo  {journal} {JHEP}\ }\textbf {\bibinfo {volume} {10}},\ \bibinfo {pages} {073} (\bibinfo {year} {2006})},\ \Eprint {http://arxiv.org/abs/hep-ph/0607290} {arXiv:hep-ph/0607290} \BibitemShut {NoStop}%
\bibitem [{\citenamefont {Llanes-Estrada}\ \emph {et~al.}(2012)\citenamefont {Llanes-Estrada}, \citenamefont {Pavlova},\ and\ \citenamefont {Williams}}]{Llanes-Estrada:2011gwu}%
  \BibitemOpen
  \bibfield  {author} {\bibinfo {author} {\bibfnamefont {F.~J.}\ \bibnamefont {Llanes-Estrada}}, \bibinfo {author} {\bibfnamefont {O.~I.}\ \bibnamefont {Pavlova}}, \ and\ \bibinfo {author} {\bibfnamefont {R.}~\bibnamefont {Williams}},\ }\href {\doibase 10.1140/epjc/s10052-012-2019-9} {\bibfield  {journal} {\bibinfo  {journal} {Eur. Phys. J. C}\ }\textbf {\bibinfo {volume} {72}},\ \bibinfo {pages} {2019} (\bibinfo {year} {2012})},\ \Eprint {http://arxiv.org/abs/1111.7087} {arXiv:1111.7087 [hep-ph]} \BibitemShut {NoStop}%
\bibitem [{\citenamefont {Carlson}\ \emph {et~al.}(2015)\citenamefont {Carlson}, \citenamefont {Gandolfi}, \citenamefont {Pederiva}, \citenamefont {Pieper}, \citenamefont {Schiavilla}, \citenamefont {Schmidt},\ and\ \citenamefont {Wiringa}}]{Carlson:2014vla}%
  \BibitemOpen
  \bibfield  {author} {\bibinfo {author} {\bibfnamefont {J.}~\bibnamefont {Carlson}}, \bibinfo {author} {\bibfnamefont {S.}~\bibnamefont {Gandolfi}}, \bibinfo {author} {\bibfnamefont {F.}~\bibnamefont {Pederiva}}, \bibinfo {author} {\bibfnamefont {S.~C.}\ \bibnamefont {Pieper}}, \bibinfo {author} {\bibfnamefont {R.}~\bibnamefont {Schiavilla}}, \bibinfo {author} {\bibfnamefont {K.~E.}\ \bibnamefont {Schmidt}}, \ and\ \bibinfo {author} {\bibfnamefont {R.~B.}\ \bibnamefont {Wiringa}},\ }\href {\doibase 10.1103/RevModPhys.87.1067} {\bibfield  {journal} {\bibinfo  {journal} {Rev. Mod. Phys.}\ }\textbf {\bibinfo {volume} {87}},\ \bibinfo {pages} {1067} (\bibinfo {year} {2015})},\ \Eprint {http://arxiv.org/abs/1412.3081} {arXiv:1412.3081 [nucl-th]} \BibitemShut {NoStop}%
\bibitem [{\citenamefont {Yan}\ and\ \citenamefont {Blume}(2017)}]{Yan_2017}%
  \BibitemOpen
  \bibfield  {author} {\bibinfo {author} {\bibfnamefont {Y.}~\bibnamefont {Yan}}\ and\ \bibinfo {author} {\bibfnamefont {D.}~\bibnamefont {Blume}},\ }\href {\doibase 10.1088/1361-6455/aa8d7f} {\bibfield  {journal} {\bibinfo  {journal} {Journal of Physics B: Atomic, Molecular and Optical Physics}\ }\textbf {\bibinfo {volume} {50}},\ \bibinfo {pages} {223001} (\bibinfo {year} {2017})}\BibitemShut {NoStop}%
\bibitem [{\citenamefont {Gandolfi}\ \emph {et~al.}(2020)\citenamefont {Gandolfi}, \citenamefont {Lonardoni}, \citenamefont {Lovato},\ and\ \citenamefont {Piarulli}}]{Gandolfi:2020pbj}%
  \BibitemOpen
  \bibfield  {author} {\bibinfo {author} {\bibfnamefont {S.}~\bibnamefont {Gandolfi}}, \bibinfo {author} {\bibfnamefont {D.}~\bibnamefont {Lonardoni}}, \bibinfo {author} {\bibfnamefont {A.}~\bibnamefont {Lovato}}, \ and\ \bibinfo {author} {\bibfnamefont {M.}~\bibnamefont {Piarulli}},\ }\href {\doibase 10.3389/fphy.2020.00117} {\bibfield  {journal} {\bibinfo  {journal} {Front. in Phys.}\ }\textbf {\bibinfo {volume} {8}},\ \bibinfo {pages} {117} (\bibinfo {year} {2020})},\ \Eprint {http://arxiv.org/abs/2001.01374} {arXiv:2001.01374 [nucl-th]} \BibitemShut {NoStop}%
\bibitem [{\citenamefont {{Hylleraas}}(1928)}]{Hylleraas:1928}%
  \BibitemOpen
  \bibfield  {author} {\bibinfo {author} {\bibfnamefont {E.~A.}\ \bibnamefont {{Hylleraas}}},\ }\href {\doibase 10.1007/BF01340013} {\bibfield  {journal} {\bibinfo  {journal} {Zeitschrift fur Physik}\ }\textbf {\bibinfo {volume} {48}},\ \bibinfo {pages} {469} (\bibinfo {year} {1928})}\BibitemShut {NoStop}%
\bibitem [{\citenamefont {{Hylleraas}}(1930)}]{Hylleraas:1930}%
  \BibitemOpen
  \bibfield  {author} {\bibinfo {author} {\bibfnamefont {E.~A.}\ \bibnamefont {{Hylleraas}}},\ }\href {\doibase 10.1007/BF01397032} {\bibfield  {journal} {\bibinfo  {journal} {Zeitschrift fur Physik}\ }\textbf {\bibinfo {volume} {65}},\ \bibinfo {pages} {209} (\bibinfo {year} {1930})}\BibitemShut {NoStop}%
\bibitem [{\citenamefont {{Hylleraas}}(1931)}]{Hylleraas:1931}%
  \BibitemOpen
  \bibfield  {author} {\bibinfo {author} {\bibfnamefont {E.~A.}\ \bibnamefont {{Hylleraas}}},\ }\href {\doibase 10.1007/BF01344443} {\bibfield  {journal} {\bibinfo  {journal} {Zeitschrift fur Physik}\ }\textbf {\bibinfo {volume} {71}},\ \bibinfo {pages} {739} (\bibinfo {year} {1931})}\BibitemShut {NoStop}%
\bibitem [{\citenamefont {James}\ and\ \citenamefont {Coolidge}(1933)}]{James:1933}%
  \BibitemOpen
  \bibfield  {author} {\bibinfo {author} {\bibfnamefont {H.~M.}\ \bibnamefont {James}}\ and\ \bibinfo {author} {\bibfnamefont {A.~S.}\ \bibnamefont {Coolidge}},\ }\href {https://api.semanticscholar.org/CorpusID:54209263} {\bibfield  {journal} {\bibinfo  {journal} {Journal of Chemical Physics}\ }\textbf {\bibinfo {volume} {1}},\ \bibinfo {pages} {825} (\bibinfo {year} {1933})}\BibitemShut {NoStop}%
\bibitem [{\citenamefont {Anderson}(1975)}]{Anderson:1975}%
  \BibitemOpen
  \bibfield  {author} {\bibinfo {author} {\bibfnamefont {J.~B.}\ \bibnamefont {Anderson}},\ }\href {\doibase 10.1063/1.431514} {\bibfield  {journal} {\bibinfo  {journal} {The Journal of Chemical Physics}\ }\textbf {\bibinfo {volume} {63}},\ \bibinfo {pages} {1499} (\bibinfo {year} {1975})},\ \Eprint {http://arxiv.org/abs/https://pubs.aip.org/aip/jcp/article-pdf/63/4/1499/18897623/1499\_1\_online.pdf} {https://pubs.aip.org/aip/jcp/article-pdf/63/4/1499/18897623/1499\_1\_online.pdf} \BibitemShut {NoStop}%
\bibitem [{\citenamefont {Mentch}\ and\ \citenamefont {Anderson}(1981)}]{Mentch:1981}%
  \BibitemOpen
  \bibfield  {author} {\bibinfo {author} {\bibfnamefont {F.}~\bibnamefont {Mentch}}\ and\ \bibinfo {author} {\bibfnamefont {J.~B.}\ \bibnamefont {Anderson}},\ }\href {\doibase 10.1063/1.441022} {\bibfield  {journal} {\bibinfo  {journal} {The Journal of Chemical Physics}\ }\textbf {\bibinfo {volume} {74}},\ \bibinfo {pages} {6307} (\bibinfo {year} {1981})},\ \Eprint {http://arxiv.org/abs/https://pubs.aip.org/aip/jcp/article-pdf/74/11/6307/18928646/6307\_1\_online.pdf} {https://pubs.aip.org/aip/jcp/article-pdf/74/11/6307/18928646/6307\_1\_online.pdf} \BibitemShut {NoStop}%
\bibitem [{\citenamefont {Lee}\ \emph {et~al.}(1983)\citenamefont {Lee}, \citenamefont {Vashishta},\ and\ \citenamefont {Kalia}}]{Lee:1983}%
  \BibitemOpen
  \bibfield  {author} {\bibinfo {author} {\bibfnamefont {M.~A.}\ \bibnamefont {Lee}}, \bibinfo {author} {\bibfnamefont {P.}~\bibnamefont {Vashishta}}, \ and\ \bibinfo {author} {\bibfnamefont {R.~K.}\ \bibnamefont {Kalia}},\ }\href {\doibase 10.1103/PhysRevLett.51.2422} {\bibfield  {journal} {\bibinfo  {journal} {Phys. Rev. Lett.}\ }\textbf {\bibinfo {volume} {51}},\ \bibinfo {pages} {2422} (\bibinfo {year} {1983})}\BibitemShut {NoStop}%
\bibitem [{\citenamefont {Anzai}\ \emph {et~al.}(2013)\citenamefont {Anzai}, \citenamefont {Prausa}, \citenamefont {Smirnov}, \citenamefont {Smirnov},\ and\ \citenamefont {Steinhauser}}]{Anzai:2013tja}%
  \BibitemOpen
  \bibfield  {author} {\bibinfo {author} {\bibfnamefont {C.}~\bibnamefont {Anzai}}, \bibinfo {author} {\bibfnamefont {M.}~\bibnamefont {Prausa}}, \bibinfo {author} {\bibfnamefont {A.~V.}\ \bibnamefont {Smirnov}}, \bibinfo {author} {\bibfnamefont {V.~A.}\ \bibnamefont {Smirnov}}, \ and\ \bibinfo {author} {\bibfnamefont {M.}~\bibnamefont {Steinhauser}},\ }\href {\doibase 10.1103/PhysRevD.88.054030} {\bibfield  {journal} {\bibinfo  {journal} {Phys. Rev. D}\ }\textbf {\bibinfo {volume} {88}},\ \bibinfo {pages} {054030} (\bibinfo {year} {2013})},\ \Eprint {http://arxiv.org/abs/1308.1202} {arXiv:1308.1202 [hep-ph]} \BibitemShut {NoStop}%
\bibitem [{\citenamefont {Brambilla}\ \emph {et~al.}(1999{\natexlab{b}})\citenamefont {Brambilla}, \citenamefont {Pineda}, \citenamefont {Soto},\ and\ \citenamefont {Vairo}}]{Pineda:1998kn}%
  \BibitemOpen
  \bibfield  {author} {\bibinfo {author} {\bibfnamefont {N.}~\bibnamefont {Brambilla}}, \bibinfo {author} {\bibfnamefont {A.}~\bibnamefont {Pineda}}, \bibinfo {author} {\bibfnamefont {J.}~\bibnamefont {Soto}}, \ and\ \bibinfo {author} {\bibfnamefont {A.}~\bibnamefont {Vairo}},\ }\href {\doibase 10.1103/PhysRevD.60.091502} {\bibfield  {journal} {\bibinfo  {journal} {Phys. Rev. D}\ }\textbf {\bibinfo {volume} {60}},\ \bibinfo {pages} {091502} (\bibinfo {year} {1999}{\natexlab{b}})},\ \Eprint {http://arxiv.org/abs/hep-ph/9903355} {arXiv:hep-ph/9903355} \BibitemShut {NoStop}%
\bibitem [{\citenamefont {Kniehl}\ and\ \citenamefont {Penin}(1999)}]{Kniehl:1999ud}%
  \BibitemOpen
  \bibfield  {author} {\bibinfo {author} {\bibfnamefont {B.~A.}\ \bibnamefont {Kniehl}}\ and\ \bibinfo {author} {\bibfnamefont {A.~A.}\ \bibnamefont {Penin}},\ }\href {\doibase 10.1016/S0550-3213(99)00564-7} {\bibfield  {journal} {\bibinfo  {journal} {Nucl. Phys. B}\ }\textbf {\bibinfo {volume} {563}},\ \bibinfo {pages} {200} (\bibinfo {year} {1999})},\ \Eprint {http://arxiv.org/abs/hep-ph/9907489} {arXiv:hep-ph/9907489} \BibitemShut {NoStop}%
\bibitem [{\citenamefont {Toulouse}\ and\ \citenamefont {Umrigar}(2007)}]{toulouse2007optimization}%
  \BibitemOpen
  \bibfield  {author} {\bibinfo {author} {\bibfnamefont {J.}~\bibnamefont {Toulouse}}\ and\ \bibinfo {author} {\bibfnamefont {C.~J.}\ \bibnamefont {Umrigar}},\ }\href {\doibase 10.1063/1.2437215} {\bibfield  {journal} {\bibinfo  {journal} {The Journal of chemical physics}\ }\textbf {\bibinfo {volume} {126}} (\bibinfo {year} {2007}),\ 10.1063/1.2437215}\BibitemShut {NoStop}%
\bibitem [{\citenamefont {Hylleraas}\ and\ \citenamefont {Ore}(1947)}]{Hylleraas:1947zza}%
  \BibitemOpen
  \bibfield  {author} {\bibinfo {author} {\bibfnamefont {E.~A.}\ \bibnamefont {Hylleraas}}\ and\ \bibinfo {author} {\bibfnamefont {A.}~\bibnamefont {Ore}},\ }\href {\doibase 10.1103/PhysRev.71.493} {\bibfield  {journal} {\bibinfo  {journal} {Phys. Rev.}\ }\textbf {\bibinfo {volume} {71}},\ \bibinfo {pages} {493} (\bibinfo {year} {1947})}\BibitemShut {NoStop}%
\bibitem [{\citenamefont {Born}\ and\ \citenamefont {Oppenheimer}(1927)}]{Born:1927rpw}%
  \BibitemOpen
  \bibfield  {author} {\bibinfo {author} {\bibfnamefont {M.}~\bibnamefont {Born}}\ and\ \bibinfo {author} {\bibfnamefont {R.}~\bibnamefont {Oppenheimer}},\ }\href {\doibase 10.1002/andp.19273892002} {\bibfield  {journal} {\bibinfo  {journal} {Annalen Phys.}\ }\textbf {\bibinfo {volume} {389}},\ \bibinfo {pages} {457} (\bibinfo {year} {1927})}\BibitemShut {NoStop}%
\bibitem [{\citenamefont {Huang}\ \emph {et~al.}(2021)\citenamefont {Huang}, \citenamefont {Zhao},\ and\ \citenamefont {Zhuang}}]{Huang:2020dci}%
  \BibitemOpen
  \bibfield  {author} {\bibinfo {author} {\bibfnamefont {G.}~\bibnamefont {Huang}}, \bibinfo {author} {\bibfnamefont {J.}~\bibnamefont {Zhao}}, \ and\ \bibinfo {author} {\bibfnamefont {P.}~\bibnamefont {Zhuang}},\ }\href {\doibase 10.1103/PhysRevD.103.054014} {\bibfield  {journal} {\bibinfo  {journal} {Phys. Rev. D}\ }\textbf {\bibinfo {volume} {103}},\ \bibinfo {pages} {054014} (\bibinfo {year} {2021})},\ \Eprint {http://arxiv.org/abs/2012.14845} {arXiv:2012.14845 [hep-ph]} \BibitemShut {NoStop}%
\bibitem [{\citenamefont {Wheeler}(1946)}]{Wheeler:1946xth}%
  \BibitemOpen
  \bibfield  {author} {\bibinfo {author} {\bibfnamefont {J.~A.}\ \bibnamefont {Wheeler}},\ }\href {\doibase 10.1111/j.1749-6632.1946.tb31764.x} {\bibfield  {journal} {\bibinfo  {journal} {Annals N. Y. Acad. Sci.}\ }\textbf {\bibinfo {volume} {48}},\ \bibinfo {pages} {219} (\bibinfo {year} {1946})}\BibitemShut {NoStop}%
\bibitem [{\citenamefont {Sharma}(1968)}]{Sharma:1968}%
  \BibitemOpen
  \bibfield  {author} {\bibinfo {author} {\bibfnamefont {R.~R.}\ \bibnamefont {Sharma}},\ }\href {\doibase 10.1103/PhysRev.171.36} {\bibfield  {journal} {\bibinfo  {journal} {Phys. Rev.}\ }\textbf {\bibinfo {volume} {171}},\ \bibinfo {pages} {36} (\bibinfo {year} {1968})}\BibitemShut {NoStop}%
\bibitem [{\citenamefont {Ho}(1986)}]{Ho:1986zz}%
  \BibitemOpen
  \bibfield  {author} {\bibinfo {author} {\bibfnamefont {Y.~K.}\ \bibnamefont {Ho}},\ }\href {\doibase 10.1103/PhysRevA.33.3584} {\bibfield  {journal} {\bibinfo  {journal} {Phys. Rev. A}\ }\textbf {\bibinfo {volume} {33}},\ \bibinfo {pages} {3584} (\bibinfo {year} {1986})}\BibitemShut {NoStop}%
\bibitem [{\citenamefont {Frolov}\ and\ \citenamefont {Smith}(1996)}]{Frolov:1996kq}%
  \BibitemOpen
  \bibfield  {author} {\bibinfo {author} {\bibfnamefont {A.~M.}\ \bibnamefont {Frolov}}\ and\ \bibinfo {author} {\bibfnamefont {V.~H.}\ \bibnamefont {Smith}},\ }\href@noop {} {\bibfield  {journal} {\bibinfo  {journal} {J. Phys. B}\ }\textbf {\bibinfo {volume} {29}},\ \bibinfo {pages} {L433} (\bibinfo {year} {1996})}\BibitemShut {NoStop}%
\bibitem [{\citenamefont {Czarnecki}(2009)}]{Czarnecki:2009yeo}%
  \BibitemOpen
  \bibfield  {author} {\bibinfo {author} {\bibfnamefont {A.}~\bibnamefont {Czarnecki}},\ }\href {\doibase 10.1016/j.nuclphysa.2009.05.118} {\bibfield  {journal} {\bibinfo  {journal} {Nucl. Phys. A}\ }\textbf {\bibinfo {volume} {827}},\ \bibinfo {pages} {541c} (\bibinfo {year} {2009})}\BibitemShut {NoStop}%
\bibitem [{\citenamefont {Aslam}\ \emph {et~al.}(2021)\citenamefont {Aslam}, \citenamefont {Chen}, \citenamefont {Czarnecki}, \citenamefont {Mir},\ and\ \citenamefont {Mubasher}}]{Aslam:2021uqu}%
  \BibitemOpen
  \bibfield  {author} {\bibinfo {author} {\bibfnamefont {M.~J.}\ \bibnamefont {Aslam}}, \bibinfo {author} {\bibfnamefont {W.}~\bibnamefont {Chen}}, \bibinfo {author} {\bibfnamefont {A.}~\bibnamefont {Czarnecki}}, \bibinfo {author} {\bibfnamefont {S.~R.}\ \bibnamefont {Mir}}, \ and\ \bibinfo {author} {\bibfnamefont {M.}~\bibnamefont {Mubasher}},\ }\href {\doibase 10.1103/PhysRevA.104.052803} {\bibfield  {journal} {\bibinfo  {journal} {Phys. Rev. A}\ }\textbf {\bibinfo {volume} {104}},\ \bibinfo {pages} {052803} (\bibinfo {year} {2021})},\ \Eprint {http://arxiv.org/abs/2108.06785} {arXiv:2108.06785 [hep-ph]} \BibitemShut {NoStop}%
\bibitem [{\citenamefont {Cassidy}\ \emph {et~al.}(2005)\citenamefont {Cassidy}, \citenamefont {Deng}, \citenamefont {Greaves}, \citenamefont {Maruo}, \citenamefont {Nishiyama}, \citenamefont {Snyder}, \citenamefont {Tanaka},\ and\ \citenamefont {Mills}}]{Cassidy:2005tb}%
  \BibitemOpen
  \bibfield  {author} {\bibinfo {author} {\bibfnamefont {D.~B.}\ \bibnamefont {Cassidy}}, \bibinfo {author} {\bibfnamefont {S.~H.~M.}\ \bibnamefont {Deng}}, \bibinfo {author} {\bibfnamefont {R.~G.}\ \bibnamefont {Greaves}}, \bibinfo {author} {\bibfnamefont {T.}~\bibnamefont {Maruo}}, \bibinfo {author} {\bibfnamefont {N.}~\bibnamefont {Nishiyama}}, \bibinfo {author} {\bibfnamefont {J.~B.}\ \bibnamefont {Snyder}}, \bibinfo {author} {\bibfnamefont {H.~K.~M.}\ \bibnamefont {Tanaka}}, \ and\ \bibinfo {author} {\bibfnamefont {A.~P.}\ \bibnamefont {Mills}, \bibfnamefont {Jr}},\ }\href {\doibase 10.1103/PhysRevLett.95.195006} {\bibfield  {journal} {\bibinfo  {journal} {Phys. Rev. Lett.}\ }\textbf {\bibinfo {volume} {95}},\ \bibinfo {pages} {195006} (\bibinfo {year} {2005})}\BibitemShut {NoStop}%
\bibitem [{\citenamefont {Cassidy}\ \emph {et~al.}(2012)\citenamefont {Cassidy}, \citenamefont {Hisakado}, \citenamefont {Tom},\ and\ \citenamefont {Mills}}]{Cassidy:2012}%
  \BibitemOpen
  \bibfield  {author} {\bibinfo {author} {\bibfnamefont {D.~B.}\ \bibnamefont {Cassidy}}, \bibinfo {author} {\bibfnamefont {T.~H.}\ \bibnamefont {Hisakado}}, \bibinfo {author} {\bibfnamefont {H.~W.~K.}\ \bibnamefont {Tom}}, \ and\ \bibinfo {author} {\bibfnamefont {A.~P.}\ \bibnamefont {Mills}},\ }\href {\doibase 10.1103/PhysRevLett.108.133402} {\bibfield  {journal} {\bibinfo  {journal} {Phys. Rev. Lett.}\ }\textbf {\bibinfo {volume} {108}},\ \bibinfo {pages} {133402} (\bibinfo {year} {2012})}\BibitemShut {NoStop}%
\bibitem [{\citenamefont {Emami-Razavi}\ and\ \citenamefont {Darewych}(2021)}]{emami2021review}%
  \BibitemOpen
  \bibfield  {author} {\bibinfo {author} {\bibfnamefont {M.}~\bibnamefont {Emami-Razavi}}\ and\ \bibinfo {author} {\bibfnamefont {J.~W.}\ \bibnamefont {Darewych}},\ }\href@noop {} {\bibfield  {journal} {\bibinfo  {journal} {The European Physical Journal D}\ }\textbf {\bibinfo {volume} {75}},\ \bibinfo {pages} {1} (\bibinfo {year} {2021})}\BibitemShut {NoStop}%
\bibitem [{\citenamefont {Dyson}(1952)}]{dyson1952divergence}%
  \BibitemOpen
  \bibfield  {author} {\bibinfo {author} {\bibfnamefont {F.~J.}\ \bibnamefont {Dyson}},\ }\href@noop {} {\bibfield  {journal} {\bibinfo  {journal} {Physical Review}\ }\textbf {\bibinfo {volume} {85}},\ \bibinfo {pages} {631} (\bibinfo {year} {1952})}\BibitemShut {NoStop}%
\bibitem [{\citenamefont {Pineda}\ and\ \citenamefont {Soto}(1998{\natexlab{c}})}]{pineda1998potential}%
  \BibitemOpen
  \bibfield  {author} {\bibinfo {author} {\bibfnamefont {A.}~\bibnamefont {Pineda}}\ and\ \bibinfo {author} {\bibfnamefont {J.}~\bibnamefont {Soto}},\ }\href@noop {} {\bibfield  {journal} {\bibinfo  {journal} {Physical Review D}\ }\textbf {\bibinfo {volume} {59}},\ \bibinfo {pages} {016005} (\bibinfo {year} {1998}{\natexlab{c}})}\BibitemShut {NoStop}%
\bibitem [{\citenamefont {Akaike}(1974)}]{AkaikeAIC}%
  \BibitemOpen
  \bibfield  {author} {\bibinfo {author} {\bibfnamefont {H.}~\bibnamefont {Akaike}},\ }\href {\doibase 10.1109/TAC.1974.1100705} {\bibfield  {journal} {\bibinfo  {journal} {IEEE Transactions on Automatic Control}\ }\textbf {\bibinfo {volume} {19}},\ \bibinfo {pages} {716} (\bibinfo {year} {1974})}\BibitemShut {NoStop}%
\bibitem [{\citenamefont {Jay}\ and\ \citenamefont {Neil}(2021)}]{Jay:2020jkz}%
  \BibitemOpen
  \bibfield  {author} {\bibinfo {author} {\bibfnamefont {W.~I.}\ \bibnamefont {Jay}}\ and\ \bibinfo {author} {\bibfnamefont {E.~T.}\ \bibnamefont {Neil}},\ }\href {\doibase 10.1103/PhysRevD.103.114502} {\bibfield  {journal} {\bibinfo  {journal} {Phys. Rev. D}\ }\textbf {\bibinfo {volume} {103}},\ \bibinfo {pages} {114502} (\bibinfo {year} {2021})},\ \Eprint {http://arxiv.org/abs/2008.01069} {arXiv:2008.01069 [stat.ME]} \BibitemShut {NoStop}%
\bibitem [{\citenamefont {Richard}\ \emph {et~al.}(1993)\citenamefont {Richard}, \citenamefont {Frohlich}, \citenamefont {Graf},\ and\ \citenamefont {Seifert}}]{Richard:1993zx}%
  \BibitemOpen
  \bibfield  {author} {\bibinfo {author} {\bibfnamefont {J.~M.}\ \bibnamefont {Richard}}, \bibinfo {author} {\bibfnamefont {J.}~\bibnamefont {Frohlich}}, \bibinfo {author} {\bibfnamefont {G.~M.}\ \bibnamefont {Graf}}, \ and\ \bibinfo {author} {\bibfnamefont {M.}~\bibnamefont {Seifert}},\ }\href {\doibase 10.1103/PhysRevLett.71.1332} {\bibfield  {journal} {\bibinfo  {journal} {Phys. Rev. Lett.}\ }\textbf {\bibinfo {volume} {71}},\ \bibinfo {pages} {1332} (\bibinfo {year} {1993})},\ \Eprint {http://arxiv.org/abs/nucl-th/9305013} {arXiv:nucl-th/9305013} \BibitemShut {NoStop}%
\bibitem [{\citenamefont {Zyla}\ \emph {et~al.}(2020)\citenamefont {Zyla} \emph {et~al.}}]{ParticleDataGroup:2020ssz}%
  \BibitemOpen
  \bibfield  {author} {\bibinfo {author} {\bibfnamefont {P.~A.}\ \bibnamefont {Zyla}} \emph {et~al.} (\bibinfo {collaboration} {Particle Data Group}),\ }\href {\doibase 10.1093/ptep/ptaa104} {\bibfield  {journal} {\bibinfo  {journal} {PTEP}\ }\textbf {\bibinfo {volume} {2020}},\ \bibinfo {pages} {083C01} (\bibinfo {year} {2020})}\BibitemShut {NoStop}%
\bibitem [{\citenamefont {Bigi}\ \emph {et~al.}(1986)\citenamefont {Bigi}, \citenamefont {Dokshitzer}, \citenamefont {Khoze}, \citenamefont {Kuhn},\ and\ \citenamefont {Zerwas}}]{Bigi:1986jk}%
  \BibitemOpen
  \bibfield  {author} {\bibinfo {author} {\bibfnamefont {I.~I.~Y.}\ \bibnamefont {Bigi}}, \bibinfo {author} {\bibfnamefont {Y.~L.}\ \bibnamefont {Dokshitzer}}, \bibinfo {author} {\bibfnamefont {V.~A.}\ \bibnamefont {Khoze}}, \bibinfo {author} {\bibfnamefont {J.~H.}\ \bibnamefont {Kuhn}}, \ and\ \bibinfo {author} {\bibfnamefont {P.~M.}\ \bibnamefont {Zerwas}},\ }\href {\doibase 10.1016/0370-2693(86)91275-X} {\bibfield  {journal} {\bibinfo  {journal} {Phys. Lett. B}\ }\textbf {\bibinfo {volume} {181}},\ \bibinfo {pages} {157} (\bibinfo {year} {1986})}\BibitemShut {NoStop}%
\bibitem [{\citenamefont {Braaten}\ \emph {et~al.}(2021)\citenamefont {Braaten}, \citenamefont {He},\ and\ \citenamefont {Mohapatra}}]{Braaten:2020nwp}%
  \BibitemOpen
  \bibfield  {author} {\bibinfo {author} {\bibfnamefont {E.}~\bibnamefont {Braaten}}, \bibinfo {author} {\bibfnamefont {L.-P.}\ \bibnamefont {He}}, \ and\ \bibinfo {author} {\bibfnamefont {A.}~\bibnamefont {Mohapatra}},\ }\href {\doibase 10.1103/PhysRevD.103.016001} {\bibfield  {journal} {\bibinfo  {journal} {Phys. Rev. D}\ }\textbf {\bibinfo {volume} {103}},\ \bibinfo {pages} {016001} (\bibinfo {year} {2021})},\ \Eprint {http://arxiv.org/abs/2006.08650} {arXiv:2006.08650 [hep-ph]} \BibitemShut {NoStop}%
\bibitem [{\citenamefont {Berwein}\ \emph {et~al.}(2024)\citenamefont {Berwein}, \citenamefont {Brambilla}, \citenamefont {Mohapatra},\ and\ \citenamefont {Vairo}}]{Berwein:2024ztx}%
  \BibitemOpen
  \bibfield  {author} {\bibinfo {author} {\bibfnamefont {M.}~\bibnamefont {Berwein}}, \bibinfo {author} {\bibfnamefont {N.}~\bibnamefont {Brambilla}}, \bibinfo {author} {\bibfnamefont {A.}~\bibnamefont {Mohapatra}}, \ and\ \bibinfo {author} {\bibfnamefont {A.}~\bibnamefont {Vairo}},\ }\href@noop {} {\  (\bibinfo {year} {2024})},\ \Eprint {http://arxiv.org/abs/2408.04719} {arXiv:2408.04719 [hep-ph]} \BibitemShut {NoStop}%
\bibitem [{\citenamefont {Bondar}\ \emph {et~al.}(2012)\citenamefont {Bondar} \emph {et~al.}}]{Belle:2011aa}%
  \BibitemOpen
  \bibfield  {author} {\bibinfo {author} {\bibfnamefont {A.}~\bibnamefont {Bondar}} \emph {et~al.} (\bibinfo {collaboration} {Belle}),\ }\href {\doibase 10.1103/PhysRevLett.108.122001} {\bibfield  {journal} {\bibinfo  {journal} {Phys. Rev. Lett.}\ }\textbf {\bibinfo {volume} {108}},\ \bibinfo {pages} {122001} (\bibinfo {year} {2012})},\ \Eprint {http://arxiv.org/abs/1110.2251} {arXiv:1110.2251 [hep-ex]} \BibitemShut {NoStop}%
\bibitem [{\citenamefont {Ablikim}\ \emph {et~al.}(2022)\citenamefont {Ablikim} \emph {et~al.}}]{BESIII:2022joj}%
  \BibitemOpen
  \bibfield  {author} {\bibinfo {author} {\bibfnamefont {M.}~\bibnamefont {Ablikim}} \emph {et~al.} (\bibinfo {collaboration} {(BESIII),, BESIII}),\ }\href {\doibase 10.1088/1674-1137/ac945c} {\bibfield  {journal} {\bibinfo  {journal} {Chin. Phys. C}\ }\textbf {\bibinfo {volume} {46}},\ \bibinfo {pages} {111002} (\bibinfo {year} {2022})},\ \Eprint {http://arxiv.org/abs/2204.07800} {arXiv:2204.07800 [hep-ex]} \BibitemShut {NoStop}%
\bibitem [{\citenamefont {Aaij}\ \emph {et~al.}(2017)\citenamefont {Aaij} \emph {et~al.}}]{LHCb:2016axx}%
  \BibitemOpen
  \bibfield  {author} {\bibinfo {author} {\bibfnamefont {R.}~\bibnamefont {Aaij}} \emph {et~al.} (\bibinfo {collaboration} {LHCb}),\ }\href {\doibase 10.1103/PhysRevLett.118.022003} {\bibfield  {journal} {\bibinfo  {journal} {Phys. Rev. Lett.}\ }\textbf {\bibinfo {volume} {118}},\ \bibinfo {pages} {022003} (\bibinfo {year} {2017})},\ \Eprint {http://arxiv.org/abs/1606.07895} {arXiv:1606.07895 [hep-ex]} \BibitemShut {NoStop}%
\bibitem [{\citenamefont {Aaij}\ \emph {et~al.}(2020)\citenamefont {Aaij} \emph {et~al.}}]{LHCb:2020bwg}%
  \BibitemOpen
  \bibfield  {author} {\bibinfo {author} {\bibfnamefont {R.}~\bibnamefont {Aaij}} \emph {et~al.} (\bibinfo {collaboration} {LHCb}),\ }\href {\doibase 10.1016/j.scib.2020.08.032} {\bibfield  {journal} {\bibinfo  {journal} {Sci. Bull.}\ }\textbf {\bibinfo {volume} {65}},\ \bibinfo {pages} {1983} (\bibinfo {year} {2020})},\ \Eprint {http://arxiv.org/abs/2006.16957} {arXiv:2006.16957 [hep-ex]} \BibitemShut {NoStop}%
\bibitem [{\citenamefont {Hayrapetyan}\ \emph {et~al.}(2024)\citenamefont {Hayrapetyan} \emph {et~al.}}]{CMS:2023owd}%
  \BibitemOpen
  \bibfield  {author} {\bibinfo {author} {\bibfnamefont {A.}~\bibnamefont {Hayrapetyan}} \emph {et~al.} (\bibinfo {collaboration} {CMS}),\ }\href {\doibase 10.1103/PhysRevLett.132.111901} {\bibfield  {journal} {\bibinfo  {journal} {Phys. Rev. Lett.}\ }\textbf {\bibinfo {volume} {132}},\ \bibinfo {pages} {111901} (\bibinfo {year} {2024})},\ \Eprint {http://arxiv.org/abs/2306.07164} {arXiv:2306.07164 [hep-ex]} \BibitemShut {NoStop}%
\bibitem [{\citenamefont {Eichten}\ and\ \citenamefont {Liu}(2017)}]{Eichten:2017ual}%
  \BibitemOpen
  \bibfield  {author} {\bibinfo {author} {\bibfnamefont {E.}~\bibnamefont {Eichten}}\ and\ \bibinfo {author} {\bibfnamefont {Z.}~\bibnamefont {Liu}},\ }\href@noop {} {\  (\bibinfo {year} {2017})},\ \Eprint {http://arxiv.org/abs/1709.09605} {arXiv:1709.09605 [hep-ph]} \BibitemShut {NoStop}%
\bibitem [{\citenamefont {Esposito}\ \emph {et~al.}(2021)\citenamefont {Esposito}, \citenamefont {Manzari}, \citenamefont {Pilloni},\ and\ \citenamefont {Polosa}}]{Esposito:2021ptx}%
  \BibitemOpen
  \bibfield  {author} {\bibinfo {author} {\bibfnamefont {A.}~\bibnamefont {Esposito}}, \bibinfo {author} {\bibfnamefont {C.~A.}\ \bibnamefont {Manzari}}, \bibinfo {author} {\bibfnamefont {A.}~\bibnamefont {Pilloni}}, \ and\ \bibinfo {author} {\bibfnamefont {A.~D.}\ \bibnamefont {Polosa}},\ }\href {\doibase 10.1103/PhysRevD.104.114029} {\bibfield  {journal} {\bibinfo  {journal} {Phys. Rev. D}\ }\textbf {\bibinfo {volume} {104}},\ \bibinfo {pages} {114029} (\bibinfo {year} {2021})},\ \Eprint {http://arxiv.org/abs/2109.10359} {arXiv:2109.10359 [hep-ph]} \BibitemShut {NoStop}%
\bibitem [{\citenamefont {Vega-Morales}\ and\ \citenamefont {Vega-Morales}(2017)}]{Vega-Morales:2017pmm}%
  \BibitemOpen
  \bibfield  {author} {\bibinfo {author} {\bibfnamefont {R.}~\bibnamefont {Vega-Morales}}\ and\ \bibinfo {author} {\bibfnamefont {R.}~\bibnamefont {Vega-Morales}},\ }\href@noop {} {\  (\bibinfo {year} {2017})},\ \Eprint {http://arxiv.org/abs/1710.02738} {arXiv:1710.02738 [hep-ph]} \BibitemShut {NoStop}%
\bibitem [{\citenamefont {Manohar}\ and\ \citenamefont {Wise}(1993)}]{Manohar:1992nd}%
  \BibitemOpen
  \bibfield  {author} {\bibinfo {author} {\bibfnamefont {A.~V.}\ \bibnamefont {Manohar}}\ and\ \bibinfo {author} {\bibfnamefont {M.~B.}\ \bibnamefont {Wise}},\ }\href {\doibase 10.1016/0550-3213(93)90614-U} {\bibfield  {journal} {\bibinfo  {journal} {Nucl. Phys. B}\ }\textbf {\bibinfo {volume} {399}},\ \bibinfo {pages} {17} (\bibinfo {year} {1993})},\ \Eprint {http://arxiv.org/abs/hep-ph/9212236} {arXiv:hep-ph/9212236} \BibitemShut {NoStop}%
\bibitem [{\citenamefont {Eichten}\ and\ \citenamefont {Quigg}(2017)}]{Eichten:2017ffp}%
  \BibitemOpen
  \bibfield  {author} {\bibinfo {author} {\bibfnamefont {E.~J.}\ \bibnamefont {Eichten}}\ and\ \bibinfo {author} {\bibfnamefont {C.}~\bibnamefont {Quigg}},\ }\href {\doibase 10.1103/PhysRevLett.119.202002} {\bibfield  {journal} {\bibinfo  {journal} {Phys. Rev. Lett.}\ }\textbf {\bibinfo {volume} {119}},\ \bibinfo {pages} {202002} (\bibinfo {year} {2017})},\ \Eprint {http://arxiv.org/abs/1707.09575} {arXiv:1707.09575 [hep-ph]} \BibitemShut {NoStop}%
\bibitem [{\citenamefont {Beneke}(1999)}]{Beneke:1998ui}%
  \BibitemOpen
  \bibfield  {author} {\bibinfo {author} {\bibfnamefont {M.}~\bibnamefont {Beneke}},\ }\href {\doibase 10.1016/S0370-1573(98)00130-6} {\bibfield  {journal} {\bibinfo  {journal} {Phys. Rept.}\ }\textbf {\bibinfo {volume} {317}},\ \bibinfo {pages} {1} (\bibinfo {year} {1999})},\ \Eprint {http://arxiv.org/abs/hep-ph/9807443} {arXiv:hep-ph/9807443} \BibitemShut {NoStop}%
\bibitem [{\citenamefont {Beneke}(1998)}]{Beneke:1998rk}%
  \BibitemOpen
  \bibfield  {author} {\bibinfo {author} {\bibfnamefont {M.}~\bibnamefont {Beneke}},\ }\href {\doibase 10.1016/S0370-2693(98)00741-2} {\bibfield  {journal} {\bibinfo  {journal} {Phys. Lett. B}\ }\textbf {\bibinfo {volume} {434}},\ \bibinfo {pages} {115} (\bibinfo {year} {1998})},\ \Eprint {http://arxiv.org/abs/hep-ph/9804241} {arXiv:hep-ph/9804241} \BibitemShut {NoStop}%
\bibitem [{\citenamefont {Brambilla}\ \emph {et~al.}(2018)\citenamefont {Brambilla}, \citenamefont {Komijani}, \citenamefont {Kronfeld},\ and\ \citenamefont {Vairo}}]{Brambilla:2017hcq}%
  \BibitemOpen
  \bibfield  {author} {\bibinfo {author} {\bibfnamefont {N.}~\bibnamefont {Brambilla}}, \bibinfo {author} {\bibfnamefont {J.}~\bibnamefont {Komijani}}, \bibinfo {author} {\bibfnamefont {A.~S.}\ \bibnamefont {Kronfeld}}, \ and\ \bibinfo {author} {\bibfnamefont {A.}~\bibnamefont {Vairo}} (\bibinfo {collaboration} {TUMQCD}),\ }\href {\doibase 10.1103/PhysRevD.97.034503} {\bibfield  {journal} {\bibinfo  {journal} {Phys. Rev. D}\ }\textbf {\bibinfo {volume} {97}},\ \bibinfo {pages} {034503} (\bibinfo {year} {2018})},\ \Eprint {http://arxiv.org/abs/1712.04983} {arXiv:1712.04983 [hep-ph]} \BibitemShut {NoStop}%
\bibitem [{\citenamefont {Kronfeld}(2024)}]{Kronfeld:2024qao}%
  \BibitemOpen
  \bibfield  {author} {\bibinfo {author} {\bibfnamefont {A.~S.}\ \bibnamefont {Kronfeld}},\ }\href {\doibase 10.22323/1.453.0341} {\bibfield  {journal} {\bibinfo  {journal} {PoS}\ }\textbf {\bibinfo {volume} {LATTICE2023}},\ \bibinfo {pages} {341} (\bibinfo {year} {2024})},\ \Eprint {http://arxiv.org/abs/2401.07380} {arXiv:2401.07380 [hep-ph]} \BibitemShut {NoStop}%
\bibitem [{\citenamefont {Pineda}\ and\ \citenamefont {Soto}(2000)}]{Pineda:2000gza}%
  \BibitemOpen
  \bibfield  {author} {\bibinfo {author} {\bibfnamefont {A.}~\bibnamefont {Pineda}}\ and\ \bibinfo {author} {\bibfnamefont {J.}~\bibnamefont {Soto}},\ }\href {\doibase 10.1016/S0370-2693(00)01261-2} {\bibfield  {journal} {\bibinfo  {journal} {Phys. Lett. B}\ }\textbf {\bibinfo {volume} {495}},\ \bibinfo {pages} {323} (\bibinfo {year} {2000})},\ \Eprint {http://arxiv.org/abs/hep-ph/0007197} {arXiv:hep-ph/0007197} \BibitemShut {NoStop}%
\end{thebibliography}%

\clearpage

\end{document}